\documentclass[a4paper,10pt]{article}
\pdfoutput=1
\usepackage{jheppub}
\usepackage{slashed}
\usepackage[dvipsnames,table]{xcolor}
\usepackage{hyperref} 
\usepackage{xspace}
\usepackage[tight]{subfigure}
\usepackage{amsmath}
\usepackage{amssymb}
\usepackage{amsfonts}
\usepackage{mathrsfs}
\usepackage{comment}
\usepackage{afterpage}
\usepackage{verbatim}
\usepackage{booktabs}
\usepackage{array}
\usepackage[title,titletoc]{appendix}
\usepackage{tabularx}
\newcolumntype{C}[1]{>{\centering\arraybackslash}p{#1}}

\def\refeq#1{\mbox{(\ref{#1})}}
\def\reffi#1{\mbox{Figure~\ref{#1}}}
\def\reffis#1{\mbox{Figures~\ref{#1}}}
\def\refta#1{\mbox{Table~\ref{#1}}}

\def\refse#1{\mbox{Section~\ref{#1}}}
\def\refses#1{\mbox{Sections~\ref{#1}}}

\def\citere#1{\mbox{Ref.~\cite{#1}}}
\def\citeres#1{\mbox{Refs.~\cite{#1}}}

\newcommand{\newc}{\newcommand}
\newc{\beq}{\begin{equation}}
\newc{\eeq}{\end{equation}}
\newc{\bit}{\begin{itemize}}
\newc{\eit}{\end{itemize}}
\newc{\ben}{\begin{enumerate}}
\newc{\een}{\end{enumerate}}
\newc{\bce}{\begin{center}}
\newc{\ece}{\end{center}}
\newc{\bfi}{\begin{figure}}
\newc{\efi}{\end{figure}}

\newcommand{\ri}{\mathrm i}

\newcommand{\rd}{\mathrm d}

\newcommand{\rT}{{\mathrm{T}}}
\newcommand{\rR}{{\mathrm{R}}}
\newcommand{\rL}{{\mathrm{L}}}
\newcommand{\rF}{{\mathrm{F}}}

\newcommand{\ie}{\emph{i.e.}\ }
\newcommand{\eg}{\emph{e.g.}\ }

\newcommand{\GeV}{\ensuremath{\,\text{GeV}}\xspace}

\newcommand{\PH}{\ensuremath{\text{H}}\xspace}

\newcommand{\Pp}{\ensuremath{\text{p}}}
\newcommand{\Pe}{\ensuremath{\text{e}}\xspace}
\newcommand{\Pb}{\ensuremath{\text{b}}\xspace}
\newcommand{\Pq}{\ensuremath{q}}
\newcommand{\Pt}{\ensuremath{\text{t}}\xspace}

\newcommand{\Pg}{\ensuremath{\text{g}}}

\newcommand{\PW}{\ensuremath{\text{W}}\xspace}
\newcommand{\PZ}{\ensuremath{\text{Z}}\xspace}

\newcommand{\Mt}{\ensuremath{m_\Pt}\xspace}
\newcommand{\MH}{\ensuremath{M_\PH}\xspace}
\newcommand{\MWOS}{\ensuremath{M_\PW^\text{OS}}\xspace}
\newcommand{\MW}{\ensuremath{M_\PW}\xspace}
\newcommand{\MZOS}{\ensuremath{M_\PZ^\text{OS}}\xspace}
\newcommand{\MZ}{\ensuremath{M_\PZ}\xspace}

\newcommand{\GZ}{\ensuremath{\Gamma_\PZ}\xspace}
\newcommand{\GZOS}{\ensuremath{\Gamma_\PZ^\text{OS}}\xspace}

\newcommand{\GWOS}{\ensuremath{\Gamma_\PW^\text{OS}}\xspace}

\newcommand{\GF}{\ensuremath{G_\mu}}

\newcommand{\sw}{s_{\mathrm{w}}}
\newcommand{\cw}{c_{\mathrm{w}}}

\newcommand{\MVOS}{\ensuremath{M_{V}^\text{OS}}\xspace}%
\newcommand{\GVOS}{\ensuremath{\Gamma_{V}^\text{OS}}\xspace}%

\newcommand{\recola}{{\sc Recola}\xspace}

\newcommand{\mocanlo}{{\sc MoCaNLO}\xspace}
\newcommand{\collier}{{\sc Collier}\xspace}

\newcommand{\madgraph}{{\sc\small MadGraph}\xspace}

\newcommand{\phantommc}{{\sc\small PHANTOM}\xspace}

\newcolumntype{.}{D{.}{.}{-1}}
\newcolumntype{d}[1]{D{.}{.}{#1}}
\colorlet{tableoverheadcolor}{gray!37.5}
\colorlet{tableheadcolor}{gray!25}
\colorlet{tablerowcolor}{gray!12.5}

\def\draftdate{\relax}
\def\mda{\relax}
\def\mua{\relax}
\def\mla{\relax}
\def\draft{
\def\thtystars{******************************}
\def\sixtystars{\thtystars\thtystars}
\typeout{}
\typeout{\sixtystars**}
\typeout{* Draft mode!
         For final version remove \protect\draft\space in source file *}
\typeout{\sixtystars**}
\typeout{}
\def\draftdate{\today}
\def\mua{\marginpar[\boldmath\hfil$\uparrow$]%
                   {\boldmath$\uparrow$\hfil}\color{black}%
                    \typeout{marginpar: $\uparrow$}\ignorespaces}
\def\mda{\color{red}\marginpar[\boldmath\hfil$\downarrow$]%
                   {\boldmath$\downarrow$\hfil}%
                    \typeout{marginpar: $\downarrow$}\ignorespaces}
\def\mla{\marginpar[\boldmath\hfil$\rightarrow$]%
                   {\boldmath$\leftarrow $\hfil}%
                    \typeout{marginpar: $\leftrightarrow$}\ignorespaces}
\def\Mua{\marginpar[\boldmath\hfil$\Uparrow$]%
                   {\boldmath$\Uparrow$\hfil}\color{black}%
                    \typeout{marginpar: $\uparrow$}\ignorespaces}
\def\Mda{\color{red}\marginpar[\boldmath\hfil$\Downarrow$]%
                   {\boldmath$\Downarrow$\hfil}%
                    \typeout{marginpar: $\downarrow$}\ignorespaces}
\def\Mla{\marginpar[\boldmath\hfil\textcolor{red}{$\Rightarrow$}]%
                   {\boldmath\textcolor{red}{$\Leftarrow $}\hfil}%
                    \typeout{marginpar: $\leftrightarrow$}\ignorespaces}
\overfullrule 5pt
\oddsidemargin 15mm
\marginparwidth 29mm
}

\newcommand{\mc}{\mathcal}
\newcommand{\as}{\alpha_{\textrm{s}}}
\newcommand{\pt}[1]{p_{\rT,{#1}}}
\newcommand{\stiny}[1]{{\scalebox{0.6}{$#1$}}}
\newcommand{\nnb}{\nonumber}
\newcommand{\NLOew}{NLO$_{\rm EW}$\xspace}
\newcommand{\dpatwotwo}{$\rm DPA^{(2,2)}$\xspace}
\newcommand{\dpathreetwo}{$\rm DPA^{(3,2)}$\xspace}
\newcommand{\dpatwothree}{$\rm DPA^{(2,3)}$\xspace}
\newcommand{\tl}{\theta^*_{\ell^+}}

\title{NLO EW and QCD corrections to polarized $\PZ\PZ$~production in the
  four-charged-lepton channel at the LHC} 
\author{Ansgar Denner and}
\author{Giovanni Pelliccioli}
\affiliation{Universit\"at W\"urzburg, Institut f\"ur Theoretische Physik und Astrophysik, 97074 W\"urzburg, Germany}
\emailAdd{ansgar.denner@physik.uni-wuerzburg.de}
\emailAdd{giovanni.pelliccioli@physik.uni-wuerzburg.de}

\abstract{
  Measuring the polarization of electroweak bosons at the LHC allows for important tests of the
  electroweak-symmetry-breaking mechanism that is realized in nature. Therefore,
  precise Standard Model predictions are needed for the production of polarized
  bosons in the presence of realistic kinematic selections.
  We formulate a method for the calculation of polarized
  cross-sections at NLO that relies
  on the pole approximation and the separation of polarized matrix elements at
  the amplitude level. 
  In this framework, we compute NLO-accurate cross-sections for the
  production of two polarized $\PZ$~bosons at the LHC, including 
  for the first time NLO EW corrections and combining them with NLO QCD corrections and
  contributions from the gluon-induced process.
}


\begin{document}

\strut\hfill

\maketitle
\section{Introduction}
The production of four charged leptons at the LHC 
provides an optimal framework to study $\PZ\PZ$~production as well as to investigate Higgs-boson decays, both with
the aim of probing the Standard Model (SM) of fundamental interactions and with the purpose of searching
for new-physics effects.
The presence of four charged leptons in the final state renders this signature one of the cleanest at the LHC,
\ie it features a high signal-to-background ratio, and this motivates the great interest
both from the experimental and from the theoretical side in investigating this process.

Four-charged-lepton production has been measured by ATLAS and CMS with
13-TeV data to constrain the decay width of the Higgs boson
\cite{Aaboud:2018puo,Sirunyan:2019twz}
and its anomalous couplings to SM particles
\cite{Sirunyan:2017tqd,Sirunyan:2019twz,ATLAS:2020wny}.
The production of $\PZ$~pairs in the four-charged-lepton channel
has been employed to measure the SM cross-section
\cite{Khachatryan:2016txa,Sirunyan:2017zjc}
and to constrain anomalous gauge-boson couplings
\cite{Aaboud:2017rwm,Sirunyan:2017zjc,Sirunyan:2020pub}.
The same channel has been used to measure the four-lepton decay of a single $\PZ$~boson
\cite{Khachatryan:2016txa,Sirunyan:2017zjc}.
A complete study of the full spectrum of the four-lepton invariant mass has been performed,
including the measurement of single-$\PZ$ and Higgs production, $\PZ\PZ$ on-shell production and
interference effects \cite{Aaboud:2019lxo,ATLAS:2021kog}.

The next-to-leading-order (NLO) QCD corrections to the production of $\PZ$-boson pairs have been known
for almost thirty years \cite{Mele:1990bq,Ohnemus:1990za,Ohnemus:1994ff,Dixon:1998py,Dixon:1999di}.
The NLO electroweak (EW) corrections have been calculated for on-shell \cite{Accomando:2004de,Bierweiler:2013dja}
and for off-shell \PZ~bosons in the four-charged-lepton channel \cite{Biedermann:2016yvs,Biedermann:2016lvg}.
NLO EW and QCD corrections have been combined in the context of the SM
\cite{Baglio:2013toa,Kallweit:2017khh} and in the presence of anomalous couplings \cite{Chiesa:2018lcs}.
The next-to-next-to-leading-order (NNLO) QCD corrections to the production of four leptons have been recently computed
\cite{Cascioli:2014yka,Grazzini:2015hta,Heinrich:2017bvg,Kallweit:2018nyv}, and
studies have been dedicated to the NLO QCD corrections to the loop-induced gluon--gluon
channel \cite{Caola:2015psa,Caola:2016trd,Grazzini:2018owa,Grazzini:2021iae}.
The combination of state-of-the-art perturbative corrections has been performed,
including NNLO QCD and NLO EW corrections to off-shell $\PZ\PZ$ production \cite{Kallweit:2019zez}.
The matching to a parton shower (PS) has been achieved at NLO QCD for
the quark-induced \cite{Melia:2011tj,Nason:2013ydw}  and for the gluon-induced channel \cite{Alioli:2016xab,Alioli:2021wpn}.
Very recently the complete NLO corrections have been matched to QCD and QED PS \cite{Chiesa:2020ttl},
and the NNLO QCD corrections have been matched to QCD PS \cite{Alioli:2021egp}.

Thanks to the possibility of reconstructing all particles in the final state,
the production of four charged leptons provides a large number of angular
observables that are sensitive to the spin structure of the underlying resonances,
both in the case of the Higgs and single-$\PZ$~decay into four leptons, and
in the case of the leptonic decay of two $\PZ$~bosons.
In fact, the Higgs decay into four leptons has been widely studied
\cite{Soni:1993jc,Chang:1993jy,Skjold:1993jd,Arens:1994wd,Buszello:2002uu,
  Choi:2002jk,Berge:2015jra,Hagiwara:2009wt,Gao:2010qx,Bolognesi:2012mm}
and then measured with Run-2 data
 \cite{Sirunyan:2017tqd,Sirunyan:2017exp,Aaboud:2017vzb,Aaboud:2017oem,Aaboud:2018puo,Sirunyan:2019twz,Aad:2020mkp,ATLAS:2020wny} 
with the purpose of determining the spin and parity of the Higgs boson
as well as its CP properties via angular and energy correlations.

For the process $\Pp\Pp\to\PZ\PZ\to\Pe^+\Pe^-\mu^+\mu^-$, the
possibility to fully reconstruct the final state allows
to access observables that are sensitive to the polarization of $\PZ$~bosons.
This can be done in the Higgs-resonance region
\cite{Bruni:2019xwu,Maina:2020rgd,Maina:2021xpe}, where
the polarization of the weak bosons may give hints of new-physics effects
in the coupling of the Higgs to longitudinally- or transversely-polarized bosons.
The polarization structure can also be studied in the so-called on-shell region \cite{Bierweiler:2013dja},
where two $\PZ$~bosons are produced and then decay into charged lepton pairs:
this is the target of this paper. 

The polarization of $\PW$ and $\PZ$~bosons has been measured at the LHC
in $\PW+\rm jet$ and $\PZ+\rm jet$ production
\cite{Chatrchyan:2011ig,ATLAS:2012au,Khachatryan:2015paa,Aad:2016izn}
and in the decay of top quarks
\cite{ATLAS:2016fbc,Khachatryan:2016fky,CMS:2020ezf}
with Run-1 data, and more recently also in $\PW\PZ$ production \cite{Aaboud:2019gxl}
and in same-sign $\PW\PW$ scattering \cite{Sirunyan:2020gvn} with the 13-TeV data.
The upcoming high-luminosity LHC run is expected to enhance the sensitivity to
the polarization of $\PZ$~bosons produced in $\PZ\PZ$ inclusive production and
scattering \cite{CMS:2018mbt,Azzi:2019yne}.

The increasing experimental interest in measuring
the weak-boson polarization in multi-boson processes has triggered a number
of phenomenological studies for the LHC environment. The seminal works concerned the production
of weak bosons in association with jets \cite{Bern:2011ie}
and a number of other processes \cite{Stirling:2012zt}.
The effect of lepton cuts on polarization extraction was highlighted in
\citeres{Bern:2011ie,Belyaev:2013nla}. More recently, a study of polarization-sensitive
coefficients has been performed for $\PW^\pm\PZ$ inclusive production \cite{Baglio:2018rcu,Baglio:2019nmc}.
The separation of polarization states at the amplitude level at leading order (LO)
has been automated and investigated in vector-boson scattering
\cite{Ballestrero:2017bxn,Ballestrero:2019qoy,Ballestrero:2020qgv} with the \phantommc Monte
Carlo \cite{Ballestrero:2007xq} and in Higgs decays to four leptons \cite{Maina:2020rgd,Maina:2021xpe}.
The generation of polarized events in general multi-boson processes
is available at LO in the \madgraph framework \cite{Alwall:2014hca}
via the spin-correlated narrow-width approximation
\cite{Uhlemann:2008pm,Artoisenet:2012st,BuarqueFranzosi:2019boy}.
The separation of polarized signals at the amplitude level has been extended to
NLO QCD \cite{Denner:2020bcz,Denner:2020eck} and to NNLO QCD \cite{Poncelet:2021jmj}
for di-boson production in the fully-leptonic channel.
The growing interest in the polarization structure of di-boson production
and the increased sensitivity to polarization observables have
motivated us to extend these polarization studies to NLO EW.

The aim of this work is twofold. On the one hand, we present a
general, well-behaved
definition of polarized $\PZ\PZ$ signals (both bosons have definite polarization state)
including the decay modelling at NLO EW accuracy,
which is achieved via the extension of the double-pole approximation
(DPA) used in
\citeres{Denner:2020bcz,Denner:2020eck} to real radiation from leptonic decay products,
and via the separation of polarization at the amplitude level in all parts of the
calculation.
On the other hand, we investigate the LHC phenomenology of polarized $\PZ\PZ$
production including all LO and NLO contributions (both of QCD and of EW origin)
in the presence of realistic kinematic selections, in order to address possible
experimental analyses in this di-boson channel.

This paper is organized as follows. In \refse{sec:details}, we point out
the main difficulties that arise when separating resonant $\PZ\PZ$ contributions
from the non-doubly-resonant ones at NLO, and we describe in detail (\refses{subsec:nloqcd}
and \ref{subsec:nloew}) our method to achieve this goal. The separation of
polarizations is described in \refse{subsec:pol}. The SM
input parameters and the kinematic setups that are used in the numerical simulations are
described in \refses{subsec:input} and \ref{subsec:setup}, respectively.
The integrated and differential results for polarized $\PZ\PZ$ production
at NLO EW are discussed in \refse{subsec:inc} for an inclusive setup.
The combined NLO QCD and EW corrections, as well as the loop-induced gluonic
corrections to the polarization structure of $\PZ\PZ$ production are 
presented in \refse{subsec:fid} for a fiducial region that mimics the latest
experimental selections for this process. In \refse{sub:con} we draw our conclusions.

\section{Calculation details}\label{sec:details}
\begin{figure}
  \centering
    \includegraphics[scale=0.48]{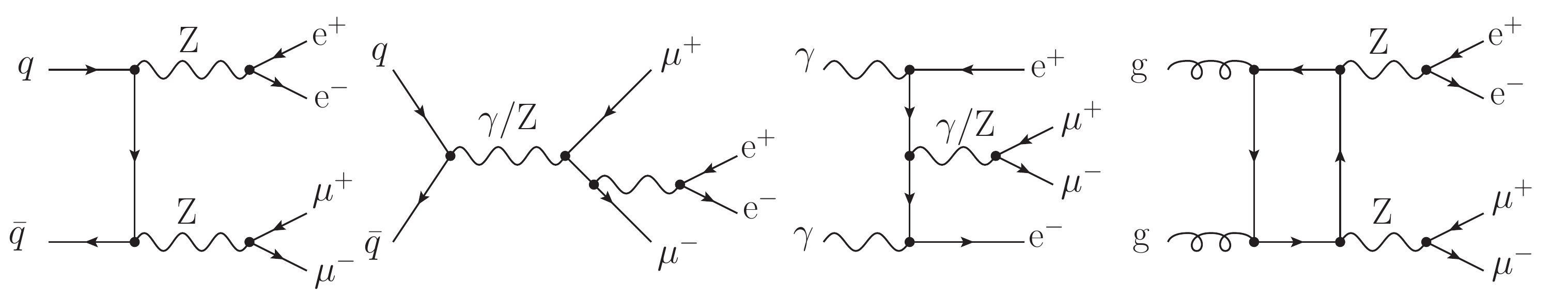}   
  \caption{
    Sample LO Feynman diagrams for four-charged-lepton production. From left to right: $\PZ\PZ$-resonant diagram in the
    $q\bar{q}$ channel, non-doubly-resonant diagram in the $q\bar{q}$ channel, non-doubly-resonant diagram in the
    $\gamma\gamma$ channel, $\PZ\PZ$-resonant loop-induced diagram in the $\Pg\Pg$ channel.
  }\label{fig:diags}
\end{figure}

We consider the production of a $\PZ\PZ$ pair in the fully-leptonic channel,
\beq
\Pp\Pp\rightarrow\Pe^+\Pe^-\mu^+\mu^-\,+\,X\,,
\eeq
at NLO accuracy, including both QCD and EW radiative corrections.
At tree-level [$\mc O(\alpha^4)$] only quark--antiquark and
photon--photon partonic channels are present,
while at NLO both photon--quark and gluon--quark channels contribute to the
real radiation at $\mc O(\alpha^5)$ and $\mc O(\alpha^4\as)$, respectively.
Since the final state is charge neutral, the loop-induced gluon--gluon partonic channel [formally of order
$\mc O(\alpha^4\as^2)$] gives a non-negligible contribution to this process thanks to large gluon luminosity
in the proton. Sample LO diagrams are shown in \reffi{fig:diags}.
Part of LO contributions to the $q\bar{q}$ and $\Pg\Pg$ channels
involve two $\PZ$~bosons that are produced and then separately decay
into charged lepton pairs:
we refer to these contributions as {\em doubly resonant} or {\em $\PZ\PZ$ resonant}.
However, the full SM amplitude receives contributions also from diagrams that
do not allow for the interpretation in terms of production and decay of two $\PZ$~bosons:
we refer to them as \emph{non-doubly-resonant}.
In order to separate the polarization modes of $\PZ$~bosons, only $\PZ\PZ$-resonant diagrams must be selected.
Nonetheless, this selection cannot be performed simply dropping non-doubly-resonant diagrams,
because the EW gauge invariance would be lost. In order to recover gauge invariance,
we use a pole-approximation technique
\cite{Bardin:1988xt,Stuart:1991cc,Stuart:1991xk,Denner:2005fg,Denner:2000bj,Denner:2019vbn},
whose details are discussed in \refses{subsec:nloqcd} and \ref{subsec:nloew}.
Another method that is often used to simulate the production and decay of unstable
particles is provided by the narrow-width approximation
\cite{Uhlemann:2008pm,Artoisenet:2012st,BuarqueFranzosi:2019boy,Poncelet:2021jmj}.

In the $\gamma\gamma$ partonic channel, at most one $s$-channel $\PZ$~propagator can be
present in the SM amplitude. Therefore, this partonic process cannot contribute
to the $\PZ\PZ$-resonant production neither at LO nor at NLO EW.
It contributes to the full calculation only, and we have
computed it including also its complete NLO EW corrections.

\begin{figure}
  \centering  
    \includegraphics[scale=0.5]{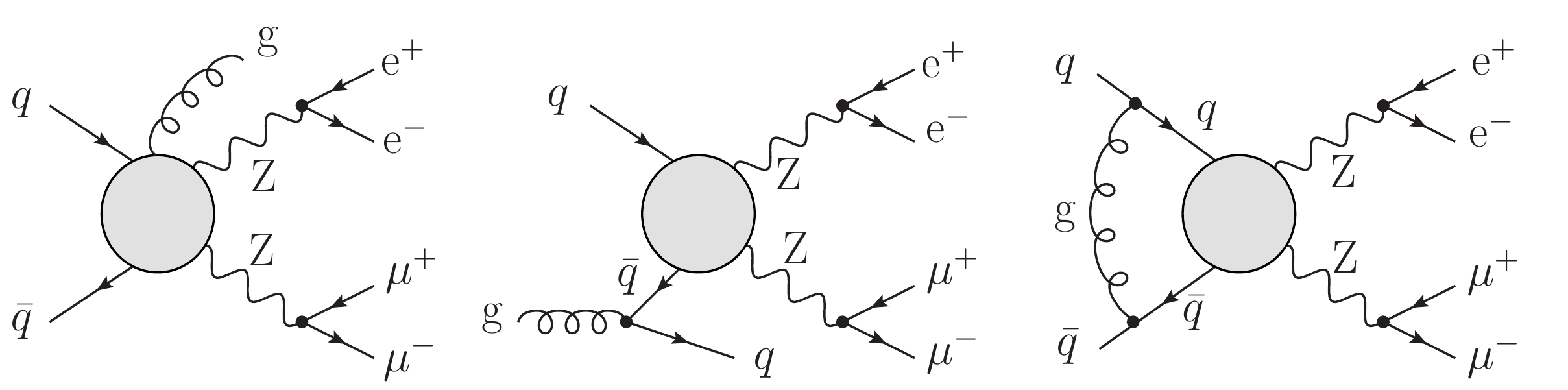} 
  \caption{   
    Sample tree-level and one-loop contributions
    to NLO QCD real and virtual corrections to
    resonant $\PZ\PZ$ production. The shaded circle denotes
    arbitrary tree-level sub-diagrams.
  }\label{fig:diagsQCD}
\end{figure}

The selection of $\PZ\PZ$-resonant diagrams and a prescription to recover gauge invariance
is needed also for NLO radiative corrections of virtual and real origin.
Since we consider a final state with colour-less particles, both real and virtual QCD
corrections of order $\mc O(\alpha^4\as)$ only modify the structure of the initial state.
This enables a simple separation between doubly-resonant and non-doubly-resonant contributions, exactly
in the same fashion as at LO. Sample real and virtual diagrams for QCD
corrections to resonant $\PZ\PZ$
production are shown in \reffi{fig:diagsQCD}.
The details of the DPA used for NLO QCD corrections
are described in \refse{subsec:nloqcd}.

The NLO EW corrections of order $\mc O(\alpha^5)$ to the production of
four charged leptons at the LHC are more involved than the QCD ones, as
both real and virtual contributions modify the structure of the initial
state as well as the structure of the final state, allowing for
additional EW propagators (either on-shell or off-shell) connecting the
initial and the final state. This non-factorized structure implies a
more involved selection of $\PZ\PZ$-resonant diagrams, which is needed for the
separation of polarization states.

\begin{figure}
  \centering
  \subfigure[Box correction to the $\PZ\PZ$ production process\label{fig:Visr}]{\includegraphics[scale=0.37]{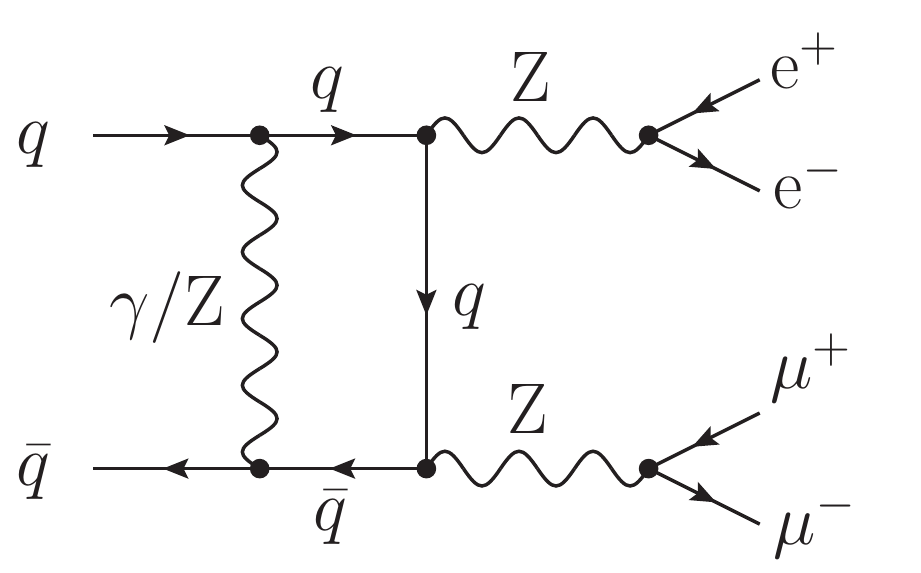}}\quad
  \subfigure[Vertex correction to the decay $\PZ\rightarrow \Pe^+\Pe^-$\label{fig:Vfsr}]{\includegraphics[scale=0.37]{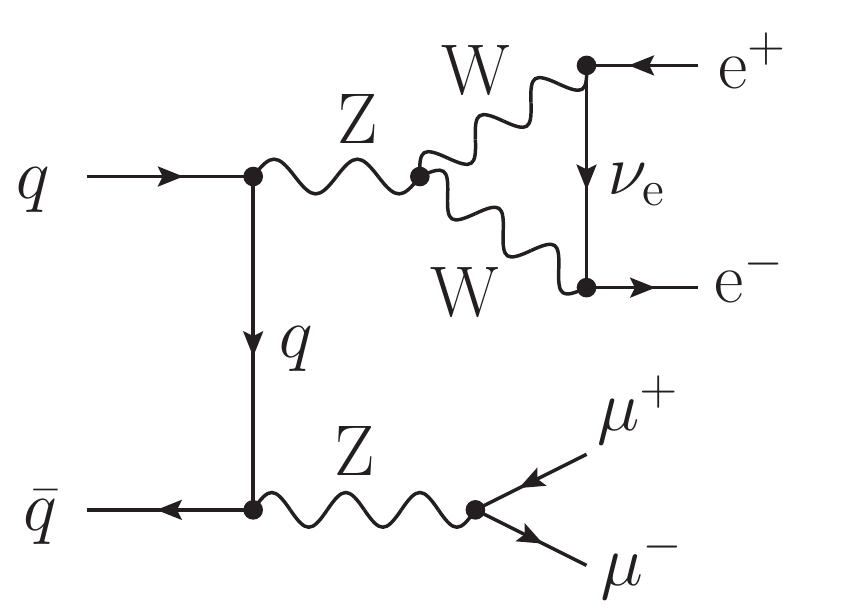}}\quad
  \subfigure[Initial--final non-fac\-to\-ri\-zable correction\label{fig:Vnfif}]{\includegraphics[scale=0.37]{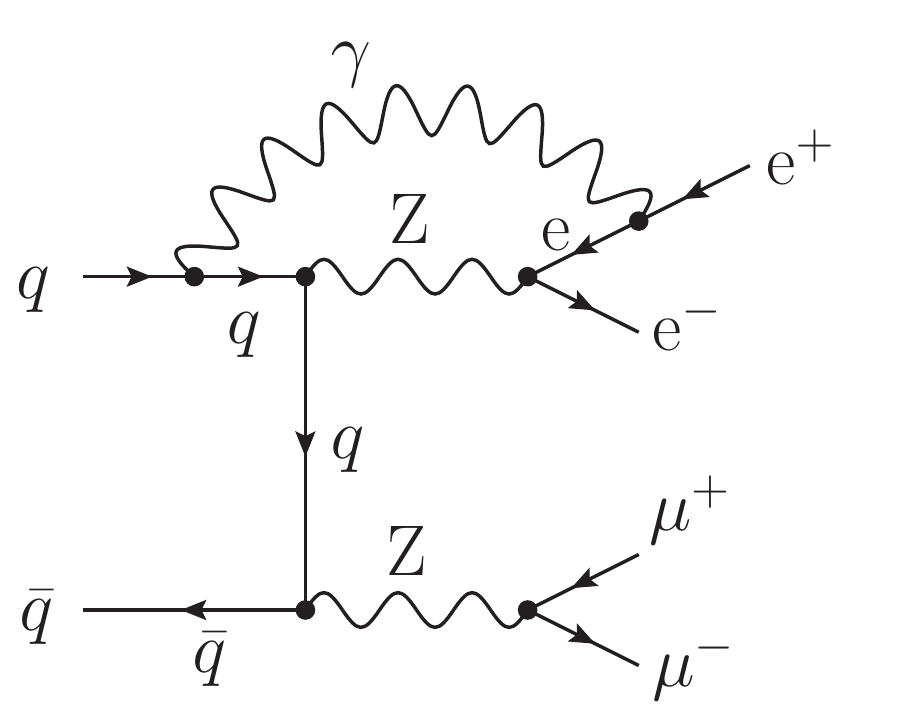}}\quad
  \subfigure[Final--final non-fac\-to\-r\-izable correction\label{fig:Vnfff}]{\includegraphics[scale=0.37]{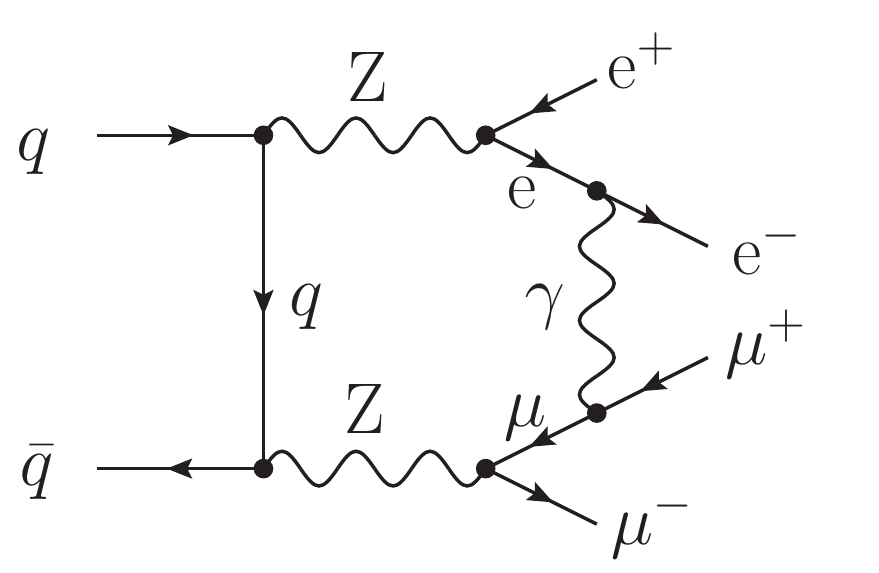}}
  \caption{
    Sample one-loop diagrams for four-charged-lepton production
    in the $q\bar{q}$ partonic channel at NLO EW. 
  }\label{fig:diagsVew}
\end{figure}

Sample one-loop diagrams are shown in \reffi{fig:diagsVew}.
In order to extract $\PZ\PZ$-resonant contributions, only loop diagrams that can be interpreted
as production and decay of two $\PZ$~bosons, called \emph{factorizable}, must be selected.
They include both corrections
to the $\PZ\PZ$~production process, like the one of \reffi{fig:Visr}, and corrections to the two
decays $\PZ\rightarrow \ell^+\ell^-$, $\ell=\Pe,\mu$, like the one in \reffi{fig:Vfsr}.

Also contributions with a virtual soft photon connecting either initial- and final-state
particles [\reffi{fig:Vnfif}], or two final-state particles coming from different
$\PZ$-boson decays [\reffi{fig:Vnfff}], called
\emph{non-factorizable}, are doubly resonant.
In fact, such corrections are characterized by a universal structure that factorizes
the doubly-resonant LO squared amplitude \cite{Denner:1997ia,Beenakker:1997ir,Dittmaier:2015bfe}.
However, the impact of these corrections is expected to be small thanks to cancellations
against the corresponding non-factorizable real-radiation
contributions \cite{Denner:1997ia,Beenakker:1997ir}. Therefore,
we exclude both virtual and real non-factorizable corrections from the DPA calculation.

All other one-loop, non-doubly-resonant contributions need to be dropped.
The virtual EW corrections to the $\gamma\gamma$ partonic channel involve
doubly-resonant contributions (box and triangle fermion-loop diagrams),
which however need to be interfered with tree-level diagrams which
can only be non doubly resonant. Therefore, these loop diagrams with a $\gamma\gamma$ initial state
enter the DPA calculation only at order $\mc O(\alpha^6)$, in the
same fashion as the $\Pg\Pg$ channel gives doubly-resonant contributions at order $\mc O(\alpha^4\as^2)$.

\begin{figure}
  \centering  
    \subfigure[Initial-state radiation \hfil\strut\hfil (ISR)\label{fig:Risr}]{\includegraphics[scale=0.5]{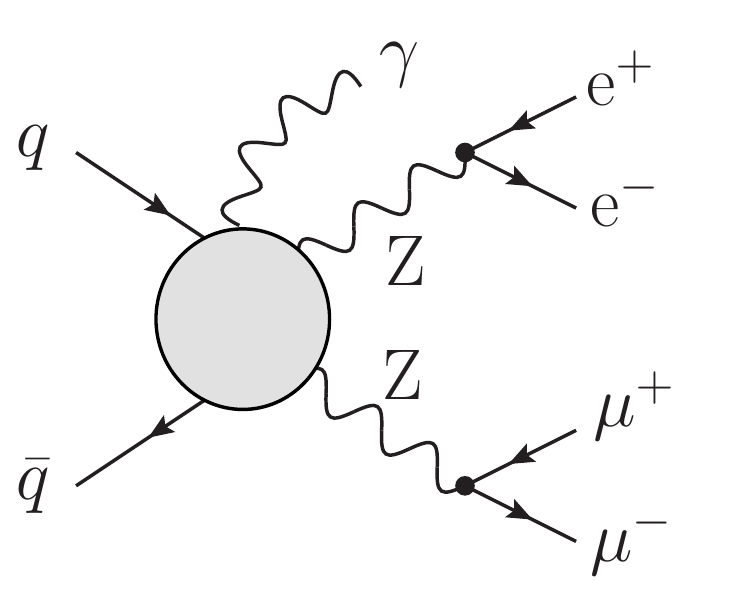}}\quad
    \subfigure[Final-state radiation from \hfil\strut\hfil $\PZ\rightarrow \Pe^+\Pe^-$ (FSR1)\label{fig:Rfsr1}]{\includegraphics[scale=0.5]{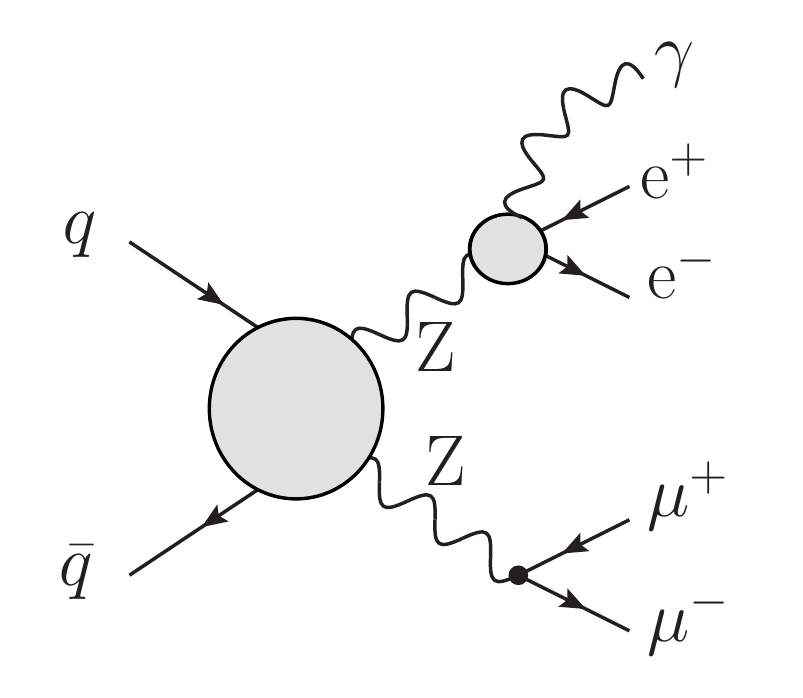}}\quad
    \subfigure[Final-state radiation from  \hfil\strut\hfil $\PZ\rightarrow \mu^+\mu^-$ (FSR2)\label{fig:Rfsr2}]{\includegraphics[scale=0.5]{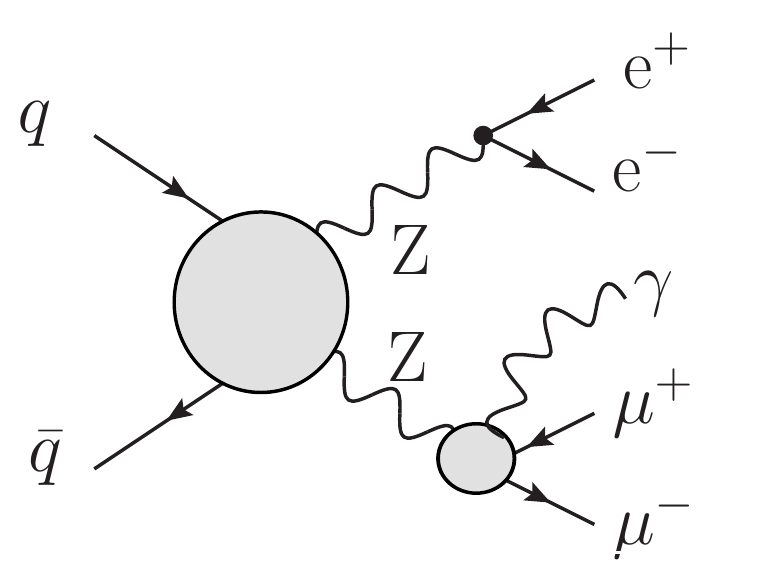}}
  \caption{
    Real-radiation contributions to resonant $\PZ\PZ$ production
    in the $q\bar{q}$ partonic channel at NLO EW. 
  }\label{fig:diagsRew}
\end{figure}

The real-radiation contributions to resonant $\PZ\PZ$ production in the quark--antiquark
channel, shown in \reffi{fig:diagsRew}, can be divided in three classes.
The first one concerns the radiation off the initial state [ISR,
\reffi{fig:Risr}], the second and the third ones are characterized by
the final-state radiation (FSR1, FSR2) of a photon off the
$\Pe^+\Pe^-$ and $\mu^+\mu^-$ pair, respectively [see
\reffis{fig:Rfsr1}--\ref{fig:Rfsr2}].  It is important to note that
squaring the sum of real-radiation contributions as in
\reffi{fig:diagsRew} returns also non-factorizable corrections,
arising from the interference of diagrams of different types. The
soft singularities embedded in such contributions are cancelled by
virtual non-factorizable contributions as those shown in
\reffis{fig:Vnfif}--\ref{fig:Vnfff}.  After combining real and virtual
contributions, the non-factorizable corrections are expected to
be smaller than the factorizable ones, given by the incoherent sum of
squared ISR, FSR1 and FSR2 corrections
\cite{Denner:1997ia,Beenakker:1997ir}. 

In order to cancel infrared (IR) singularities properly, a careful
selection of subtraction counter\-terms is needed, retaining only those
(both integrated and unintegrated) that cure the singularities of
doubly-resonant real corrections.  In the full computation, also non-doubly-resonant
structures contribute to real and virtual corrections, leading
to a larger number of singular configurations that need to be
subtracted.  We perform the subtraction of IR divergences in the
Catani--Seymour (CS) dipole formalism \cite{Catani:1996vz,Dittmaier:1999mb,Catani:2002hc}. 

\begin{figure}
  \centering  
   \subfigure[$\PZ\PZ$-resonant contribution\label{fig:aRisr1}]{ \includegraphics[scale=0.5]{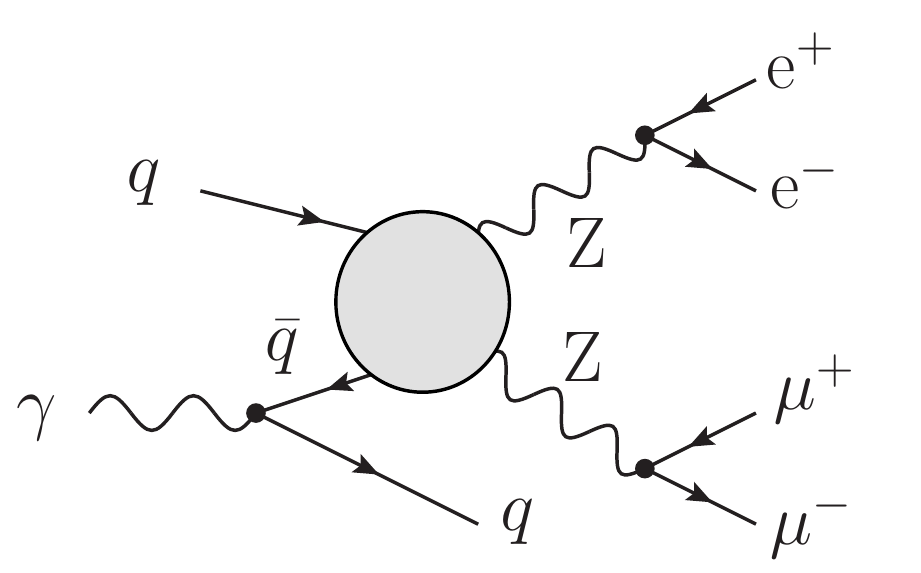} }\qquad
   \subfigure[Non-doubly-resonant contribution\label{fig:aRisr2}]{ \includegraphics[scale=0.5]{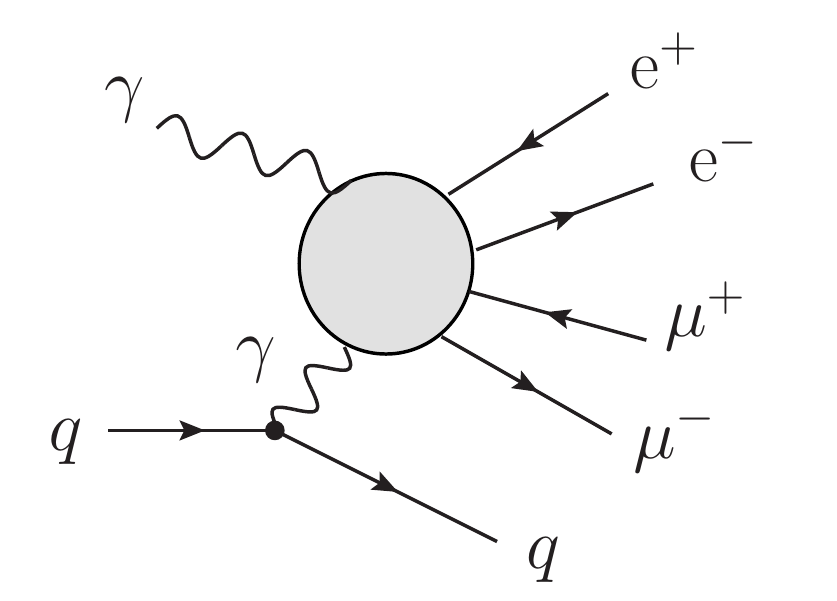} }
  \caption{
    Real-radiation contributions to four-charged-lepton production
    in the $\gamma\Pq$ partonic channel at NLO EW.
  }\label{fig:diagsRew2}
\end{figure}

Photon-induced partonic channels open up at NLO EW. Sample diagrams
for these real corrections are shown in \reffi{fig:diagsRew2}.
The selection of $\PZ\PZ$-resonant diagrams in this case is much easier,
as the final state (anti)quark cannot come from the decay of a
$\PZ$~boson and the doubly-resonant contributions are of ISR type only.
The diagram topologies shown in \reffi{fig:aRisr1} and \ref{fig:aRisr2}
can embed singular configurations in which the emitted (anti)quark is
collinear either to the initial-state photon [\reffi{fig:aRisr1}] or
to the initial-state (anti)quark 
[\reffi{fig:aRisr2}]. It is important to select only the subtraction counterterms that absorb
the first kind of configurations, since the underlying LO
contribution of the second configuration is non doubly resonant ($\gamma\gamma$~induced).

The selection of doubly-resonant diagrams in Born-level, virtual, and real-radiation
contributions is essential to enable the separation of
polarizations in $\PZ$-boson propagators.
In the DPA \cite{Denner:2005fg,Denner:2000bj},
the two resonant bosons must be projected on their mass shell in order to
restore EW gauge invariance. This is achieved via suitable
on-shell projections that are described in \refses{subsec:nloqcd}
and~\ref{subsec:nloew}.

Before going into the details of the DPA used for this calculation, we
point out that the neutral electric charge of the $\PZ$~boson
simplifies the selection of doubly-resonant structures both in real
and in virtual corrections, as photons cannot be radiated off the
resonant $\PZ$-bosons. In the case of $\PW$ bosons, additional care
must be taken in the separation of initial-state and final-state
radiation contributions, in particular concerning the definition of
subtraction counterterms.  We postpone the treatment of charged
resonances at NLO EW to future work.

\subsection{A double-pole approximation for Born-like doubly-resonant
  structures}\label{subsec:nloqcd}
After selecting doubly-resonant diagrams, it is crucial to apply an on-shell projection
to all LO and NLO matrix elements for $\PZ\PZ$ production, retaining the off-shell-ness
of the Breit--Wigner modulation in the denominator of $\PZ$-boson propagators.
This has to be pursued with special
attention to the subtraction counterterms, devising suitable projections that do not
spoil the cancellation of IR singularities.
A general NLO (either QCD or EW) observable receives the following contributions in the dipole formalism
\cite{Catani:1996vz,Dittmaier:1999mb,Catani:2002hc},
\begin{align}\label{eq:subtrNLO}
  \left(\frac{\rm \rd\sigma}{\rm \rd\xi}\right)_{\rm NLO} ={}&
  \frac{1}{2s}
  \int\! {\rm d}\Phi^{(4)}_n \,\,\mc B(\Phi^{(4)}_n)\,\delta(\xi-\xi_n)\,\nnb\\
  &{}+  \frac{1}{2s}
  \int\! \rd\Phi^{(4)}_n \left[\,\mc V(\Phi^{(D)}_n) + \mc C(\Phi^{(D)}_n) +\int \!\rd\Phi^{(D)}_{\rm rad}
    \,\,{\mc D}({\Phi}^{(D)}_{n},\Phi^{(D)}_{\rm rad}) \,\right]_{D=4}\,\delta(\xi-\xi_n)\nnb\\
  &{}+   \frac{1}{2s}
\int \rd\Phi^{(4)}_{n+1}\,\,\left[\,\mc R(\Phi^{(4)}_{n+1})\,\delta(\xi-\xi_{n+1})-\mc D(\bar{\Phi}^{(4)}_{n},\Phi^{(4)}_{\rm rad})\,\delta(\xi-\xi_n)\,\right]\,,
\end{align}
where $\mc B,\,\mc V,\,\mc R$ are Born, virtual and real
squared matrix elements, respectively,
including spin and colour averages as well as symmetry
factors, while $1/(2s)$ is the flux factor. The $\mc D$ term
is the sum of subtraction dipoles that appears both in its 4-dimensional, unintegrated version
and in its $D$-dimensional, integrated version.
The quantity $\mc C$ represents the collinear counterterm that
cancels left-over collinear singularities of initial-state type.
The arguments of the $\delta$ symbols indicate whether the observable $\xi$ is evaluated with $n$-body or with $(n+1)$-body
kinematics (with $D=4$). Note that the subtraction dipoles depend on the mapped $n$-body kinematics $\bar{\Phi}_n$,
which is derived from $(n+1)$-body kinematics via CS mappings \cite{Catani:1996vz}, and on
the radiation phase-space measure $\Phi_{\rm rad}$. While we keep $n$
variable in the formulas, $n=4$ for $\PZ\PZ$ production with leptonic $\PZ$~decays.

Let us consider all NLO contributions which embed a doubly-resonant
sub-amplitude with two two-body decays, namely 
$\PZ\,(\rightarrow \Pe^+\Pe^-)\,\PZ\,(\rightarrow\mu^+\mu^-)$.
This is the case for Born matrix elements, virtual corrections,
integrated dipoles, and real-subtracted initial-state radiation contributions.

We label the pole approximation applied to these contributions with $\rm DPA^{(2,2)}$.
Let us call $q_1,q_2$ the two off-shell $\PZ$-boson momenta in the laboratory system,
and $\tilde{q}_1,\tilde{q}_2$ the corresponding on-shell-projected momenta.
We choose $\tilde{q}_1,\tilde{q}_2$ such that
\beq
\tilde{q}_1+\tilde{q}_2 = {q_1}+{q_2},\qquad \tilde{q}_1^2 =\tilde{q}_1^2=\MZ^2
\eeq
and the spatial direction of the two bosons in the di-boson centre-of-mass
frame is preserved \cite{Denner:2000bj}. This procedure is valid for
$M_{\Pe^+\Pe^-\mu^+\mu^-}>2\MZ$. 
In the four-charged-lepton channel, this constraint can be achieved also
experimentally thanks to the reconstructable final state \cite{ATLAS:2021kog}.\\
We are left with the modification of the momenta of the four final-state leptons, according
to the on-shell kinematics of the two projected bosons.
Considering the $\PZ\,(q_1)\rightarrow\Pe^+(k_1)\,\Pe^-(k_2)$ decay, we choose to preserve the
energy fraction and the spatial direction of the electron (and of the positron)
in the corresponding $\PZ$~rest frame:
\begin{itemize}
\item[-] we boost $k_1,k_2$ into the rest frame of the off-shell $\PZ$
  boson (momentum $q_1$),
  where the two lepton momenta have $\hat{n}_{k_1}$ and $\hat{n}_{k_2}$ as spatial
  directions;
\item[-] we rescale the lepton energies according to the on-shell-ness of the projected
  $\PZ$~boson, \emph{i.e.} we multiply the off-shell energies by $\MZ/\sqrt{q_1^2}$; 
\item[-] we set the spatial directions in the rest frame of the
    on-shell $\PZ$~bosons equal to the off-shell ones ($\hat{n}_{k_1}$,
  $\hat{n}_{k_2}$); it is easy to see that in this reference frame the two projected
  momenta of the leptons read $\tilde{k}_1 = (1,\hat{n}_{k_1}) {\MZ}/{2}$ and
  $\tilde{k}_2 =(1,-\hat{n}_{k_1}) {\MZ}/{2}$;
\item[-] we boost back $\tilde{k}_1,\tilde{k}_2$ into the laboratory frame
  according to the projected $\PZ$-boson kinematics (momentum $\tilde{q}_1$).
\end{itemize}
The same procedure is then applied to the decay products of the second $\PZ$~boson
($\PZ\rightarrow\mu^+\mu^-$).

The described  DPA can be safely applied to factorizable virtual corrections and integrated dipoles,
as well as to the real-subtracted corrections of ISR type. Since the QCD corrections only
modify the $\PZ\PZ$ production process, all contributions at order $\mc O(\alpha^4\as)$ can be treated
with \dpatwotwo. Note that only two subtraction dipoles are needed 
to cancel IR singularities of QCD origin, and they are both of initial--initial type, therefore
the modification of the momenta of final-state leptons (colourless) does not interfere
with the phase-space mappings used in subtraction counterterms.
Both subtraction terms $\mc D$ and initial-state collinear counterterms $\mc C$ [in Eq.~\eqref{eq:subtrNLO}]
need to be computed with \dpatwotwo. 
This DPA has already been used to compute NLO QCD corrections to polarized $\PW^+\PW^-$ and
$\PW^+\PZ$ production \cite{Denner:2020bcz,Denner:2020eck}.

In the computation of real EW corrections, only ISR contributions
[see \reffi{fig:Risr}] can be computed with \dpatwotwo,
following the same procedure that has been used in the NLO QCD case,
but taking care of one additional subtlety.
Differently from QCD corrections, the spectator for the ISR subtraction dipoles
may be an initial-state or a final-state particle.
However, only an initial-state spectator must be selected to properly cancel the IR singularities
of ISR type only in \dpatwotwo. Furthermore, treating the $(n+1)$-body kinematics and the mapped kinematics
(in the sense of CS mappings) with the same DPA technique ensures that the subtraction is not spoiled.
Adapting the notation of Eq.~\eqref{eq:subtrNLO}, the real-subtracted contribution reads (with off-shell kinematics),
\beq
\mc R_{\rm ISR}(\Phi_{n+1})\,\delta(\xi-\xi_{n+1}) 
- \mc D^{I_1i,I_2}(\bar{\Phi}_n,\Phi_{\rm rad}(k_i))\,\delta(\xi-\xi_{n})  
- \mc D^{I_2i,I_1}(\bar{\Phi}_n,\Phi_{\rm rad}(k_i))\,\delta(\xi-\xi_{n})\,,
\eeq
where $\bar{\Phi}_n$ is the mapped $n$-body kinematics, according to an initial--initial CS
mapping, while $k_i$ is the radiated-photon momentum. Both the emitter and the
spectator are initial-state particles ($I_1,\,I_2$). After the application of \dpatwotwo,
\beq
\mc R_{\rm ISR}({\tilde{\Phi}_{n+1}})\,\delta(\xi-\xi_{n+1})
- \mc D^{I_1i,I_2}(\tilde{\bar{\Phi}}_n,\Phi_{\rm rad}(k_i))\,\delta(\xi-\xi_{n})
- \mc D^{I_2i,I_1}(\tilde{\bar{\Phi}}_n,\Phi_{\rm rad}(k_i))\,\delta(\xi-\xi_{n})\,,
\eeq
where $\tilde{\bar{\Phi}}_n$ and $\tilde{\Phi}_{n+1}$ represent
the on-shell-projected momenta 
of the $n$- and $(n+1)$-body phase space. The \dpatwotwo does not modify the photon radiation,
neither in the $(n+1)$-body kinematics, nor in the radiation phase-space that
appears in the dipoles. Since the initial--initial CS mappings modify all final-state
particles with the same boost, the application of \dpatwotwo to the mapped kinematics
does not interfere with the CS-mapping procedure, as the total momentum
of the four leptons is preserved in the DPA, \ie the on-shell
projection of \dpatwotwo and the initial--initial CS mappings commute. 

Note that the \dpatwotwo can also be used to treat the photon-induced
real contributions, since the corresponding doubly-resonant
contributions do not allow for final-state radiation.

We stress that, since also integrated dipoles are treated with \dpatwotwo,
the needed correspondence between subtraction and integrated dipoles is maintained for ISR
contributions.

\subsection{A double-pole approximation for final-state radiation}\label{subsec:nloew}
The treatment of final-state-radiation contributions to EW real corrections
is more delicate than the treatment of ISR ones.
Let us consider the case in which a $\PZ$~boson decays into $\Pe^+(k_1)\,\Pe^-(k_2)\,\gamma(k_5)$,
while the other one decays into $\mu^+(k_3)\,\mu^-(k_4)$, corresponding to the diagram
topology shown in \reffi{fig:Rfsr1}. 
For such corrections, we have devised an on-shell projection (labelled \dpathreetwo) that
differs from \dpatwotwo\ but follows the same driving idea in terms of preserved quantities.
For the projection of the momenta of $\PZ$~bosons, we apply the same method as for $\rm DPA^{(2,2)}$,
with the difference that the momentum of the first $\PZ$~boson is now given by the sum of three momenta,
\beq
q_1 = k_1+k_2+k_5,\qquad q_2 = k_3+k_4\,.
\eeq
After this step, we have $\tilde{q}_1^2=\tilde{q}_2^2=\MZ^2$ and $\tilde{q}_1+\tilde{q}_2 = q_1+q_2$.
The momenta $k_3,k_4$ can be projected following the prescription of $\rm DPA^{(2,2)}$ (see \refse{subsec:nloqcd}), as
a two-body decay is understood for the second $\PZ$~boson.
For the three-body decay, 
\begin{itemize}
\item[-] we boost $k_1,k_2,k_5$ into the rest frame of the off-shell $\PZ$~boson (momentum $q_1$),
  where the three particles have $\hat{n}_{k_1},\hat{n}_{k_2},\hat{n}_{k_5}$ as spatial
  directions;
\item[-] we rescale the lepton and photon energies according to the on-shell-ness of the projected
  $\PZ$~boson, \emph{i.e.} we multiply the off-shell energies by $\MZ/\sqrt{q_1^2}$; 
\item[-] we set the spatial directions of the final-state momenta in the rest frame of the
    on-shell $\PZ$~boson equal to the off-shell ones ($\hat{n}_{k_1}$, 
  $\hat{n}_{k_2}$, $\hat{n}_{k_5}$), which allows to preserve the relative angles among the
  three particles;
\item[-] we boost back $\tilde{k}_1,\tilde{k}_2,\tilde{k}_5$ into the laboratory frame
  according to the projected $\PZ$-boson kinematics (momentum $\tilde{q}_1$).
\end{itemize}
It is easy to check that this prescription does not violate  momentum-conservation and preserves
the ratio of Lorentz invariants constructed with the three decay products,
\beq\label{dpaproperty1}
\frac{\tilde{k}_a\cdot\tilde{k}_{b}}{\tilde{k}_a\cdot\tilde{k}_{c}}=
\frac{k_a\cdot k_b}{k_a\cdot k_c}\quad\textrm{ for any }\,\,
a,b,c=1,2,5\, \textrm{ and }\, a\neq b\neq c\,.
\eeq
It is crucial that the energy fractions and the relative angles are preserved
in the three-body decay, as the radiated photon may be soft and/or collinear to the electron
or to the positron. The \dpathreetwo guarantees that the
soft-ness and/or collinearity of the photon is preserved.

The subtracted FSR1 real corrections read [with off-shell $(n+1)$-body
kinematics]:
\beq\label{eq:subtrFSR1}
\mc R_{\rm FSR1}({\Phi}_{n+1})\,\delta(\xi-\xi_{n+1})
- \mc D^{15,2}(\bar{\Phi}_n,\Phi_{\rm rad}(k_5))\,\delta(\xi-\xi_{n})
- \mc D^{25,1}(\bar{\Phi}_n,\Phi_{\rm rad}(k_5))\,\delta(\xi-\xi_{n})\,,
\eeq
where $\bar{\Phi}_n$ is the mapped phase-space according to a final--final CS mapping.
The contributing dipoles involve as emitter and spectator either the
electron or the positron, which form with the photon the decay products
of the first $\PZ$~boson.
At variance with the \dpatwotwo, the photon momentum $k_5$ is modified by the \dpathreetwo.
Therefore the same DPA must be applied to the radiation phase-space and to the
factorized phase-space in the CS-mapped process to guarantee the proper subtraction
of IR singularities. This could spoil the correspondence between the
subtraction dipoles appearing in Eq.~\eqref{eq:subtrFSR1} and their integrated counterparts,
as the former are projected with $\rm DPA^{(3,2)}$, while the latter can only be
projected with $\rm DPA^{(2,2)}$ (as they feature $n$-body kinematics). We will see
in the following that this is not the case.\\
If no DPA is applied, the CS mapping for $\mc D^{15,2}$ reads,
\beq \label{eq:off32map}
\bar{k}_1  = k_1+k_5 - \frac{s_{15}}{s_{12}+s_{25}}\,k_2\,,\quad  \bar{k}_2  =
\frac{s_{125}}{s_{12}+s_{25}}\,k_2\,,\quad \bar{k}_3=k_3\,,\quad \bar{k}_4=k_4\,,
\eeq
where $s_{ab}=2k_a\cdot k_b$ and $s_{125}=s_{12}+s_{25}+s_{15}$.
In order to use the DPA for the subtraction dipoles, we cannot simply apply the $\rm DPA^{(2,2)}$ to the
mapped momenta, otherwise we would implicitly assume that the photon momentum is untouched,
which is not the case when applying $\rm DPA^{(3,2)}$ to FSR1 real contributions.
A consistent procedure is to apply the CS mapping to the $(n+1)$-body kinematics that has already
been projected with the \dpathreetwo, 
\beq\label{eq:on32map}
\bar{\tilde{k}}_1  = \tilde{k}_1+\tilde{k}_5 -
\frac{\tilde{s}_{15}}{\tilde{s}_{12}+\tilde{s}_{25}}\,\tilde{k}_2\,,\quad  \bar{\tilde{k}}_2  =
\frac{\tilde{s}_{125}}{\tilde{s}_{12}+\tilde{s}_{25}}\,\tilde{k}_2\,,\quad \bar{\tilde{k}}_3=
\tilde{k}_3\,,\quad \bar{\tilde{k}}_4=\tilde{k}_4\,.
\eeq
The final--final CS mapping does not modify the virtuality of the $\PZ$~boson
[$\tilde{q}_1^2 = (\tilde{k}_1+\tilde{k}_2+\tilde{k}_5)^2=\MZ^2$]. Furthermore,
according to Eq.~\eqref{dpaproperty1}, the Lorentz invariants that
appear in Eq.~\eqref{eq:on32map} coincide with those of Eq.~\eqref{eq:off32map}.

The more involved character of the \dpathreetwo with respect to \dpatwotwo originates from
the fact that the on-shell projection of \dpathreetwo and the final--final CS mappings are not
commuting, unlike the case of \dpatwotwo application and initial--initial
CS mappings.
In order to further prove the decent behaviour of the \dpathreetwo, we check that its
application to the subtracted final-state-radiation corrections does not spoil the correspondence between
unintegrated and integrated dipoles.
Let us consider the $D$-dimensional structure of a dipole with emissus $i$ (photon),
emitter $j$ ($\Pe^+$ or $\Pe^-$) and spectator $k$ ($\Pe^-$ or $\Pe^+$), recalling
that the three particles are decay products of the same $\PZ$~resonance, 
\begin{align}\label{eq:subtrdip}
  \rd\bar{\Phi}_{n}\,&
  \rd\Phi_{\rm rad}(k_i)\,
  {\mc D}^{(ji,k)}(\bar{\Phi}_{n},\Phi_{\rm rad}(k_i))\,\nnb\\
  ={}&
  \rd\bar{\Phi}_{n}\,
  \rd y\,\rd z\,
  \frac{(\bar{s}_{jk})^{-\epsilon}}{16\pi^2}\,\frac{(4\pi)^\epsilon}{\Gamma(1-\epsilon)}\,[z\,(1-z)]^{-\epsilon}(1-y)^{1-2\epsilon}y^{-1-\epsilon}\,
  \nnb\\
  &{}\hspace{1.5cm}\times\,\left(-
  \frac{Q_{j}\,Q_k}{Q_{j}^2}
  \right)
  \big\langle \bar{k}_1,\ldots,\bar{k}_n|\,V_{ijk}\,|\bar{k}_1,\ldots,\bar{k}_n\big\rangle\,\mc B(\bar{\Phi}_{n}),
\end{align}
where $Q_j,\,Q_k$ are the electric charges (in unit of $e$) of the emitter and spectator, respectively, and the phase-space measure
depends on the variables
\beq
y = \frac{s_{ij}}{s_{ijk}} = \left({1+\frac{s_{jk}}{s_{ij}}+\frac{s_{ik}}{s_{ij}}}\right)^{-1}\,,\qquad z = \frac{s_{ik}}{s_{ijk}-s_{ij}} = \left(1+\frac{s_{jk}}{s_{ik}}\right)^{-1}\,.
\eeq
The kernel that has to be integrated is \cite{Catani:1996vz,Dittmaier:1999mb}
\beq
\langle s| V_{ijk}(y,z)|s'\rangle\,=\,8\pi\alpha\,\mu^{2\epsilon}\,Q_{j}^2
\left(\frac2{1-z(1-y)}-(1+z)-\epsilon\,(1-z)\right)\,\delta_{ss'}
\,.
\eeq
We can then re-write Eq.~\eqref{eq:subtrdip} as 
\begin{align}
\rd\bar{\Phi}_{n}\,&
  \rd\Phi_{\rm rad}(k_i)\,
  {\mc D}^{(ji,k)}(\bar{\Phi}_{n},\Phi_{\rm rad}(k_i))\,\nnb\\
  ={}&
  \rd\bar{\Phi}_{n}\,  
  \frac{\alpha}{2\pi}\,\frac{(4\pi)^\epsilon}{\Gamma(1-\epsilon)}\,\left(\frac{\mu^2}{\bar{s}_{jk}}\right)^{\epsilon}\,\mc B(\bar{\Phi}_{n})
  \nnb\\
  &\hspace{1.5cm}\times\,
  \,\,\rd y\,\rd z\,[z\,(1-z)]^{-\epsilon}(1-y)^{1-2\epsilon}y^{-1-\epsilon}\,
  \left(\frac2{1-z(1-y)}-(1+z)-\epsilon\,(1-z)\right)\,
  ,\nnb
\end{align}
where we have used $Q_jQ_k=-1$ and $Q_j^2=1$ for the $\Pe^+\Pe^-$
final state. The 4-dimensional version of the subtraction dipole
reads
\beq
\rd\bar{\Phi}_{n}\,  
  \frac{\alpha}{2\pi}\,\,\mc B(\bar{\Phi}_{n})
  \,\,\rd y\,\rd z\,\frac{1-y}y\,
  \left(\frac2{1-z(1-y)}-(1+z)\right)\,
  ,
\eeq
while its integration in $D$~dimensions ($D=4-2\epsilon$) gives
\begin{align}\label{intdip}
&&
  \rd\bar{\Phi}_{n}\,  
  \frac{\alpha}{2\pi}\,\frac{(4\pi)^\epsilon}{\Gamma(1-\epsilon)}\,\left(\frac{\mu^2}{\bar{s}_{jk}}\right)^{\epsilon}\,\left[\frac1{\epsilon^2}+\frac3{2\epsilon}+\left(5-\frac{\pi^2}2\right)+\mc O(\epsilon)\right]\,\mc B(\bar{\Phi}_{n})\,.
\end{align}
For $\PZ\PZ$ production, we have $n=4$ and the LO doubly-resonant structure can be written as
\beq
\mc B(\bar{k}^{\stiny{(j5,k)}}_1,\bar{k}^{\stiny{(j5,k)}}_2,\bar{k}^{\stiny{(j5,k)}}_3,\bar{k}^{\stiny{(j5,k)}}_4) \,=\,  \frac{\mc N_{\mc B}(\bar{k}^{\stiny{(j5,k)}}_1,\bar{k}^{\stiny{(j5,k)}}_2,\bar{k}^{\stiny{(j5,k)}}_3,\bar{k}^{\stiny{(j5,k)}}_4)}{
\left[\bigl(\bar{s}^{\stiny{(j5,k)}}_{12}-\MZ^2\bigr)^2+(\GZ\MZ)^2\right]\left[\bigl(\bar{s}^{\stiny{(j5,k)}}_{34}-\MZ^2\bigr)^2+(\GZ\MZ)^2\right]}\,,
\eeq
where the barred kinematics depends on the choice of the emitter $j$ and spectator $k$ that is used in the
dipole phase-space mapping. The mapped momenta are labelled by the
indices of the emitter--emissus pair $j5$ and the spectator $k$.
The $(n+1)$-body part of the NLO correction for the doubly-resonant FSR1 contribution reads ($k_5$ is the photon momentum):
\begin{align}\label{eq:offFSR}  
  \rd\Phi&_{n+1}\,  
  \frac{\mc N_{\mc R}({k}_1,k_2,k_3,k_4,k_5)}{
    \left[\bigl({s}_{125}-\MZ^2\bigr)^2+(\GZ\MZ)^2\right]\left[\bigl(s_{34}-\MZ^2\bigr)^2+(\GZ\MZ)^2\right]}\,\delta(\xi-\xi_{n+1})
  \nnb\\
  &-  
  \rd\bar{\Phi}^{(15,2)}_{n}\,  
  \frac{\alpha}{2\pi}\,\frac{\mc N_{\mc B}(\bar{k}^{\stiny{(15,2)}}_1,\bar{k}^{\stiny{(15,2)}}_2,\bar{k}^{\stiny{(15,2)}}_3,\bar{k}^{\stiny{(15,2)}}_4)}{
    \left[\bigl(\bar{s}^{\stiny{(15,2)}}_{12}-\MZ^2\bigr)^2+(\GZ\MZ)^2\right]\left[\bigl(\bar{s}^{\stiny{(15,2)}}_{34}-\MZ^2\bigr)^2+(\GZ\MZ)^2\right]
  }\nnb\\
  &\qquad\times
  \,\rd y\,\rd z\,\frac{1-y}y\,
  \left(\frac2{1-z(1-y)}-(1+z)\right)\,\delta(\xi-\xi_{n})\,
  \nnb\\
  &-
\rd\bar{\Phi}^{(25,1)}_{n}\,  
  \frac{\alpha}{2\pi}\,\frac{\mc N_{\mc B}(\bar{k}^{\stiny{(25,1)}}_1,\bar{k}^{\stiny{(25,1)}}_2,\bar{k}^{\stiny{(25,1)}}_3,\bar{k}^{\stiny{(25,1)}}_4)}{
    \left[\bigl(\bar{s}^{\stiny{(25,1)}}_{12}-\MZ^2\bigr)^2+(\GZ\MZ)^2\right]\left[\bigl(\bar{s}^{\stiny{(25,1)}}_{34}-\MZ^2\bigr)^2+(\GZ\MZ)^2\right]
  }  \nnb\\
  &\qquad\times\,\rd y'\,\rd z'\,\frac{1-y'}{y'}\,
  \left(\frac2{1-z'(1-y')}-(1+z')\right)\,\delta(\xi-\xi_{n})\,
  ,
\end{align}
where $y',z'$ are defined as $y,z$ up to the exchange of the momenta $k_1$ and $k_2$.
Note also that, according to the final--final CS mapping, $\bar{s}^{\stiny{(15,2)}}=\bar{s}^{\stiny{(25,1)}}=s_{125}$ and
$\bar{s}^{\stiny{(15,2)}}_{34}=\bar{s}^{\stiny{(25,1)}}_{34}={s}_{34}$.

We observe that the action of $\rm DPA^{(2,2)}$ on the mapped $n$-body kinematics would result
in the modification of the radiation measure $\rd\Phi_{\rm rad}(k_i)$ as well as of the subtraction
dipole, because the radiation momentum would be left untouched by the on-shell projection.
On the contrary, if $\rm DPA^{(3,2)}$ is applied to the factorized $(n+1)$-body kinematics,
 both the integration measure and the dipoles do not change thanks to Eq.~\eqref{dpaproperty1}.
Furthermore, this way the three partons are correctly projected to reconstruct an on-shell $\PZ$~boson.
The application of $\rm DPA^{(3,2)}$ to both the complete and the factorized $(n+1)$-body
kinematics gives the DPA-regulated version of Eq.~\eqref{eq:offFSR}:
\begin{align}\label{eq:onFSR}
  \rd\Phi&_{n+1}\,  
  \frac{\mc N_{\mc R}(\tilde{k}_1,\tilde{k}_2,\tilde{k}_3,\tilde{k}_4,\tilde{k}_5)}{
    \left[\bigl({s}_{125}-\MZ^2\bigr)^2+(\GZ\MZ)^2\right]\left[\bigl(s_{34}-\MZ^2\bigr)^2+(\GZ\MZ)^2\right]
  }\,\delta(\xi-\xi_{n+1})
  \nnb\\
  &-  
   \rd\bar{\Phi}^{(15,2)}_{n}\,  
   \frac{\alpha}{2\pi}\,\frac{\mc N_{\mc B}(
     \bar{\tilde{k}}^{\stiny{(15,2)}}_1,
     \bar{\tilde{k}}^{\stiny{(15,2)}}_2,
     \bar{\tilde{k}}^{\stiny{(15,2)}}_3,
     \bar{\tilde{k}}^{\stiny{(15,2)}}_4
     )}{
     \left[\bigl({s}_{125}-\MZ^2\bigr)^2+(\GZ\MZ)^2\right]\left[\bigl(s_{34}-\MZ^2\bigr)^2+(\GZ\MZ)^2\right]
   }\nnb\\
   &\qquad\times\,\rd y\,\rd z\,\frac{1-y}y\,
  \left(\frac2{1-z(1-y)}-(1+z)\right)\,\delta(\xi-\xi_{n})\,
  \nnb\\
  &-
   \rd\bar{\Phi}^{(25,1)}_{n}\,  
   \frac{\alpha}{2\pi}\,\frac{\mc N_{\mc B}(
     \bar{\tilde{k}}^{\stiny{(25,1)}}_1,
     \bar{\tilde{k}}^{\stiny{(25,1)}}_2,
     \bar{\tilde{k}}^{\stiny{(25,1)}}_3,
     \bar{\tilde{k}}^{\stiny{(25,1)}}_4
     )}{
     \left[\bigl({s}_{125}-\MZ^2\bigr)^2+(\GZ\MZ)^2\right]\left[\bigl(s_{34}-\MZ^2\bigr)^2+(\GZ\MZ)^2\right]
  }\nnb\\
   &\qquad\times\,\rd y'\,\rd z'\,\frac{1-y'}{y'}\,
  \left(\frac2{1-z'(1-y')}-(1+z')\right)\,\delta(\xi-\xi_{n})\,
  ,
\end{align}
where we have used the fact that $\tilde{y}=y$ and $\tilde{z}=z$, since these variables only
depend on ratios of $(n+1)$-body invariants that are preserved according to Eq.~\eqref{dpaproperty1}.
We stress again that the momenta in the numerators $\mc N_{\mc B}$ are obtained 
from $(n+1)$-body kinematics performing first the DPA and then the CS mapping (the procedures do not commute).
Owing to Eq.~\eqref{dpaproperty1} the integration measure $\rd y\,\rd z$
is left untouched by the DPA, and the integrated dipoles do not need
to be modified but give
exactly the expression of Eq.~\eqref{intdip}. This is the case because the DPA$^{(2,2)}$
applied to the $n$-body kinematics of the integrated dipoles proceeds in the same fashion as DPA$^{(3,2)}$.

The analogous procedure, which we label \dpatwothree, is applied to FSR2 real contributions
shown in \reffi{fig:Rfsr2} and the corresponding two subtraction dipoles.

The entire DPA formalism that we have presented for $\PZ\PZ$ inclusive production, in particular
the on-shell projections for Born-like and final-state radiation structures, can be straightforwardly extended
to more complicated processes involving $\PZ$~bosons, \eg $\PZ\PZ$
scattering, or other neutral resonances.

As a last comment of this section, we stress that the DPA
  can be applied rather straightforwardly to same-flavour leptonic decays,
  \eg $\Pp\Pp \rightarrow \PZ(\Pe^+\Pe^-)\PZ(\Pe^+\Pe^-)+X$. The
  intrinsic ambiguity in identifying the resonances can be solved by associating 
  the identical final-state leptons to specific \PZ~bosons and accounting for the
  proper symmetry factor in the resonant amplitudes. A DPA for a process with such 
  an ambiguity has, for instance, been used in \citere{Denner:2020orv} 
(see discussion in Section 2.2 there).

\subsection{Separating polarizations}\label{subsec:pol}
The application of the DPA to all contributions of the NLO (EW or QCD) computation
enables us to isolate in a gauge-invariant way the resonant contributions to $\PZ\PZ$
production from the non-doubly-resonant background, which can be estimated as the difference
between the full result and the doubly-resonant contributions in the DPA, and is expected
to amount to $\mc O(\GZ/\MZ)\approx 2.7\%$ of the full result.

Thanks to the DPA, it is possible to separate the polarization of
$\PZ$~bosons at the level of doubly-resonant amplitudes, following the procedure
that has already been applied to vector-boson scattering
\cite{Ballestrero:2017bxn,Ballestrero:2019qoy,Ballestrero:2020qgv},
to $\PW^+\PZ$ and $\PW^+\PW^-$ production at NLO QCD \cite{Denner:2020bcz,Denner:2020eck},
and to Higgs decays \cite{Maina:2020rgd,Maina:2021xpe}.
After selecting doubly-resonant diagrams,
the following structure (written in the `t Hooft--Feynman gauge) characterizes the SM amplitude, 
\beq\label{eq:firstseppol}
\mc A_{\rm res} = \mc P_{\mu\nu}({q}_1,{q}_2)\,\frac{-g^{\mu\alpha}}{q_1^2-\MZ^2+\ri \GZ\MZ}\,\,\frac{-g^{\nu\beta}}{q_2^2-\MZ^2+\ri \GZ\MZ}\,\mc D^{(1)}_{\alpha}({q}_1)\,\mc D^{(2)}_{\beta}({q}_2)\,,
\eeq
where $\mc P$ is the production part of the amplitude, while the $\mc D$
terms (not to be confused with the subtraction terms in \refses{subsec:nloqcd} and \ref{subsec:nloew})
are the parts corresponding to the two- or three-body decays of the $\PZ$~bosons.
Writing the metric tensors in terms of polarization vectors, Eq.~\eqref{eq:firstseppol} becomes
\begin{align}\label{eq:seppolZZ}
\mc A_{\rm res} ={}& \mc P_{\mu\nu}(q_1,q_2)\,
\frac{\big[\sum_{\lambda=\rL,\pm}\varepsilon_\lambda^{\mu\,*}(q_1)\,
    \varepsilon_\lambda^{\alpha}(q_1)\big]\,-q_1^\mu q_1^\alpha/\MZ^2\,}{q_1^2-\MZ^2+\ri\GZ\MZ}\,\,\mc D^{(1)}_{\alpha}(q_1)\nnb\\
&\hspace*{1.5cm}\times
\frac{\big[\sum_{\lambda'=\rL,\pm}\varepsilon_{\lambda'}^{\nu\,*}(q_2)\,
    \varepsilon_{\lambda'}^{\beta}(q_2)\big]\,-q_2^\nu q_2^\beta/\MZ^2\,}{q_2^2-\MZ^2+\ri\GZ\MZ}\,\,\mc D^{(2)}_{\beta}(q_2)\,\nnb\\
={}&
\sum_{\lambda=\rL,\pm}\sum_{\lambda'=\rL,\pm}
\left[\mc P_{\mu\nu}(q_1,q_2)\,\varepsilon_\lambda^{\mu\,*}(q_1)\varepsilon_{\lambda'}^{\nu\,*}(q_2)\,\right]
\frac{\big[\,
    \varepsilon_\lambda^{\alpha}(q_1)\,\mc D^{(1)}_{\alpha}(q_1)\,\big]\,}{q_1^2-\MZ^2+\ri\GZ\MZ}\,
\frac{\big[
    \varepsilon_{\lambda'}^{\beta}(q_2)\,\mc D^{(2)}_{\beta}(q_2)\,\big]\,}{q_2^2-\MZ^2+\ri\GZ\MZ}\,\,\nnb\\
\equiv{}&
\sum_{\lambda=\rL,\pm}\sum_{\lambda'=\rL,\pm}\mc A_{\lambda\lambda'}\,,
\end{align}
where for both $\PZ$~bosons the polarization sum runs over the longitudinal ($\rL$), left-handed ($-$), and right-handed states ($+$).
Since we consider massless final-state leptons, the gauge terms that are proportional to the boson momenta
vanish for all LO and NLO contributions. In fact, it is easy to see that the $\PZ$-boson propagators are always
contracted with massless fermionic currents, both in tree-level and in
one-loop doubly-resonant amplitudes. In the general case, the contributions
of the terms proportional to boson momenta in the Z-boson propagators
cancel against diagrams where the $\PZ$~bosons are replaced by the
corresponding would-be Goldstone bosons.

From Eq.~\eqref{eq:seppolZZ} it is clear that 
applying the DPA and squaring the doubly-resonant (unpolarized) amplitude,
one obtains well-defined polarized squared
matrix elements
and off-diagonal contributions which are called interference terms:
\beq\label{eq:squaredAres}
|\mc A_{\rm res}|^2 = \sum_{\lambda=\rL,\pm}\sum_{\lambda'=\rL,\pm}|\mc A_{\lambda\lambda'}|^2\,+ (\textrm{interference terms})\,.
\eeq
Replacing the unpolarized squared matrix element with a single $\{\lambda,\lambda'\}$
term in the sum of Eq.~\eqref{eq:squaredAres}, we can  generate events for the
production and decay of two $\PZ$~bosons with definite polarization states ($\lambda$
and $\lambda'$, respectively).

In order to reduce the number of different contributions and to minimize the impact of interference terms, we
have defined the transverse mode (T) as the coherent sum of the left- and right-handed
modes (including the left--right interference). 

The polarization vectors in Eq.~\eqref{eq:seppolZZ} must be defined in a certain reference frame.
Common choices for di-boson production are the laboratory (LAB) frame \cite{Baglio:2018rcu,Denner:2020bcz}
and the centre-of-mass (CM) frame of the two bosons \cite{Baglio:2019nmc,Denner:2020eck}.
The CM-frame choice has been shown \cite{Denner:2020eck} to be more natural than the LAB one
for di-boson production and has been used in the latest ATLAS analysis
of $\PW^\pm\PZ$ production \cite{Aaboud:2019gxl}.
It is essential that the same definition of polarization vectors is used in all parts of the
NLO calculation.
In the case of the CM-frame definition, for final-state-radiation contributions
  to NLO EW corrections, the CM frame is the rest frame of the system formed by the four
  charged leptons and the photon, which is crucial to have a functioning subtraction of IR singularities.
In \refse{subsec:inc} we compare the results in both the CM- and the LAB-frame definitions, while
in \refse{subsec:fid} we only consider polarizations defined in the CM frame.

Before presenting the numerical results of our calculation, it is worth recalling that
the polarization structure of $\PZ\PZ$ production at the LHC is expected to be rather
different from the one of $\PW^+\PW^-$ \cite{Denner:2020bcz,Poncelet:2021jmj},
in spite of the same contributing partonic channels.
At LO, the differential cross-sections for on-shell $\PZ\PZ$ production in $q\bar{q}$ annihilation
can be written in terms of the CM energy squared $s$ and the scattering angle $\theta$
in the CM frame,
\begin{align}\label{eq:analyticZZform}
  \frac{\rd\sigma_{\rL\rL}}{\rd\cos\theta} ={}& \frac{\pi\alpha^2(c^4_{\rm L,\Pq}+c^4_{\rm R,\Pq})}{192\,\sw^4\,\cw^4}
  \frac{2 \MZ^4 \, \left(s-4 \MZ^2\right)\,\sin^2\theta \cos^2\theta}{\sqrt{1-({4 \MZ^2}/{s})}\, \left(s \left(4
    \MZ^2-s\right)\, \cos^2\theta+\left(s-2 \MZ^2\right)^2\right)^2}\nnb\\
  ={}&
  \frac{\pi\alpha^2(c^4_{\rm L,\Pq}+c^4_{\rm R,\Pq})}{96\,\sw^4\,\cw^4}\, \frac{ {\MZ^4}\,\cos ^2\theta}{s^3 \sin^2\theta}+\mc{O}\left(\frac1{s^4}\right)\,,
  \nnb\\ 
  \frac{\rd\sigma_{\rL\rT}}{\rd\cos\theta} ={}& \frac{\pi\alpha^2(c^4_{\rm L,\Pq}+c^4_{\rm R,\Pq})}{192\,\sw^4\,\cw^4}
  \frac{16 \MZ^2 \left(s-4 \MZ^2\right) \left(\left(s-2 \MZ^2\right)^2+ \left(4 \MZ^4+4 \MZ^2 s-2 s^2\right)\cos ^2\theta +s^2 \cos ^4\theta
    \right)}{s\, \sqrt{1-({4 \MZ^2}/{s})}\, \left(s \left(4 \MZ^2-s\right)\,\cos ^2\theta +\left(s-2 \MZ^2\right)^2\right)^2}\!\!\!\nnb\\
  ={}& \frac{\pi\alpha^2(c^4_{\rm L,\Pq}+c^4_{\rm R,\Pq})}{12\,\sw^4\,\cw^4}\,\frac{\MZ^2}{s^2}+\mc O\left(\frac1{s^3}\right) \nnb\\
  \frac{\rd\sigma_{\rT\rT}}{\rd\cos\theta} ={}& \frac{\pi\alpha^2(c^4_{\rm L,\Pq}+c^4_{\rm R,\Pq})}{192\,\sw^4\,\cw^4}\,
  \frac{
     4\,\left(s-4 \MZ^2\right) \left(\left(s-2\MZ^2\right)^2+(36 \MZ^4-4\MZ^2s+s^2)\,\cos^2\theta\right)\,\sin ^2\theta
  }{ \sqrt{1-({4 \MZ^2}/{s})}\,\left(s \left(4 \MZ^2-s\right)\,\cos ^2\theta +\left(s-2 \MZ^2\right)^2\right)^2}\nnb\\
  ={}& \frac{\pi\alpha^2(c^4_{\rm L,\Pq}+c^4_{\rm R,\Pq})}{48\,\sw^4\,\cw^4}
  \frac{ 1+\cos^2\theta}{s\,\sin^2\theta}+\mc O\left(\frac1{s^2}\right)\,,
\end{align}
where $c_{\rm L,\Pq},c_{\rm R,\Pq}$ are the left- and right-handed couplings of
the $\PZ$~boson to quarks $\Pq$, and $\sw, \cw$ are the sine and cosine of the
EW mixing angle. At a given scattering angle, the expansion
in powers of $s$ gives a $\rL\rL$ cross-section that is suppressed by
$1/s^2$ w.r.t.\ the $\rT\rT$ one and by $1/s$ w.r.t.\ the mixed one.  The
suppression of the $\rL\rL$ signal can be explained using the
Goldstone-boson equivalence theorem
\cite{Cornwall:1974km,Vayonakis:1976vz,Chanowitz:1985hj,Gounaris:1986cr},
which relates the amplitudes for longitudinal vector-boson production
to those for would-be-Goldstone-boson production at high energies of
the vector bosons. Since the couplings of the Goldstone bosons to
massless fermions vanish, there are no Feynman diagrams for the
production of a pair of neutral would-be Goldstone bosons at LO.
Consequently, the amplitude for the production of a pair of
longitudinal on-shell $\PZ$~bosons vanishes in the high-energy limit
as $1/s$, and the corresponding cross-section is suppressed w.r.t.\ the
one for purely-transverse and mixed polarization states at large
energy \cite{Duncan:1986fk,Willenbrock:1987xz}.  In contrast, the amplitude for
longitudinal $\PW^+\PW^-$ production is not suppressed in the same
way, since the amplitude for the production of a pair of charged
Goldstone bosons is finite, thanks to the diagram with an $s$-channel
exchange of a neutral EW boson.  These on-shell results indicate
 a very small purely-longitudinal $\PZ\PZ$ signal,
which is expected also when including off-shell effects and
radiative corrections.

\subsection{Numerical tools and input parameters}\label{subsec:input}
We study the production of $\Pe^+\Pe^-\mu^+\mu^-$ in LHC proton--proton
collisions at a centre-of-mass energy of 13 TeV with NLO QCD and EW accuracy
within the SM.

The computation is performed with \mocanlo, a multi-channel Monte Carlo integrator
that has already been used for several multi-boson LHC processes, and
in particular for the NLO QCD corrections to polarized $\PW^+\PW^-$ and $\PW^+\PZ$ production
\cite{Denner:2020bcz,Denner:2020eck}. Tree-level and one-loop SM amplitudes are
computed with \recola \cite{Actis:2012qn,Actis:2016mpe}, tensor and scalar loop integrals are reduced and evaluated with
\collier \cite{Denner:2016kdg}. 
The five-flavour scheme is used for both LO and NLO predictions, and all contributions
with external $\Pb$~quarks (both in the initial and the final state) are included in
the computation.
All quarks and leptons are assumed to be massless, and no quark
mixing is understood (unit CKM matrix).
The pole masses $M_V$ of weak bosons are computed from the corresponding
on-shell masses $\MVOS$ \cite{Tanabashi:2018oca},
\begin{alignat}{2}\label{eq:ewmasses}
 \MWOS &= 80.379 \GeV,&\qquad \GWOS &= 2.085\GeV, \nnb\\
 \MZOS &= 91.1876 \GeV,&\qquad \GZOS &= 2.4952\GeV, 
\end{alignat}
via the well-known expressions \cite{Bardin:1988xt},
\beq
 M_V = \frac{\MVOS}{\sqrt{1+(\GVOS/\MVOS)^2}}\,,\qquad  
 \Gamma_V = \frac{\GVOS}{\sqrt{1+(\GVOS/\MVOS)^2}}.
\eeq
The EW coupling $\alpha$ is fixed in the $G_\mu$ scheme \cite{Denner:2000bj},
\beq
\alpha = \frac{\sqrt{2}}{\pi}\,G_\mu\MW^2\left(1-\frac{\MW^2}{\MZ^2}\right)\,,\qquad\,\GF = 1.16638\cdot10^{-5} \GeV^{-2}.
\eeq
The top-quark and Higgs-boson masses are set to
\beq\label{smpar2}
 \MH = 125\GeV,\qquad \Mt = 173 \GeV\,.
\eeq
The complex-mass scheme is utilized for the treatment of unstable particles
\cite{Denner:2000bj,Denner:2005fg,Denner:2006ic,Denner:2019vbn}.  The subtraction of
infrared singularities is achieved in the dipole formalism
\cite{Catani:1996vz,Dittmaier:1999mb,Catani:2002hc} both for QCD and
EW corrections.  The values of parton distribution functions
(PDFs) and the running of $\as$ are passed to \mocanlo via the
\textsc{LHAPDF6} interface \cite{Buckley:2014ana}.
We choose the \sloppy\textsc{NNPDF31\_nlo\_as\_0118\_luxQED} PDF set \cite{Ball:2017nwa,Bertone:2017bme},
which understands $\as(\MZ)=0.118$ and includes also the modelling of the photon luminosity in the proton
\cite{Bertone:2017bme}.
The same PDF set is used for both LO- and
NLO-accurate results. Initial-state collinear singularities are
absorbed in PDFs via the $\overline{\rm MS}$ scheme.  The
factorization and renormalization scales are simultaneously set to the
$\PZ$-boson pole mass, \ie $\mu_{\rR}=\mu_{\rF}=\MZ$.

\subsection{Kinematic selections}\label{subsec:setup}
Photon- and jet-clustering is performed with the anti-$k_{\rm T}$
algorithm \cite{Cacciari:2008gp}. Recombination is done for
particles with a maximum rapidity of 5. In particular,
photons are recombined with charged leptons $\ell=\Pe,\mu$ if
\beq
\Delta R_{\gamma\ell} = \sqrt{\Delta y_{\gamma\ell}^2+\Delta \phi_{\gamma\ell}^2} < 0.1.
\eeq
QCD jets can only arise from real initial-state radiation
and cannot generate singular configurations with the final-state
leptons. Their kinematics is not constrained by any angular-distance
cut.

Following a similar approach as in \citere{Denner:2020bcz},
we have considered two sets of cuts, a first one
(label INC) that is rather inclusive in the kinematics
of final-state leptons and a second one (label FID) that
mimics the fiducial region used in the latest ATLAS measurement \cite{ATLAS:2021kog}.
The INC setup just involves
\begin{itemize}
\item a technical cut on the transverse momentum of each charged
  lepton, $\pt{\ell}>0.001\GeV$;
\item an invariant-mass cut on the two pairs of opposite-sign, same-flavour leptons,
  $|M_{\ell^+\ell^-}-\MZ|<10\GeV$.
\end{itemize}
The second cut is essential to enhance the on-shell production of two
$\PZ$~bosons, \ie to
reduce the photon contamination and other non-doubly-resonant contributions.\\
The FID setup involves
\begin{itemize}
\item a minimum transverse-momentum and a maximum-rapidity cut for the electron and the positron,
  $\pt{\Pe^\pm}>7\GeV$, $|\eta_{\Pe^\pm}|<2.47$;
\item a minimum transverse-momentum and a maximum-rapidity cut for the muon and the antimuon,
  $\pt{\mu^\pm}>5\GeV$, $|\eta_{\mu^\pm}|<2.7$;
\item a minimum transverse-momentum cut on the leading, $\pt{\ell_1}>20\GeV$, and on the
  sub-leading charged lepton, $\pt{\ell_2}>10\GeV$ (sorted
according to transverse momentum);
\item a rapidity--azimuthal-angle separation between any two leptons, $\Delta R_{\ell\ell'}>0.05$;
\item an invariant-mass cut on the two pairs of opposite-sign, same-flavour leptons,
  $|M_{\ell^+\ell^-}-\MZ|<10\GeV$;
\item a minimum invariant-mass cut on the four-lepton system,
  $M_{4\ell}>180\GeV$.
\end{itemize}
We have checked numerically that the effect of the $M_{4\ell}>180\GeV$ cut
is to enhance the on-shell $\PZ\PZ$ production even in the absence of the
$|M_{\ell^+\ell^-}-\MZ|<10\GeV$ cut: under these conditions the non-doubly-resonant
background would account for 15\% of the full result. The joint effect
of $M_{4\ell}>180\GeV$ and $|M_{\ell^+\ell^-}-\MZ|<10\GeV$ cuts reduces the non-doubly-resonant
background to less than 1\% of the full cross-section.

\section{Results}\label{sec:results}
We now present the numerical results of our computation.
In \refse{subsec:inc} we consider the INC setup and NLO EW accuracy.
These results provide both a validation of the calculation and a first
assessment of how the polarization structure of $\PZ\PZ$ production is
modified by EW corrections.
The complete NLO results, combined with loop-induced contributions,
and a discussion of the relative impact of NLO EW and QCD corrections
are then shown for the FID  setup in \refse{subsec:fid}.

\subsection{Inclusive phase-space: NLO EW results}\label{subsec:inc}
The results presented in this section have been obtained in the
INC setup described in \refse{subsec:setup}.

Before tackling the polarized signals, we discuss the differences
between the full calculation and the DPA-approximated one for unpolarized production.
The proper functioning of the subtraction of
EW IR singularities in the DPA calculation has been tested by varying the
subtraction parameter $\alpha_{\rm dip}$ \cite{Nagy:1998bb}
between $10^{-2}$ and 1.  The sum of real-subtracted and
integrated-dipole contributions has been found to be independent of
$\alpha_{\rm dip}$, demonstrating that the combined
application of DPA techniques presented in \refse{subsec:nloqcd}
and \refse{subsec:nloew} works in the desired manner. In
\refta{table:sigmaunpolNLOEW} we show the various contributions to the
NLO EW cross-section, separating them into contributions of partonic channels.
\begin{table}
  \begin{center}
    \renewcommand{\arraystretch}{1.3}
    \begin{tabular}{C{2.3cm}|C{2.1cm}C{1.5cm}|C{2.1cm}C{1.5cm}|C{2.9cm}}%
      \cellcolor{blue!9}   & \cellcolor{blue!9}  full [fb]  &\cellcolor{blue!9}  ratio &\cellcolor{blue!9} DPA [fb]       & \cellcolor{blue!9}ratio &\cellcolor{blue!9}{$(\rm DPA/full-1)$}    \\
      $\sigma_{\rm LO }^{q\bar{q}}$                             &   20.993(1) & 100\%  &    20.749(1) & 100\%  &  $-1.16\%$  \\
      $\sigma_{\rm LO}^{\gamma\gamma}$                         &   0.01509(1) & 0.07\%  &   0   & 0\% &  -  \\
      \cellcolor{green!9} $\sigma_{\rm LO}$ &  \cellcolor{green!9} 21.008(1)  &\cellcolor{green!9}100.07\%  & \cellcolor{green!9}  20.749(1)    & \cellcolor{green!9} 100\% & \cellcolor{green!9}  $-1.23\%$ \\
      $\Delta\sigma_{\rm EW}^{q\bar{q}}$                &   $-2.273(1)$ & $-10.83\%$ &   $-2.2235(6)$  & $-10.72\%$  & $-2.18\%$   \\
      $\Delta\sigma_{\rm EW}^{\gamma\Pq,\gamma\bar{\Pq}}$   &   $-0.00579(8)$ & $-0.03\%$  &  0.000850(1) & $0.004\%$ &  - \\
      $\Delta\sigma_{\rm EW}^{\gamma\gamma}$            &   0.00115(2)  & $0.005\%$&   0  & 0\% & -   \\
      \cellcolor{green!9}$\Delta\sigma_{\rm EW}$ & \cellcolor{green!9} $-2.277(1)$  & \cellcolor{green!9} $-10.85\%$& \cellcolor{green!9}$-2.2226(6)$  &\cellcolor{green!9} $-10.71\%$ & \cellcolor{green!9} $-2.38\%$   \\
      \cellcolor{green!9}$\sigma_{\rm NLO_{\rm EW}}$  & \cellcolor{green!9}18.731(2) & \cellcolor{green!9} 89.22\% & \cellcolor{green!9} 18.526(1) &\cellcolor{green!9} 89.29\% & \cellcolor{green!9}$-1.09\%$ \\
    \end{tabular}
  \end{center}
  \caption{Contributions to the NLO EW cross-section for four-lepton production at the LHC
    in the full and DPA (unpolarized) computation. Ratios are computed relatively to the LO
    result for the $q\bar{q}$ partonic channel. The INC setup is understood (see \refse{subsec:setup}). Numerical errors are shown in parentheses. } 
  \label{table:sigmaunpolNLOEW}
\end{table}
The LO DPA result is roughly 1.2\% smaller than the full one, in agreement
with the expected difference of order $\GZ/\MZ=2.7\%$ (intrinsic accuracy of the DPA). Note that
when performing the DPA for $\PZ$~bosons, an on-shell constraint
($|M_{\ell^+\ell^-}-\MZ|<10\GeV$ in our case) is crucial to suppress the
photon contamination to the leptonic decays, which is included in the full but not in the DPA
result \cite{Ballestrero:2019qoy}.
At LO the full result receives a very small contribution from the $\gamma\gamma$-induced
partonic process.

The EW corrections in the $q\bar{q}$ partonic channel dominate
the total NLO EW correction, and the DPA
underestimates (in size) the full result by 2.2\%. This difference
comes mostly  from the large virtual and integrated-dipole
contributions ($-2.610$ fb in the full calculation) and is enhanced
by the relative sign between virtual and real corrections.  Note that
the real contributions are small and positive (0.337 fb in the full
calculation) and are approximated by the DPA within 0.9\%.

The $\gamma q$ and $\gamma\bar{q}$ partonic channels participate in the calculations
as real corrections that involve only initial-state singularities. A
marked difference is found between the DPA numerical result and the
full one for such partonic channels, 
since the real-subtracted contributions in the full case embed additional dipoles
compared to those that are included in the DPA, corresponding to the
non-doubly-resonant contributions with 
underlying $\gamma\gamma$ as Born partonic channel [see \reffi{fig:aRisr2}]. This
causes the presence of additional integrated dipoles from the $\gamma\gamma$
Born contribution that are absent in the DPA calculation.
Note that the EW corrections to the $\gamma\gamma$-induced process account
for less than 0.01\% of the LO cross-section.
Summing all contributions, the NLO EW cross-section in the DPA is 1.1\% smaller
than the full one, indicating a small non-doubly-resonant background
from photon contamination and missing off-shell effects, in particular in the region
between $162\GeV<M_{4\ell}<182\GeV$ where the DPA contribution
vanishes.
The overall EW corrections account for $-10.8\%$ ($-10.7\%$) of the full (DPA)
LO cross-section. This large effect is mostly due to the on-shell constraint
imposed on the final-state leptons. Similar EW corrections have been obtained with
analogous invariant-mass constraints in \citere{Chiesa:2020ttl}. If a looser invariant-mass
cut is applied, the impact of EW corrections is smaller
\cite{Biedermann:2016yvs,Biedermann:2016lvg}.

The difference between the full and the unpolarized DPA calculations implies
that the non-doubly-resonant background is very small ($\approx 1\%$), and on-shell
$\PZ\PZ$ production dominates the four-lepton production in the considered setup. Larger
effects may arise in differential distributions, as observed in $\PW^+\PW^-$ and $\PW\PZ$ production
\cite{Denner:2020bcz,Denner:2020eck} where at large transverse momentum single-resonant
contributions become sizeable, generating large non-doubly-resonant backgrounds. However, in the four-charged-lepton
channel the single-resonant contributions are much more suppressed than in the presence of final-state neutrinos,
thanks to the possibility of constraining both $\PZ$~bosons to be almost on-shell via
physical invariant-mass cuts.

We now consider the $\PZ\PZ$ signals where both bosons have definite polarization states (L or T).
In \refta{table:sigmainclNLO} we show the LO and NLO EW integrated cross-sections for the unpolarized and
polarized process. Polarizations are defined either in the CM or in the LAB frame. 
 \begin{table}
 \begin{center}
 \renewcommand{\arraystretch}{1.3}
 \begin{tabular}{C{2.6cm}C{2.cm}C{1.8cm}C{2.2cm}C{1.8cm}C{2.2cm}}%
 \cellcolor{blue!9} mode  & \cellcolor{blue!9}  $\sigma_{\rm LO}$ [fb]  &  \cellcolor{blue!9}{ $f_{\rm LO}$} &\cellcolor{blue!9} $\sigma_{\rm NLO_{\rm EW}}$ [fb]    &  \cellcolor{blue!9}{ $f_{\rm NLO_{\rm EW}}$}   & \cellcolor{blue!9}{$\delta_{\rm EW}$}\\
  full                &  21.008(1)   & -  &   18.731(2)       &  - & $-10.8\%$\\
  unpol.              &  20.749(1)   & 100\%  &   18.526(1)       & 100\%  &  $-10.7\%$\\
  \multicolumn{6}{c}{\cellcolor{green!9} polarizations defined in the CM frame}\\
  $\PZ_{\rL}\PZ_{\rL}$  &  1.2049(3)   & 5.81\%  &   1.0820(2)      & 5.84\%  &  $ -10.2\% $\\
  $\PZ_{\rL}\PZ_{\rT}$  &  2.3886(2)   & 11.51\%  &   2.1486(3)     &11.60\% &  $ -10.0\% $   \\
  $\PZ_{\rT}\PZ_{\rL}$  &  2.3887(2)   & 11.51\%  &   2.1488(3)    & 11.60\% &  $ -10.0\% $   \\
  $\PZ_{\rT}\PZ_{\rT}$  &  14.737(1)   & 71.03\% &   13.117(1)     & 70.80\%&  $ -11.0\% $  \\
  interf.             &  0.030(2)    & 0.14\%  &  0.029(2)    &  0.16\%   &  - \\
  \multicolumn{6}{c}{\cellcolor{green!9} polarizations defined in the LAB frame}\\
  $\PZ_{\rL}\PZ_{\rL}$   &  0.32525(6) & 1.57\% &    0.29156(7)    & 1.58\% &   $ -10.4\%$   \\
  $\PZ_{\rL}\PZ_{\rT}$   &  5.9542(5)  & 28.70\% &    5.3507(6)      & 28.88\%&    $ -10.1\%$ \\
  $\PZ_{\rT}\PZ_{\rL}$   &  5.9541(4)  & 28.70\% &     5.3505(6)     & 28.88\%&    $ -10.1\%$   \\
  $\PZ_{\rT}\PZ_{\rT}$   &   8.4752(5)  & 40.85\% &    7.4956(6)     & 40.46\% &   $ -11.6\%$    \\
  interf.             &  0.040(2)    & 0.19\%  &  0.038(2)    &  0.20\%   &  - \\
 \end{tabular}
 \end{center}
 \caption{Integrated cross-sections in the INC setup (see \refse{subsec:setup}) 
   for unpolarized and doubly-polarized $\PZ\PZ$ production at the LHC at LO 
   and NLO EW accuracy. The polarization fractions $f_{\rm LO},\,f_{\rm NLO_{\rm EW}}$ are computed as ratios of 
   polarized cross-sections over the unpolarized DPA one 
   at LO and at NLO EW, respectively.
   The NLO EW corrections are shown in the right-most column,
   as percentages relative to LO cross-sections ($\delta_{\rm EW}$).
   The full results are obtained including all doubly-resonant
   and non-doubly-resonant contributions, while all other results are based
   on the DPA described in \refse{subsec:nloew}. 
   Interference contributions are obtained by subtracting the sum of
   polarized results from the DPA unpolarized one. 
   Numerical errors are shown in parentheses.
 } 
 \label{table:sigmainclNLO}
 \end{table}
 The interference among polarization states is evaluated as the difference
 between the unpolarized DPA result (which, by definition, contains also interference terms)
 and the sum over doubly-polarized results. For integrated cross-sections, 
 interferences are found to be at the 0.2\% level both at LO and at NLO EW.

 It has been shown \cite{Denner:2020eck} that the spin correlations between the two bosons can be
 sizeable in the CM-frame definition of polarization and enhanced if
 both bosons are in
 a definite polarization state. Therefore, the sum over doubly-polarized modes is expected to
 differ from the unpolarized DPA result more than the sum over singly-polarized modes,
 \ie larger interference effects are present.
 
 Up to the correlation effects mentioned above, vanishing interferences
 are expected for the inclusive case at LO, where the two-body decay of each $\PZ$~boson
 enables an exact analytic expression of the unpolarized DPA cross-section in terms of a sum
 over polarized contributions \cite{Bern:2011ie,Stirling:2012zt}:
 \begin{align}\label{eq:costhetaINCL}  
   \frac{1}{\sigma}\frac{\rd\sigma}{\rd\cos\tl}={}&
   \frac{3}{4} f_{\rL}\,\,\left(1-\cos^2\tl\right) \nnb\\
   &+ \frac{3}{8} f_{-}\, \left(1+\cos^2\tl-2\cos\tl\frac{c^2_{\mathrm{L},\ell}-c^2_{\mathrm{R},\ell}}{c^2_{\mathrm{L},\ell}+c^2_{\mathrm{R},\ell}}\right)\nnb\\
   &+ \frac{3}{8} f_{+}\, \left(1+\cos^2\tl+2\cos\tl\frac{c^2_{\mathrm{L},\ell}-c^2_{\mathrm{R},\ell}}{c^2_{\mathrm{L},\ell}+c^2_{\mathrm{R},\ell}}\right)\,,
 \end{align}
 where $\tl$ is the decay angle of the antilepton in the rest frame of the corresponding $\PZ$~boson (with
 momentum equal to the sum of the two lepton momenta) computed
 w.r.t.\ the $\PZ$~direction in the CM (LAB) frame for polarization vectors defined in the CM (LAB) frame.
 The coefficients $f_{\rL}$, $f_+$, $f_-$ are polarization fractions,
 and $c_{\rm L,\ell}$, $c_{\rm R,\ell}$ are the left- and right-handed
 couplings of the $\PZ$~boson to leptons. This expression can be
 derived assuming that the complete phase-space of 
 decay products is available, such that interference terms vanish upon integration over the azimuthal decay angle
 \cite{Ballestrero:2017bxn}.
 Projecting the $\cos\tl$ distribution onto suitable polynomials in $\cos\tl$
 \cite{Bern:2011ie,Stirling:2012zt,Ballestrero:2017bxn}, it is possible to extract
 the polarization fractions for the $\PZ$~boson and compare the result with the one
 obtained by directly simulating polarized $\PZ$~bosons with the Monte Carlo.
 In the INC setup, this comparison yields a very good agreement for the LO
 predictions as well as for other $n$-body contributions (virtual, integrated dipoles)
 to the NLO EW cross-section.
While doubly-polarized polarization fractions cannot be obtained
from projections on parametrizations of unpolarized cross-sections like
Eq.~\refeq{eq:costhetaINCL}, their definition is straightforward
based on Eq.~\refeq{eq:squaredAres}.
 The LO interference estimated from the sum of the thus defined doubly-polarized Monte Carlo predictions
 is not identically zero (as it would be summing singly-polarized signals)
 due to the above-mentioned correlation effects. 
 
 In the presence of final-state real corrections, Eq.~\eqref{eq:costhetaINCL}
 does not hold anymore, as one $\PZ$~boson decays into three particles,
 therefore projecting the $\cos\tl$ distributions in the same way as at LO gives
 results that have nothing to do with polarization fractions \cite{Baglio:2018rcu}.
 Furthermore, in the presence of photons radiated off decay leptons,
 the interferences among polarization states are not vanishing anymore, even
 if the complete phase-space is available for the three-body decay.
 For example, if we consider a polarized $\PZ$~boson that
 decays into $\Pe^+\Pe^-\gamma$, the polarization fractions from Monte Carlo simulations ($f^{\rm MC}_{i}$)
 and those obtained projecting on $\cos\theta^*_{\Pe^+}$ distributions
 ($f^{\rm proj}_{i}$) read
 \begin{align}
   f^{\rm MC}_{\rL} ={}&  0.1437(3),\,\qquad & f^{\rm MC}_{\rT} ={}&  0.837(1)\,\nnb\\
   f^{\rm proj}_{\rL} ={}&  0.317(1),\,\qquad & f^{\rm proj}_{\rT} ={}& f^{\rm proj}_++f^{\rm proj}_- = 0.683(4)\,.
 \end{align}
 The discrepancy between the two evaluations of polarization fractions is large,
 indicating that real contributions to polarized signals can only be computed
 with polarized amplitudes in the Monte Carlo, even in a very inclusive setup
 like the one that is understood here.
 Note that the sum of polarization fractions extracted with $\cos\theta^*_{\Pe^+}$  projections
 gives 1 by definition, while the sum of polarization fractions (ratios of polarized cross-sections
 over the DPA unpolarized one) computed with the Monte Carlo suggests that a 2\% effect comes
 from interferences. However, although interferences in FSR1/2 corrections are at the 2\% level,
 when summing all EW corrections to $\PZ\PZ$ production this effect is hardly visible
 in the NLO EW total cross-section, since the real corrections are small compared to the virtual ones.
 This explains why even at NLO EW the overall interferences are small
 (0.16\% and 0.20\% for polarizations defined in the CM and the LAB frame, respectively).

 It is worth noting that in both polarization definitions the relative NLO EW corrections
 stay between $-10\%$ and $-12\%$ for all modes, with slighly more sizeable
 corrections for the $\rL\rL$ and $\rT\rT$ modes than for the mixed ones.
 In both definitions, the polarization fractions are hardly modified by NLO EW effects: the modes with at least one longitudinal boson are
 slightly enhanced, while the $\rT\rT$ mode is mildly diminished.
 In the CM-frame definition, the $\rL\rL$ and the mixed modes are of the same order of magnitude
 (5.8\% and 11.6\% respectively), while the $\rT\rT$ mode is dominant. This situation is completely
 changed in the LAB definition, as the $\rL\rL$ mode accounts for just 1.6\%, while the
 sum of mixed contributions gives  57.8\% of the total. This difference, in a similar fashion
 as in $\PW^+\PZ$ production \cite{Denner:2020eck}, is due to the fact that part of the $\rL\rL$
 contributions defined in the CM frame (where the polarization vectors just depend on one $\PZ$-boson momentum, thanks to
 the back-to-back kinematics) become transverse when boosting to the LAB frame, giving a more
 involved dependence on kinematic quantities. This has also the effect of reducing the $\rT\rT$
 component in favour of the mixed ones.
 
\begin{figure}
   \centering
    \subfigure[$\cos\theta^{*,\text{CM}}_{\Pe^+}$, CM\label{fig:cthcmCM}]{\includegraphics[scale=0.28]{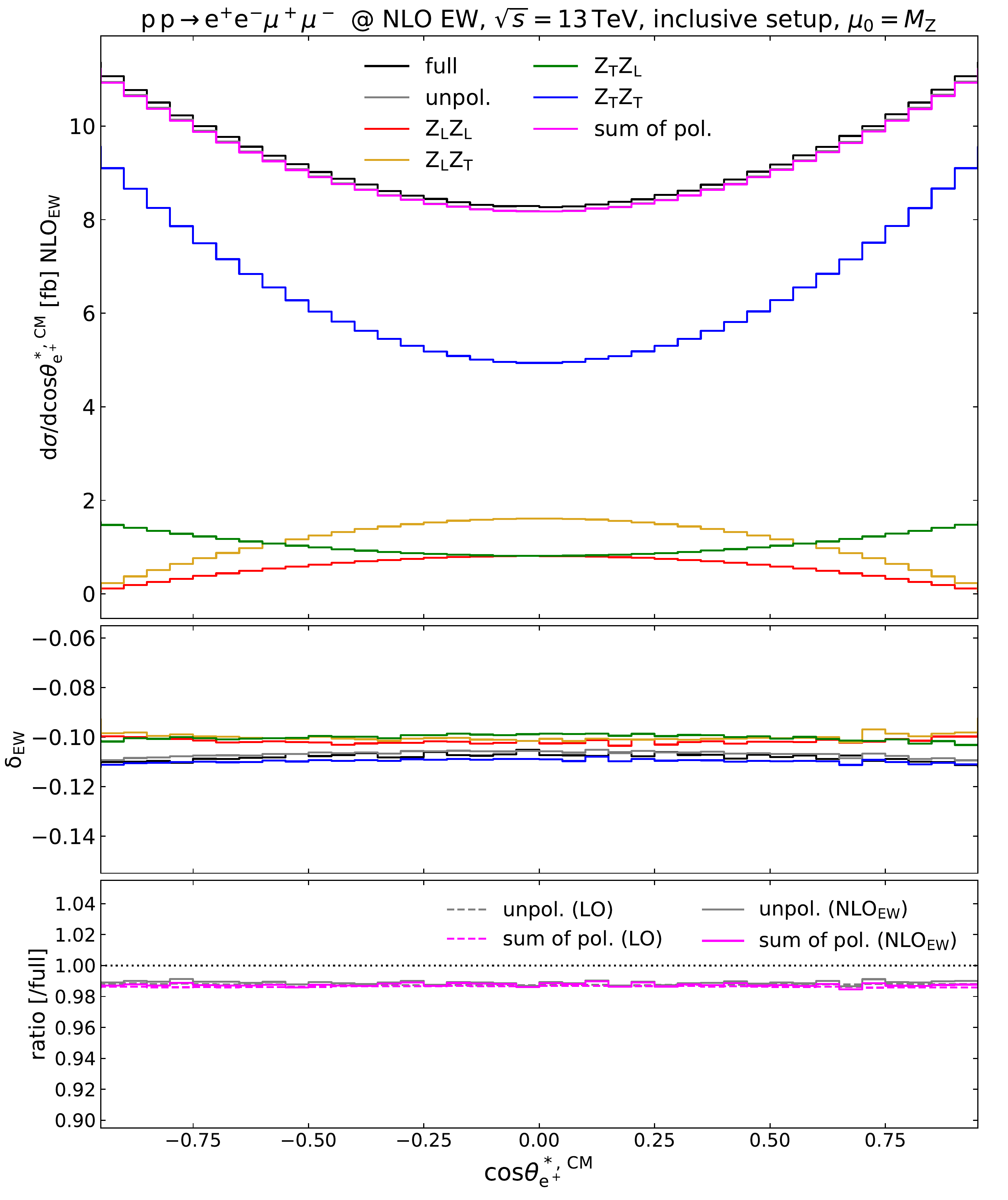}}
    \subfigure[$\cos\theta^{*,\text{CM}}_{\Pe^+}$, LAB\label{fig:cthcmLAB}]{\includegraphics[scale=0.28]{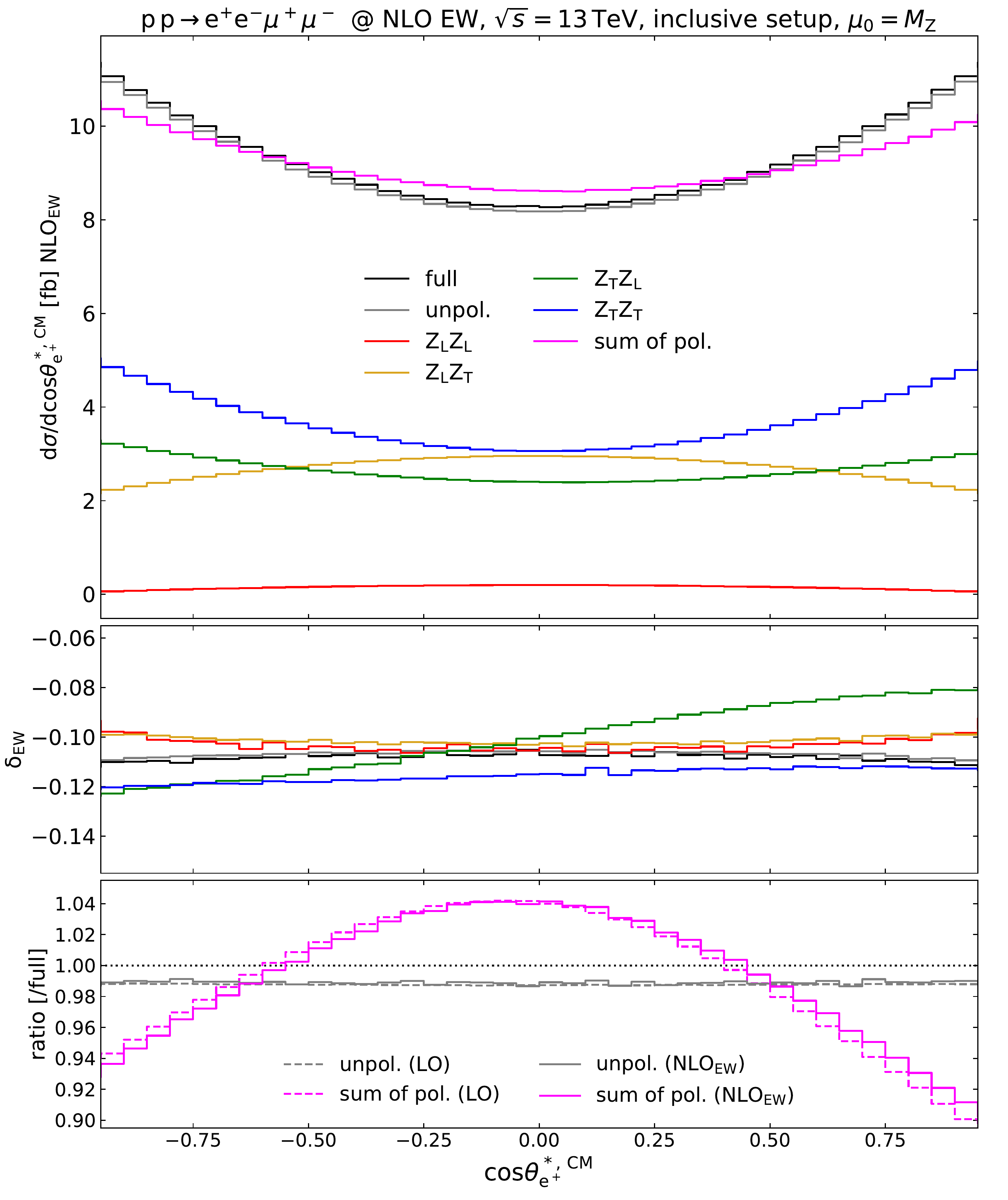}}
    \subfigure[$\cos\theta^{*,\rm LAB}_{\Pe^+}$, CM\label{fig:cthlabCM}]{\includegraphics[scale=0.28]{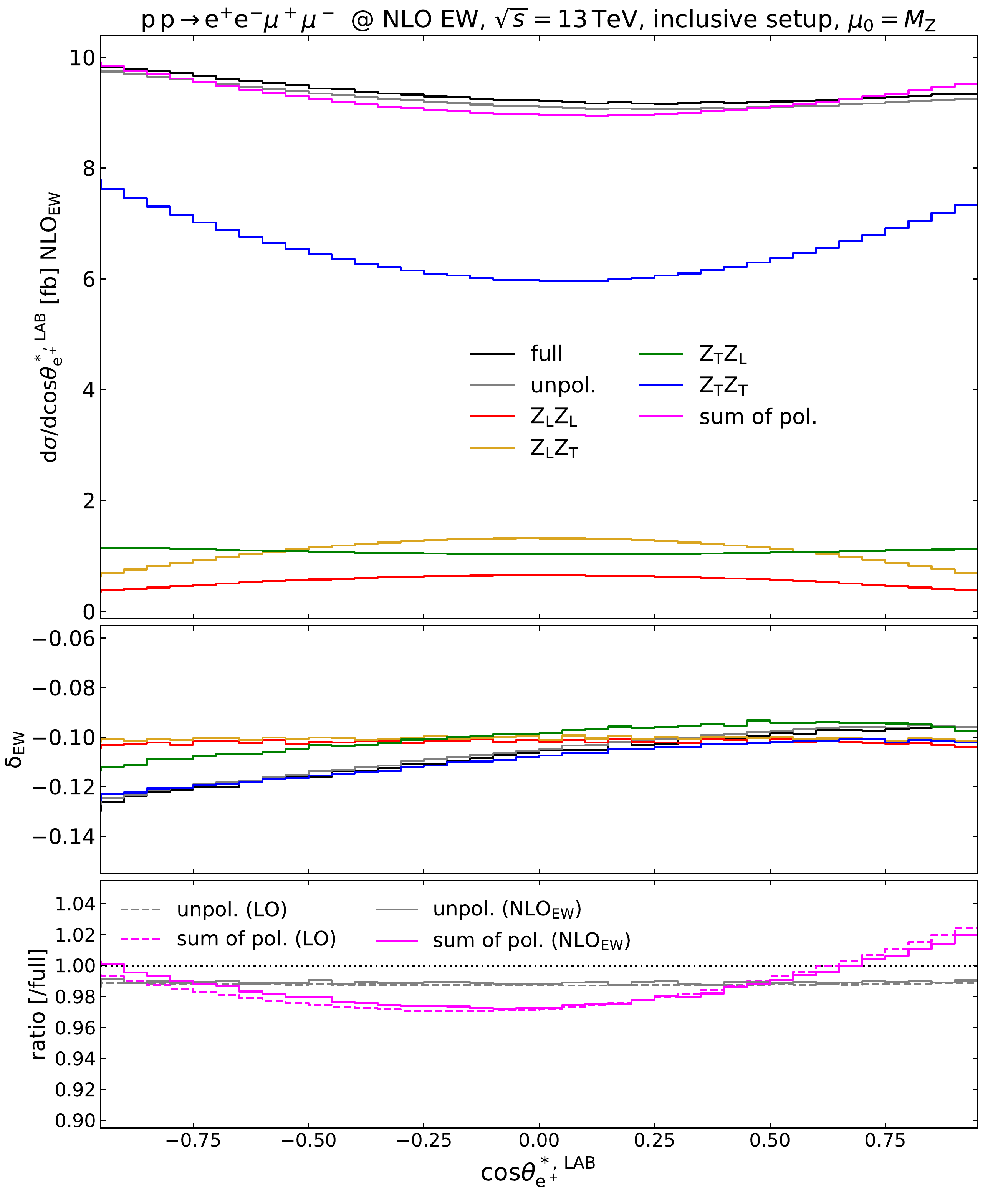}}
    \subfigure[$\cos\theta^{*,\rm LAB}_{\Pe^+}$, LAB\label{fig:cthlabLAB}]{\includegraphics[scale=0.28]{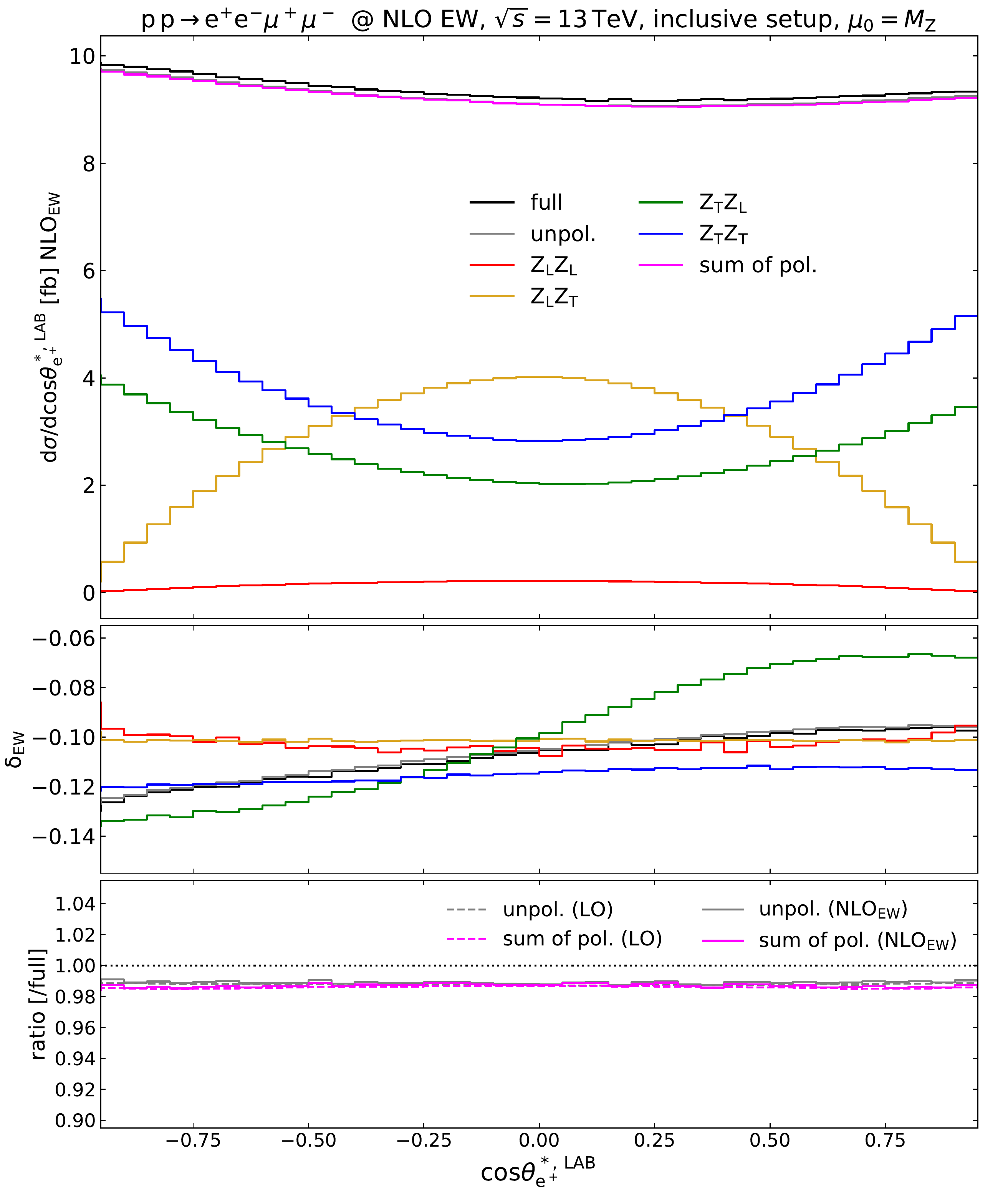}}
    \caption{Distributions in the cosine of the polar decay angle $\theta^*$ of the positron in
      its corresponding $\PZ$-boson rest frame computed w.r.t.\ the $\PZ$-boson spatial direction
      in the CM frame (a,b) and in the LAB frame (c,d). Polarizations
      are defined in the CM (a,c) and in the LAB (b,d) frame.
      Unpolarized and doubly-polarized results are shown for $\PZ\PZ$ production at the LHC
      in the inclusive setup described in \refse{subsec:setup}.
      From top down: \NLOew differential cross-sections, EW corrections relative to LO ($\delta_{\rm EW}$), and ratios of DPA over the full result.
    }\label{fig:cthcm} 
\end{figure}

\begin{figure}
  \centering
    \subfigure[$\Delta\phi_{\Pe^+\mu^-}$, CM\label{fig:incAziCM}]{\includegraphics[scale=0.28]{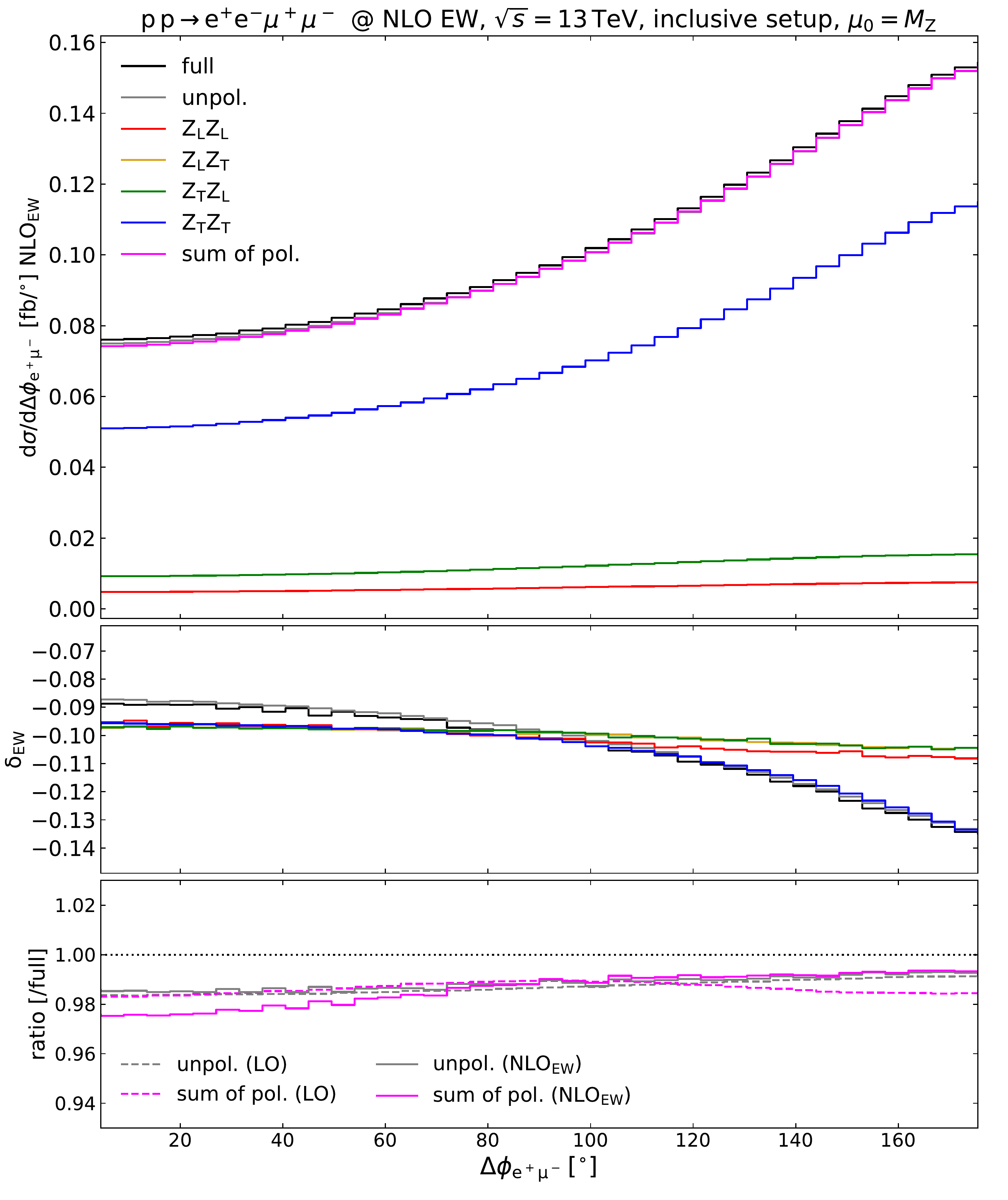}}
    \subfigure[$\Delta\phi_{\Pe^+\mu^-}$, LAB\label{fig:incAziLAB}]{\includegraphics[scale=0.28]{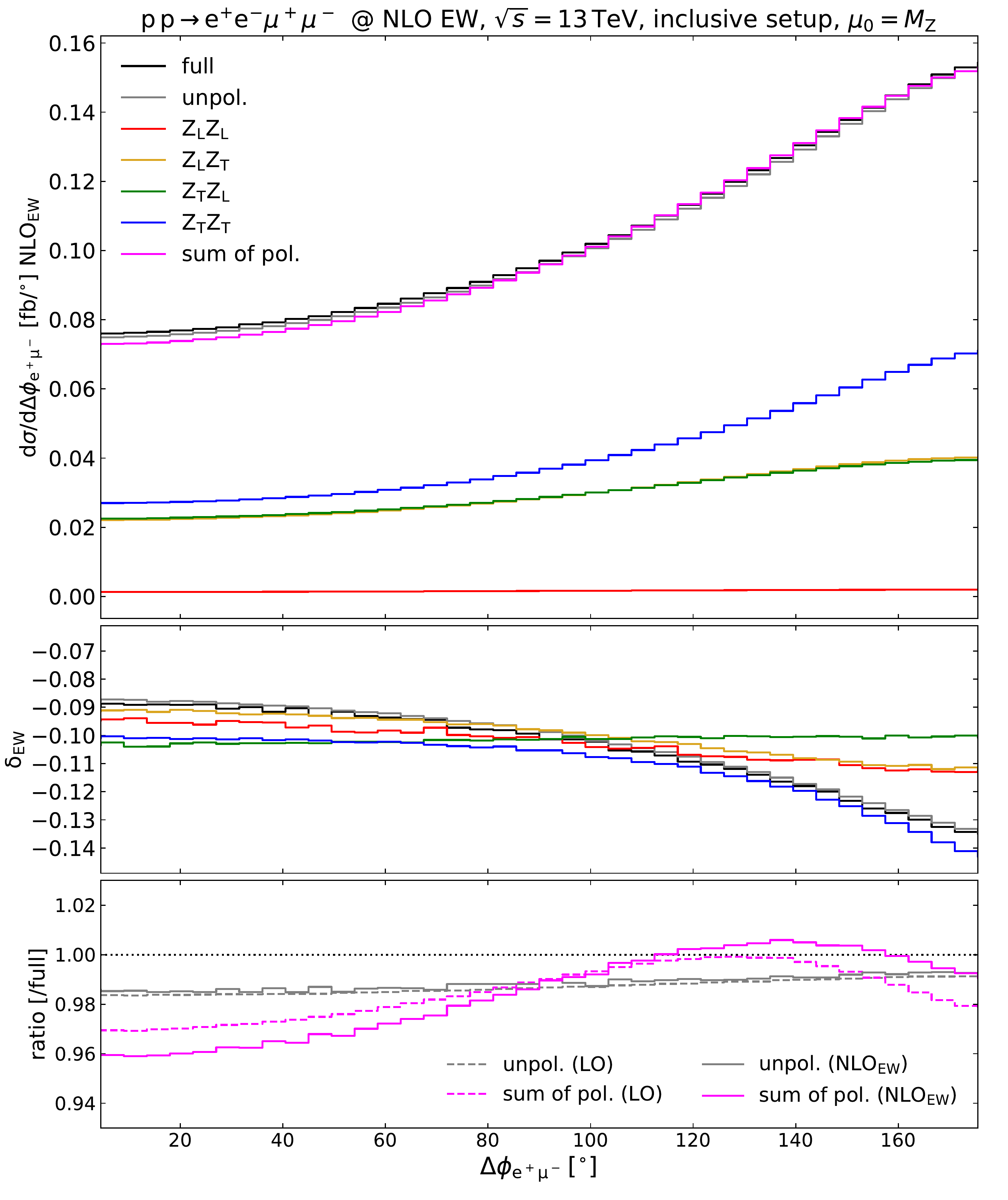}}
    \caption{Distributions in the azimuthal separation between the positron and the muon. Polarizations are defined the CM (a) and in the LAB (b) frame.   
      Same structure as in \reffi{fig:cthcm}.
    }\label{fig:incAzim4l} 
\end{figure}

 The polarized results for differential cross-sections can embed effects that are
 rather different from those obtained at the integrated level,
 in particular concerning the interference effects.
 In \reffis{fig:cthcm}--\ref{fig:incAzim4l}, we study the distributions in a number of LHC
 observables, focusing on the shapes and normalizations of differential cross-sections in top panels,
 the impact of EW corrections in middle panels, and the impact of non-doubly-resonant (grey curves)
 and interference effects (grey vs.\ magenta curves) in bottom panels.

 In \reffi{fig:cthcm} we consider the decay angle of the positron in
 the $\PZ$~rest frame, which has been introduced in Eq.~\eqref{eq:costhetaINCL}. This angular variable is
 expected to be highly sensitive to the polarization of the first \PZ~boson.
 At NLO EW the $\PZ$~boson is understood as the system of the two dressed leptons ($\Pe^+\Pe^-$),
\ie after photon recombination, in order to ensure the IR safety of the variable.
The angle $\theta^*_{\Pe^+}$ can be computed w.r.t.\ the $\PZ$-boson direction
in the CM frame or w.r.t.\ the $\PZ$-boson direction in the LAB frame.
For both definitions of $\theta^*_{\Pe^+}$, the non-doubly-resonant
background effects are flat and equal to those found at the integrated level.
If $\theta^*_{\Pe^+}$ is computed w.r.t.\ the $\PZ$-boson direction in the same 
frame where polarization vectors are defined, the analytic expression for the
differential distribution is known at LO [see Eq.~\eqref{eq:costhetaINCL}] and interferences
are expected to vanish. This is the case in \reffis{fig:cthcmCM} and \ref{fig:cthlabLAB}, where
also at NLO EW the interferences are at the few-permille level owing to
small FSR1/2 real contributions. 
Both in \reffi{fig:cthcmCM} and in \reffi{fig:cthlabLAB}, the $\rL\rL$ and $\rL\rT$ distributions
display the same shape, which has a $\sin^2\theta^*_{\Pe^+}$ functional dependence on the angle.
In \reffi{fig:cthcmCM}, the $\rT\rL$ and $\rT\rT$ distributions
exhibit the same symmetric shape, because of left--right symmetry
of $\PZ$-boson production in the CM frame.%
\footnote{Owing to Bose symmetry and the explicit form of the
    polarization vectors, left- and right-handed $\PZ$~bosons are
    produced with equal probabilities in the CM frame.}
On the contrary, in \reffi{fig:cthlabLAB} the $\rT\rL$ distribution
is not symmetric, due to a different left- and right-polarization content for the first boson
in the presence of a recoiling longitudinal boson. The difference in shape between the $\rT\rL$
and the $\rT\rT$ curves indicates that the polarization of the first boson is affected
differently by different polarization states of the second boson.

If both the polarization vectors and the $\theta^*_{\Pe^+}$ angle are defined in the
CM frame [\reffi{fig:cthcmCM}], the relative EW corrections are flat for both
polarized and unpolarized signals. 
On the contrary, for the LAB-frame definition [\reffi{fig:cthlabLAB}], the EW corrections
enhance the various polarized signals differently. They are rather uniform for the
$\rL\rL,\,\rL\rT$ and $\rT\rT$ components, while for the $\rT\rL$ component they
increase by about 6\% going from the anti-collinear configuration towards the collinear one.
Therefore, differences between the two polarization definitions are found at the level
of EW corrections, even for very inclusive variables like decay angles. Thanks to the flat behaviour
of EW corrections, the CM-frame definition seems preferable over the LAB-frame one.

If polarizations are defined in the CM (LAB) frame, the distributions in the
angle $\theta^*_{\Pe^+}$ computed w.r.t.\ the $\PZ$~direction in the LAB (CM) frame
receive large interference effects, as can be appreciated in \reffis{fig:cthcmLAB}--\ref{fig:cthlabCM}.
The interference pattern is different in the two cases: interferences are in fact negative
in the (anti)collinear configuration in \reffi{fig:cthlabCM} but positive in \reffi{fig:cthcmLAB}.
A difference in size is also evident: if polarizations are defined in the CM frame and the decay
angle is defined w.r.t.\ the $\PZ$~direction in the LAB frame [\reffi{fig:cthlabCM}], the interferences give at most a 3\%
effect, while in the opposite case they give up to 6\% effects.

The distribution in the decay angle $\theta^*_\ell$ is highly sensitive to the
polarization state of the corresponding $\PZ$~boson. However, in order to avoid
noticeable interference effects, it is essential that the definition of the angle is consistent
with the frame where polarizations are defined, as in the case of \reffis{fig:cthcmCM}
and \ref{fig:cthlabLAB}.

In \reffi{fig:incAzim4l} we consider the azimuthal distance between the
positron and the muon computed w.r.t.\ to the beam axis.
The two particles are decay products of different bosons, and tend to be produced with
$\Delta\phi_{\Pe^+\mu^-}=\pi$. For both the unpolarized and the polarized distributions
the EW corrections monotonically decrease with increasing $\Delta\phi_{\Pe^+\mu^-}$.
For unpolarized cross-sections, EW corrections range between $-9\%$ and $-13\%$.
For polarizations defined in the CM frame, all polarized
cross-sections receive the same flat corrections for
$\Delta\phi_{\Pe^+\mu^-}<\pi/2$. For $\Delta\phi_{\Pe^+\mu^-}>\pi/2$,
the EW corrections for $\rT\rT$ decrease along with the unpolarized one,
while those for other polarizations remain rather flat.
A similar behaviour is found in the LAB frame, where all polarized curves (except from the $\rT\rT$ one)
are pretty flat in the whole angular range.

The two definitions of polarizations lead to very different
interference patterns to this angular variable. In the CM-frame definition, the interferences never
exceed 0.8\%, with mild differences between the LO and
the NLO EW results. In the LAB definition, the NLO EW interferences are positive
for $\Delta\phi_{\Pe^+\mu^-}<\pi/2$ and amount to 2.5\% at $\Delta\phi_{\Pe^+\mu^-}=0$,
they are negative but less sizeable in the rest of the spectrum.

\begin{figure}
  \centering
    \subfigure[$\pt{\Pe^+}$, CM\label{fig:pteCM}]{\includegraphics[scale=0.28]{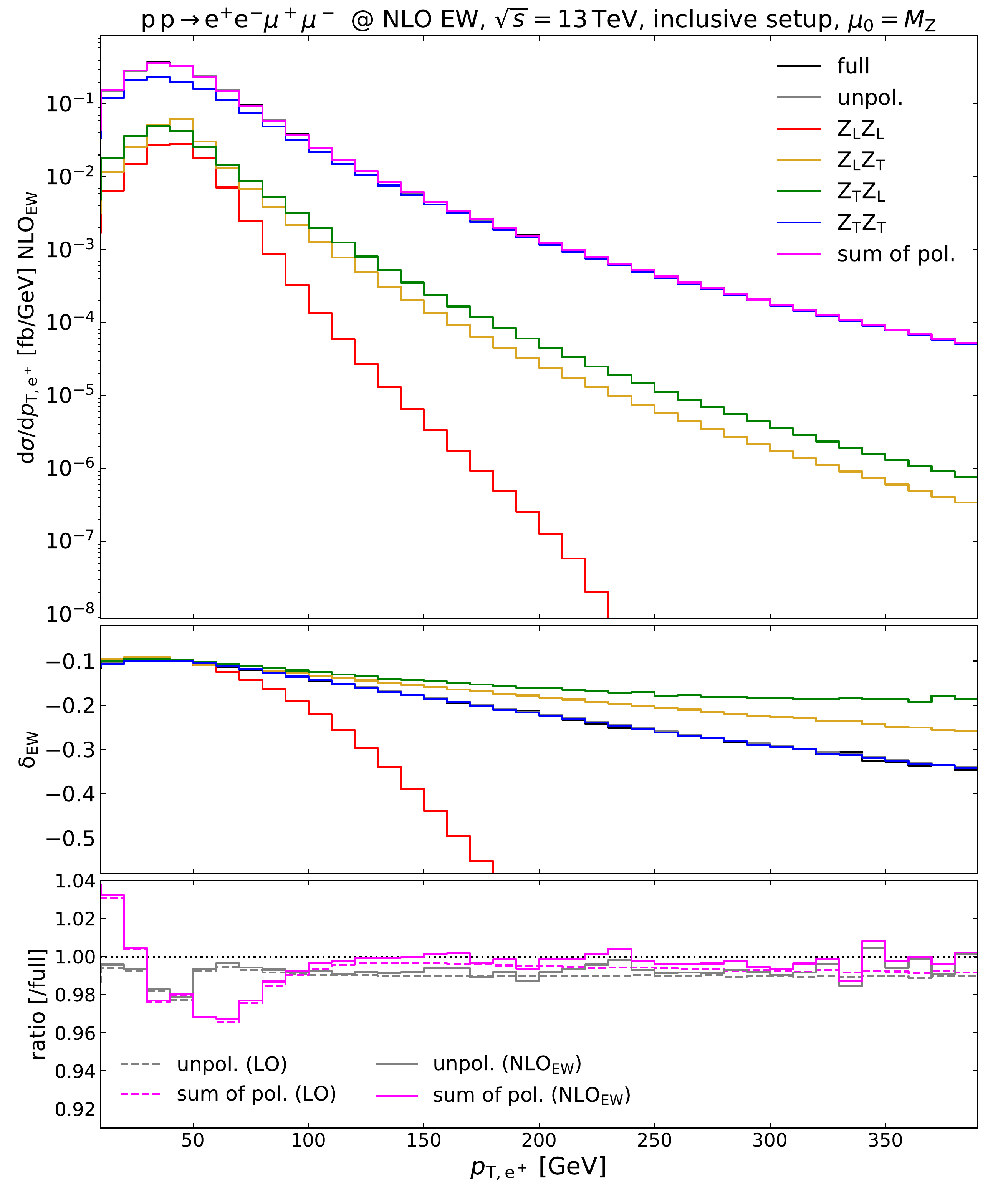}}
    \subfigure[$\pt{\Pe^+}$, LAB\label{fig:pteLAB}]{\includegraphics[scale=0.28]{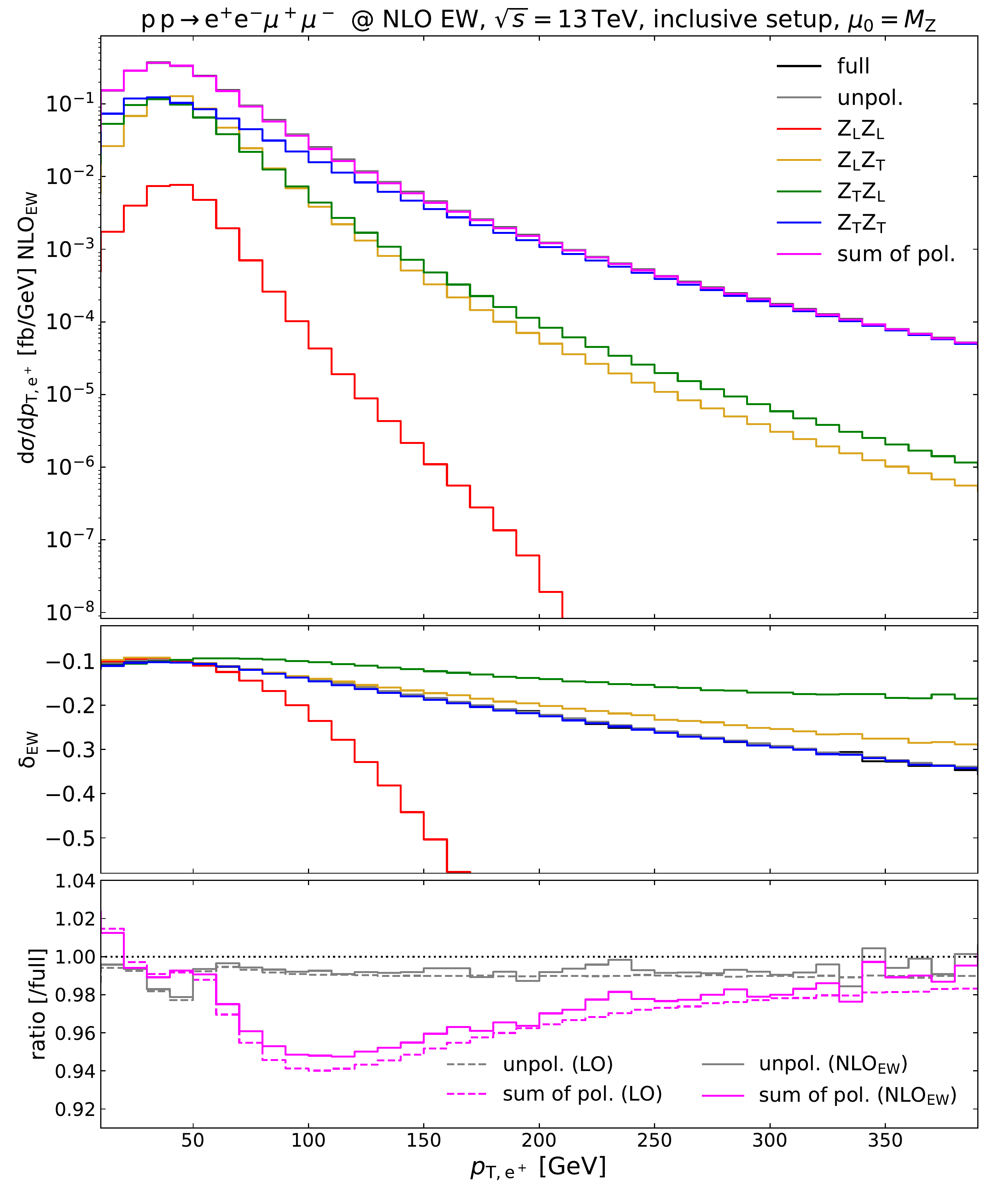}}
    \caption{Distributions in the positron transverse momentum. Polarizations are defined the CM (a) and in the LAB (b) frame.   
      Same structure as in \reffi{fig:cthcm}.
    }\label{fig:pteINC} 
\end{figure}

In \reffi{fig:pteINC} we consider differential cross-sections in the
positron transverse momentum.
The purely-longitudinal signal is strongly suppressed at large
transverse momentum already at LO \cite{Willenbrock:1987xz} and is also characterized by huge
and negatively-increasing EW corrections for both polarization definitions. Around $280\GeV$
the negative virtual corrections exceed the sum of Born and real contributions,
implying that the truncation of the perturbative series at order $\mc O(\alpha^5)$ is not sufficient
at large transverse momentum. We have checked numerically that the same effect can be reproduced
simulating the production of longitudinal on-shell bosons (no decays),
which was also addressed in Ref.~\cite{Bierweiler:2013dja}. 
The longitudinal signal is suppressed by $1/s$ and $1/s^2$ w.r.t.\ the mixed
and the $\rT\rT$ contributions, respectively, both at LO and in NLO
real corrections, which manifests itself by  vanishing
amplitudes for neutral Goldstone-boson pair production at tree
level. On the contrary, some one-loop diagrams give non-vanishing
contributions to Goldstone-boson
amplitudes, which at NLO are interfered with tree-level amplitudes. Including the squares
of such weak one-loop contributions [formally at order $\mc O(\alpha^6)$] makes the cross-section
positive even at large transverse momentum \cite{Bierweiler:2013dja}.
However, the smallness of polarized signals at large $\pt{\PZ}$
renders this region definitely out of reach even with increased LHC luminosity. Similar effects can
be found in the distributions in the lepton-pair transverse momentum, as well as in the four-lepton
invariant mass.

The $\rT\rT$~polarization state dominates already at moderate transverse momentum: in the CM-frame definition, the
mixed states are one order of magnitude smaller than the $\rT\rT$ one already at $\pt{\Pe^+}\approx 100\GeV$,
despite less sizeable EW corrections. In the LAB-frame definition, the mixed states and the $\rT\rT$ one
have similar size in the soft region of the spectrum.
It is interesting to observe the differences in the interference pattern between the two polarization
definitions. In the CM-frame definition they amount to $\pm 3\%$ in the soft region, while they are almost negligible
in the tail. In the LAB-frame definition, interferences are positive and more sizeable (at most 6\%) even at 
moderate-$p_{\rm T}$.

Although the interference effects account for 0.2\% of the integrated cross-section,
they can give more sizeable effects at the differential level, even in the case of
inclusive variables. This holds in particular when polarizations are defined in the LAB frame.
The impact of the non-doubly-resonant background on 
the considered distributions is not so different from the one found for the total cross-section.
The NLO EW corrections significantly distort the distribution shapes for both polarized and
unpolarized signals, with moderate effects in angular variables and dramatic effects in the
tails of transverse-momentum and invariant-mass variables.

\subsection{Fiducial phase-space: complete NLO results}\label{subsec:fid}
Thanks to the back-to-back kinematics at LO, the definition of
polarizations in the CM frame represents a natural choice for di-boson
production. It gives an enhanced $\rL\rL$ contribution, rather small
mixed contributions, and less sizeable interference effects than the
polarization definition in the LAB frame.  Therefore, we only present
results for the CM-frame definition in the fiducial setup.

The realistic cuts applied in the FID setup are rather loose in the lepton transverse momentum and
$R$~distance. The decrease of the inclusive cross-section due to the fiducial cuts is mostly due
to the rapidity cuts, which also introduce an asymmetry between the two $\PZ$~bosons
($|y_{\Pe^\pm}|<2.47$ and $|y_{\mu^\pm}|<2.7$). 
Once the invariant-mass cut on the four-lepton system,
$M_{4\ell}>180\GeV$, is applied,  the cuts on the leading and sub-leading lepton
do not modify much the event rate.  
The presence of lepton cuts is expected to give larger interference
effects than in the inclusive case \cite{Stirling:2012zt,Belyaev:2013nla,Ballestrero:2017bxn}.

In \refta{table:fidNLOQCD} we show the integrated cross-sections
in the FID setup, both for polarized and unpolarized signals.
 \begin{table}
 \begin{center}
 \renewcommand{\arraystretch}{1.3}
 \begin{tabular}{C{1.2cm}C{2.5cm}C{1.3cm}C{1.2cm}C{1.2cm}C{2.3cm}C{2.3cm}}%
 \cellcolor{blue!9} mode  & \cellcolor{blue!9}  $\sigma_{\rm LO}$ [fb]  & \cellcolor{blue!9}{${\delta_{\rm QCD}}$} & \cellcolor{blue!9}{$\delta_{\rm EW}$} & \cellcolor{blue!9}{$\delta_{\rm gg}$} & \cellcolor{blue!9} $\sigma_{\rm NLO_+}$ [fb] & \cellcolor{blue!9} $\sigma_{\rm NLO_\times}$ [fb]\\
  full                & $  11.1143(5) ^{+ 5.6 \%}_{ -6.8  \%} $ &  $+34.9\% $   &       $ -11.0  \% $        &     $+15.6\%$    &  $  15.505(6) ^{+ 5.7 \%}_{ -4.4 \%} $    &   $  15.076(5) ^{+ 5.5  \%}_{ -4.2 \%} $     \\
  unpol.              & $  11.0214(5) ^{+ 5.6  \%}_{ -6.8  \%} $ & $+35.0\% $  &      $ -10.9  \% $         &     $+15.7\%$    &   $  15.416(5) ^{+ 5.7  \%}_{ -4.4 \%} $    &   $  14.997(4) ^{+ 5.5  \%}_{ -4.2  \%} $    \\
  $\PZ_{\rL}\PZ_{\rL}$  & $  0.64302(5) ^{+ 6.8  \%}_{ -8.1  \%} $ & $+35.7\%$ &     $ -10.2 \% $         &    $+14.5\%$    &   $  0.9002(6) ^{+ 5.5\%}_{ -4.3  \%} $    &     $  0.8769(5) ^{+ 5.4\%}_{ -4.1 \%} $   \\
  $\PZ_{\rL}\PZ_{\rT}$  & $  1.30468(9) ^{+ 6.5  \%}_{ -7.7  \%} $ & $+45.3\%$ &     $ -9.9  \% $          &    $+2.8\%$    &    $  1.8016(9) ^{+ 4.3  \%}_{ -3.5 \%} $   &    $  1.7426(8) ^{+ 4.1  \%}_{ -3.3  \%} $   \\
  $\PZ_{\rT}\PZ_{\rL}$  & $  1.30854(9) ^{+ 6.5  \%}_{ -7.7  \%} $ & $+44.3\%$ &     $ -9.9  \% $          &    $+2.8\%$    &   $  1.7933(9) ^{+ 4.3 \%}_{ -3.4 \%} $    &   $  1.7355(8) ^{+ 4.0 \%}_{ -3.2 \%} $    \\
  $\PZ_{\rT}\PZ_{\rT}$  & $  7.6425(3) ^{+ 5.2  \%}_{ -6.4 \%} $ &  $+31.2 \%$   &      $ -11.2 \% $       &    $+20.5\%$    &   $  10.739(4) ^{+ 6.2  \%}_{ -4.7  \%} $   &     $  10.471(3) ^{+ 6.1 \%}_{ -4.6  \%} $   \\
 \end{tabular}
 \end{center}
 \caption{Integrated cross-sections in the fiducial setup (see \refse{subsec:setup}) 
   for unpolarized and doubly-polarized $\PZ\PZ$ production at the LHC at LO 
   and NLO accuracy.
   Polarizations are defined in the CM frame.
   Corrections are shown as percentages relative to LO cross-sections.
   The NLO cross-sections include NLO EW and QCD corrections
   and the loop-induced $\Pg\Pg$ contribution, combined with the additive (+) or multiplicative ($\times$) prescription.
   Theory uncertainties from 7-point scale variations are shown as
   percentages for the LO and NLO results. Parentheses contain
   integration errors.
 } 
 \label{table:fidNLOQCD}
  \end{table}
We have included NLO effects of EW
and QCD type, as well as the loop-induced contribution from the $\Pg\Pg$ partonic channel.
Starting from the LO results, we have combined the higher-order corrections both with an additive
and with a multiplicative approach \cite{Kallweit:2019zez}:
\begin{align}
 \rd \sigma_{\rm NLO_+} ={}&\rd\sigma_{\rm LO}\left(1+\delta_{\rm QCD}+\delta_{\rm EW}\right) +\rd\sigma_{\rm LO}  \delta_{\Pg\Pg}\,,\label{eq:additive}\\
 \rd \sigma_{\rm NLO_\times} ={}& \rd\sigma_{\rm LO}\left(1+\delta_{\rm QCD}\right)\left(1+\delta_{\rm EW}\right) +\rd\sigma_{\rm LO}  \delta_{\Pg\Pg}\,,\label{eq:multiplicative}
\end{align}
where we have defined the relative corrections in the following way:
\begin{align}
  \delta_{\rm EW} \,=\, \frac{\rd\Delta\sigma_{\rm EW}}{\rd\sigma_{\rm LO}}\,,\qquad
  \delta_{\rm QCD}\,=\, \frac{\rd\Delta\sigma_{\rm QCD}}{\rd\sigma_{\rm LO}}\,,\qquad
  \delta_{\rm gg} \,=\, \frac{\rd\sigma_{\rm gg}}{\rd\sigma_{\rm LO}}\,.
\end{align}
The cross-section $\sigma_{\rm LO}$ includes LO contributions in the $q\bar{q}$
and $\gamma\gamma$ partonic channels. The quantity $\Delta\sigma_{\rm EW}$
embeds all EW corrections in the $q\bar{q}$, $\gamma\gamma$, $\gamma q$ and
$\gamma \bar{q}$ channels, while $\Delta\sigma_{\rm QCD}$ furnishes the
QCD corrections in the $q\bar{q}$, $\Pg q$ and $\Pg \bar{q}$ channels. Loop-induced
contributions from the $\Pg\Pg$ partonic channel are included in $\sigma_{\Pg\Pg}$.
The combination of corrections according to Eqs.~\eqref{eq:additive} and \eqref{eq:multiplicative} is performed
on a bin-by-bin basis.
The difference between the multiplicative and the additive results ($=\delta_{\rm EW}\,\delta_{\rm QCD}\,\rd\sigma_{\rm LO}$) 
gives an approximated estimate of the mixed EW--QCD corrections of order $\mc{O}(\alpha^5\as)$.
The multiplicative approach is generally preferable at large partonic energies ($\hat{s}\gg \MZ^2$)
as the dominant EW corrections factorize w.r.t.\ the QCD effects.
In principle, for the factorization properties of EW and QCD corrections,
the multiplicative approach should be applied to the $q\bar{q}$ contributions only.
However, we have checked numerically that, due the smallness of photon-induced contributions, this
further refinement of the multiplicative prescription agrees  at the
permille level with results
obtained applying Eq.~\eqref{eq:multiplicative}, both for integrated cross-sections and for differential
distributions. This conclusion was also found when combining NNLO QCD and NLO EW corrections to $\PZ\PZ$ production \cite{Kallweit:2019zez}.
Therefore, in the following we stick to the simple multiplicative prescription of
Eq.~\eqref{eq:multiplicative}.

As in the inclusive setup (see \refta{table:sigmainclNLO}), the EW corrections
are negative and sizeable, accounting for about $-10\%$ of the LO results for
all polarized signals with at least one longitudinal boson, about $-11\%$ for the
purely-transverse and unpolarized signals.
The QCD corrections give the largest correction to LO cross-sections,
ranging between $+30\%$ and $+45\%$ for the various polarized and unpolarized
signals. Such corrections are particularly large for the mixed states, an effect
that characterizes also $\PW^+\PZ$ production \cite{Denner:2020eck}.
The loop-induced contribution from the gluon--gluon partonic channel 
gives a sizeable enhancement to the $\rL\rL$ and $\rT\rT$ signals of 15\%
and 20\%, respectively. For mixed polarization modes, this gluon-initiated
contribution is only at the level of 3\%.

The additive combination of NLO corrections and loop-induced contributions
results in an enhancement of the LO cross-section of about 40\% for all
signals, while the multiplicative combination gives a smaller overall correction
of about 35\% for all signals.
At the level of total cross-sections the difference between the additive
and multiplicative results is about 2.8\%, which is smaller than
NLO scale uncertainties.

The uncertainties of the fiducial NLO cross-section from 7-point
scale variations are strongly affected by the gluon-induced partonic
channel. In fact, the noticeable reduction of scale uncertainties from
LO to NLO QCD (from order 5\% to order 2\%) is counterbalanced by 25\%
scale uncertainties of the gluon-induced contribution, resulting in
only a mild reduction of the combined scale uncertainties. The most
noticeable scale-uncertainty reduction is found for the mixed
polarization states (from 7\% at LO to 4\% at NLO), owing to the small
size of loop-induced contributions.

Thanks to the joint effect of the four-lepton
and lepton-pair invariant mass cuts, the non-doubly-resonant
background accounts for just 0.8\% of the full 
result at LO and even less when including radiative corrections (0.5\% for NLO$_\times$).
The interferences are larger than in the inclusive setup. The polarization fractions
and the impact of the interferences are shown in \refta{table:fidNLOfractions}, as relative
percentages of the DPA unpolarized result.
\begin{table}
  \begin{center}
    \renewcommand{\arraystretch}{1.3}
    \begin{tabular}{C{1.cm}C{1.8cm}C{1.8cm}C{1.8cm}C{1.8cm}C{1.8cm}C{1.8cm}}%
      \cellcolor{blue!9} mode  & \cellcolor{blue!9}  LO  & \cellcolor{blue!9} ${\rm NLO_{\rm EW}}$ & \cellcolor{blue!9}{${\rm NLO_{\rm QCD}}$} & \cellcolor{blue!9}{${\rm gg}$} & \cellcolor{blue!9} ${\rm NLO_+}$  & \cellcolor{blue!9} ${\rm NLO_\times}$ \\
      unpol.              & 100\% & 100\% & 100\% & 100\% & 100\% & 100\%\\
      $\PZ_{\rL}\PZ_{\rL}$  & 5.8\% & 5.9\% & 5.9\% & 5.4\% & 5.8\% & 5.8\%\\ 
      $\PZ_{\rL}\PZ_{\rT}$  & 11.8\% & 11.9\% & 12.7\% & 2.1\%  & 11.7\% & 11.6\%\\ 
      $\PZ_{\rT}\PZ_{\rL}$  & 11.9\% & 12.0\% & 12.6\% & 2.1\%  & 11.6\% & 11.6\%\\ 
      $\PZ_{\rT}\PZ_{\rT}$  & 69.3\% & 69.1\% & 67.4\% & 90.4\% & 69.7\% & 69.8\%\\ 
      interf.  & 1.1\% & 1.1\% & 1.3\%  & $-0.02\%$ & 1.2\% & 1.2\%
    \end{tabular}
  \end{center}
  \caption{Polarization fractions in the fiducial setup (see \refse{subsec:setup}) 
    for $\PZ\PZ$ production at the LHC at LO and NLO accuracy, shown as
      percentages relative to the DPA unpolarized result.
    Polarizations are defined in the CM frame. 
  } 
  \label{table:fidNLOfractions}
\end{table}
Interference effects are at the level of 1\% both at LO and at NLO, but larger
effects could be present in differential results. They are particularly small
for the gluon-induced process.
The LO polarization fractions are roughly conserved when including NLO EW
corrections, with the $\rL\rL$ contribution accounting for 6\% of the total,
which is dominated by purely-transverse states (70\%).
The NLO QCD fractions show a $0.8\%$ increase in the mixed cases, and correspondingly
a 3\% decrease in the $\rT\rT$ one. The polarization fractions for the loop-induced
contribution differ greatly from those found in $q\bar{q}$ channels, with
very small mixed contributions of about 2\% each and a longitudinal fraction
of about 5\%, which mostly comes from box diagrams involving
top loops, and a 90\% $\rT\rT$ contribution.
The combined NLO fractions are roughly the same as the LO ones,
with basically no difference between the additive and multiplicative approaches.

\begin{figure}
  \centering
  \includegraphics[scale=0.28]{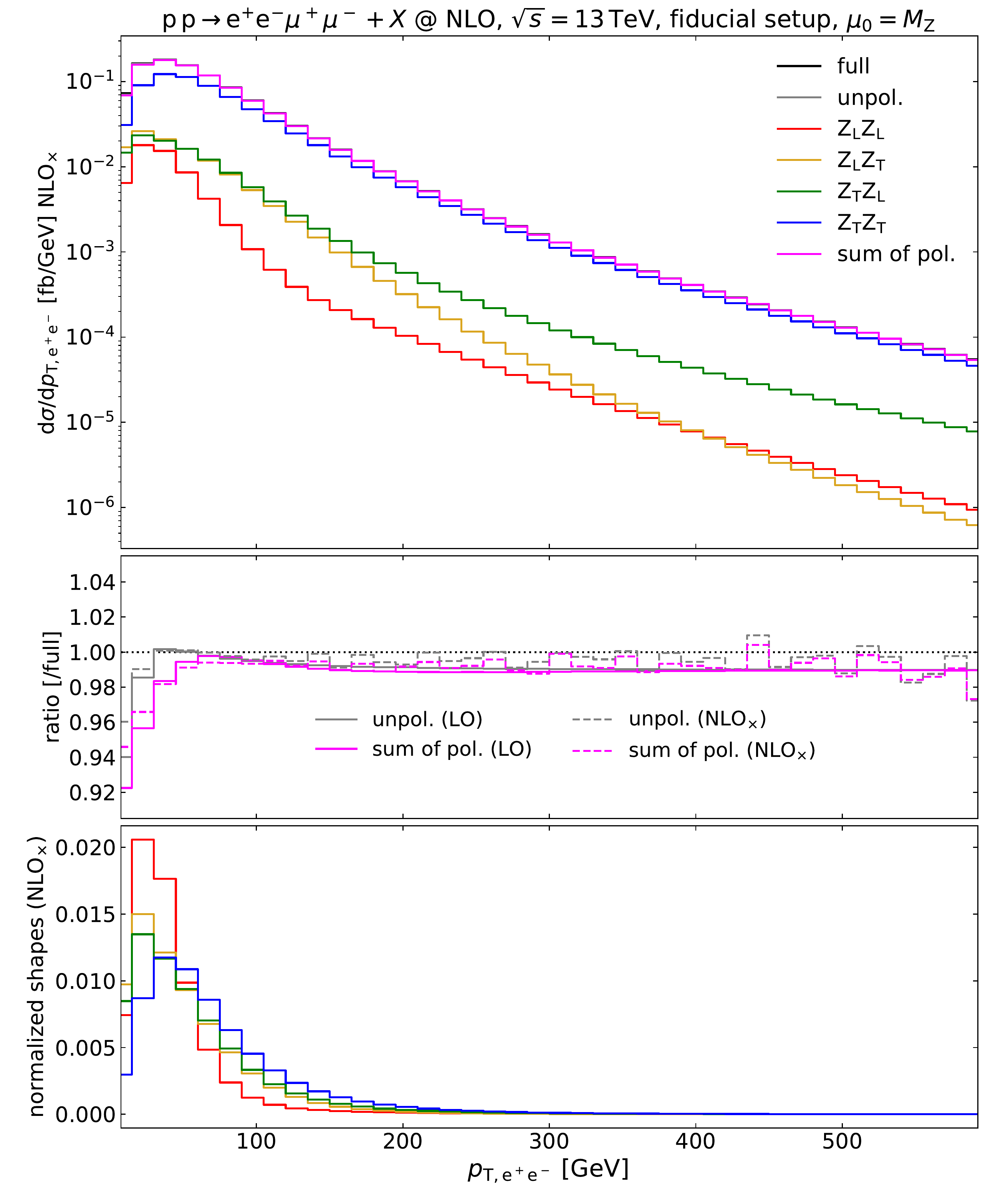}
  \includegraphics[scale=0.28]{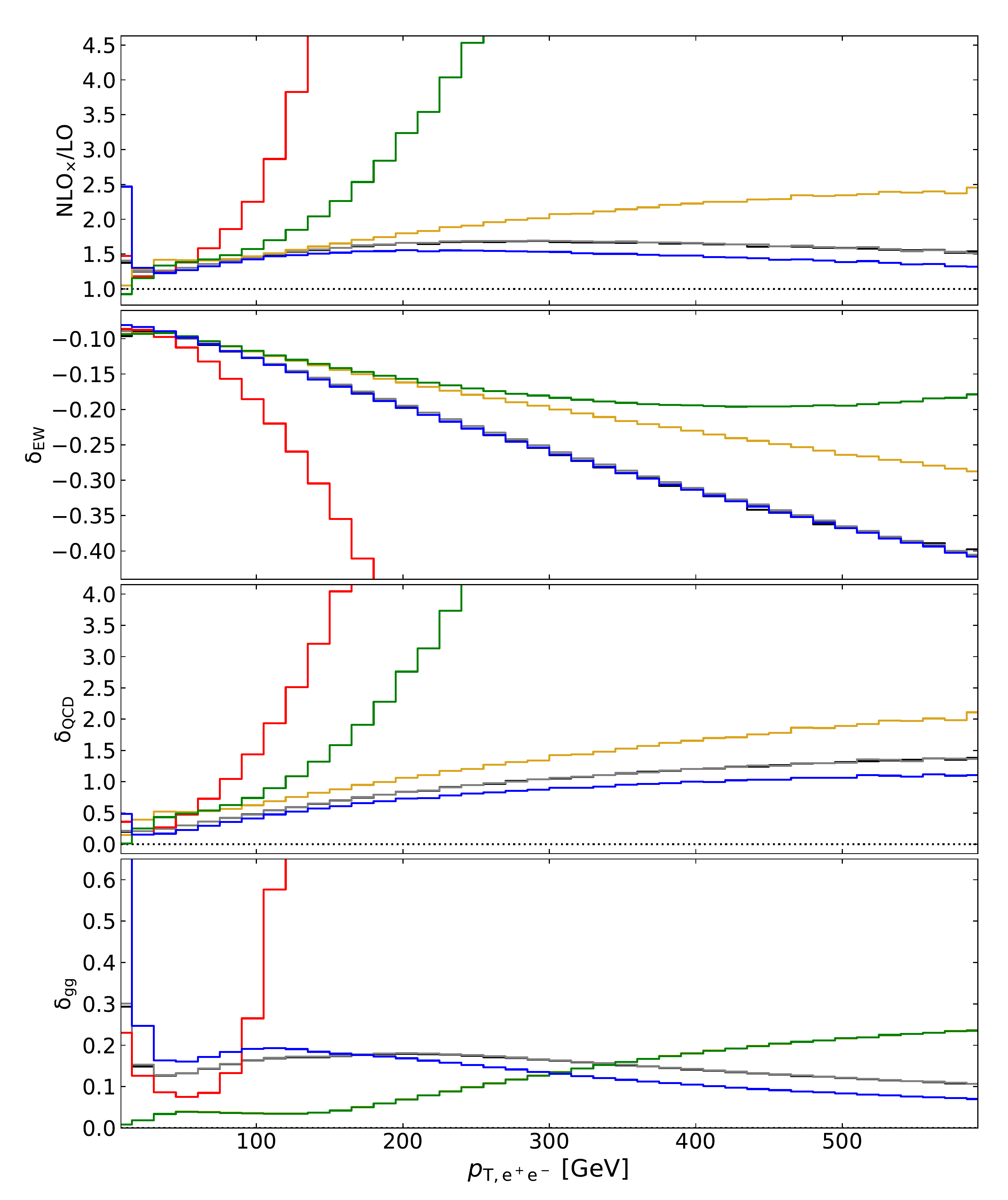}
  \caption{
    Distributions in the transverse momentum of the $\Pe^+\Pe^-$ system.
    Unpolarized and doubly-polarized results are shown for $\PZ\PZ$ production at the LHC
    in the fiducial setup described in \refse{subsec:setup} with polarizations defined the CM frame.
    Left panel from top down: NLO$_{\times}$ differential
    cross-sections, ratios of DPA over the full result,
    normalized NLO$_{\times}$ shapes (unit integral). Right panel from top down:
    NLO$_{\times}$ $K$-factor, NLO EW corrections $\delta_{\rm EW}$, NLO QCD correction $\delta_{\rm QCD}$,
    $\Pg\Pg$ contribution $\delta_{\rm gg}$. All curves in the right panel are relative to the LO cross-sections.
  }\label{fig:fid_first} 
\end{figure}

\begin{figure}
  \centering
  \includegraphics[scale=0.28]{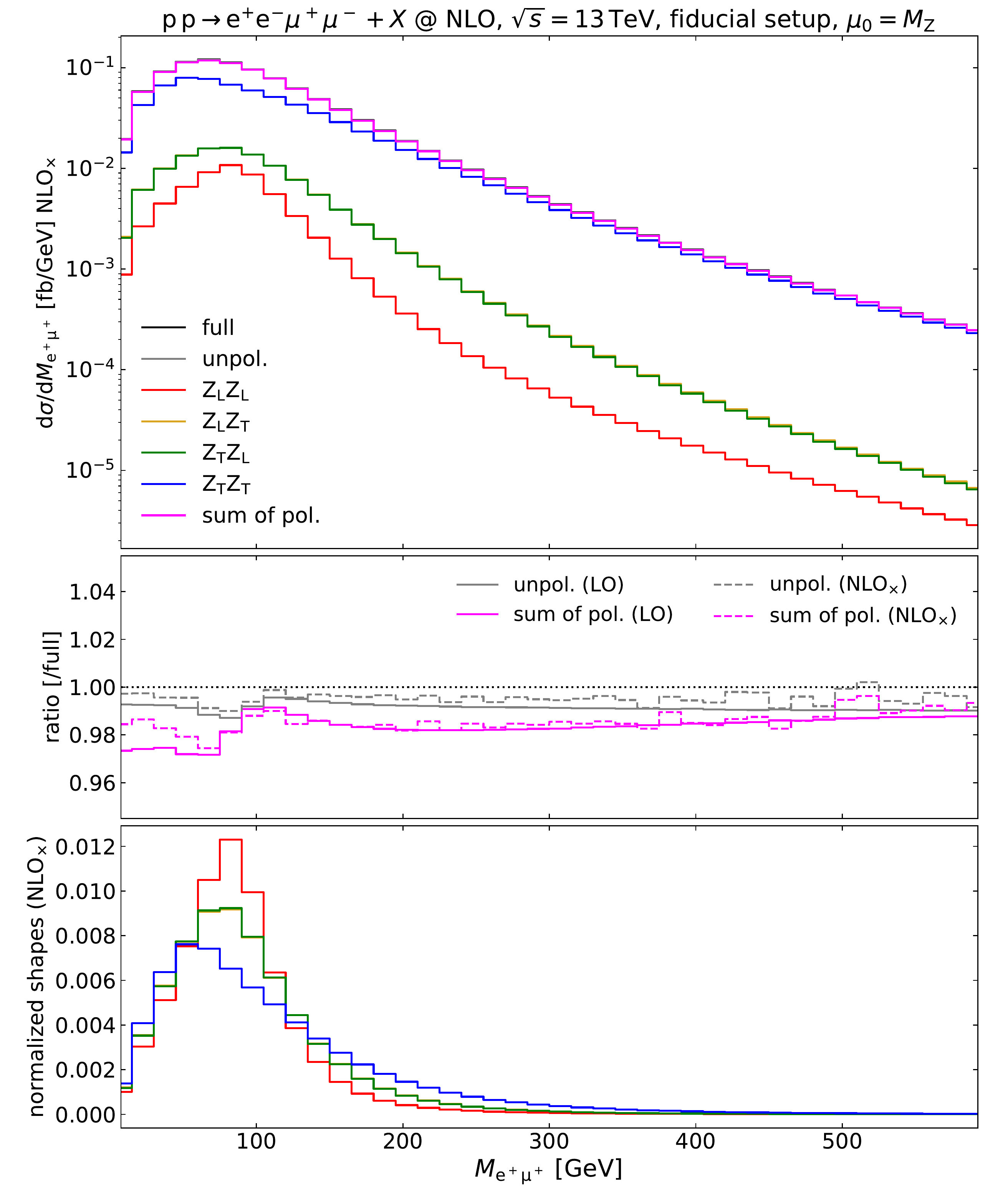}
  \includegraphics[scale=0.28]{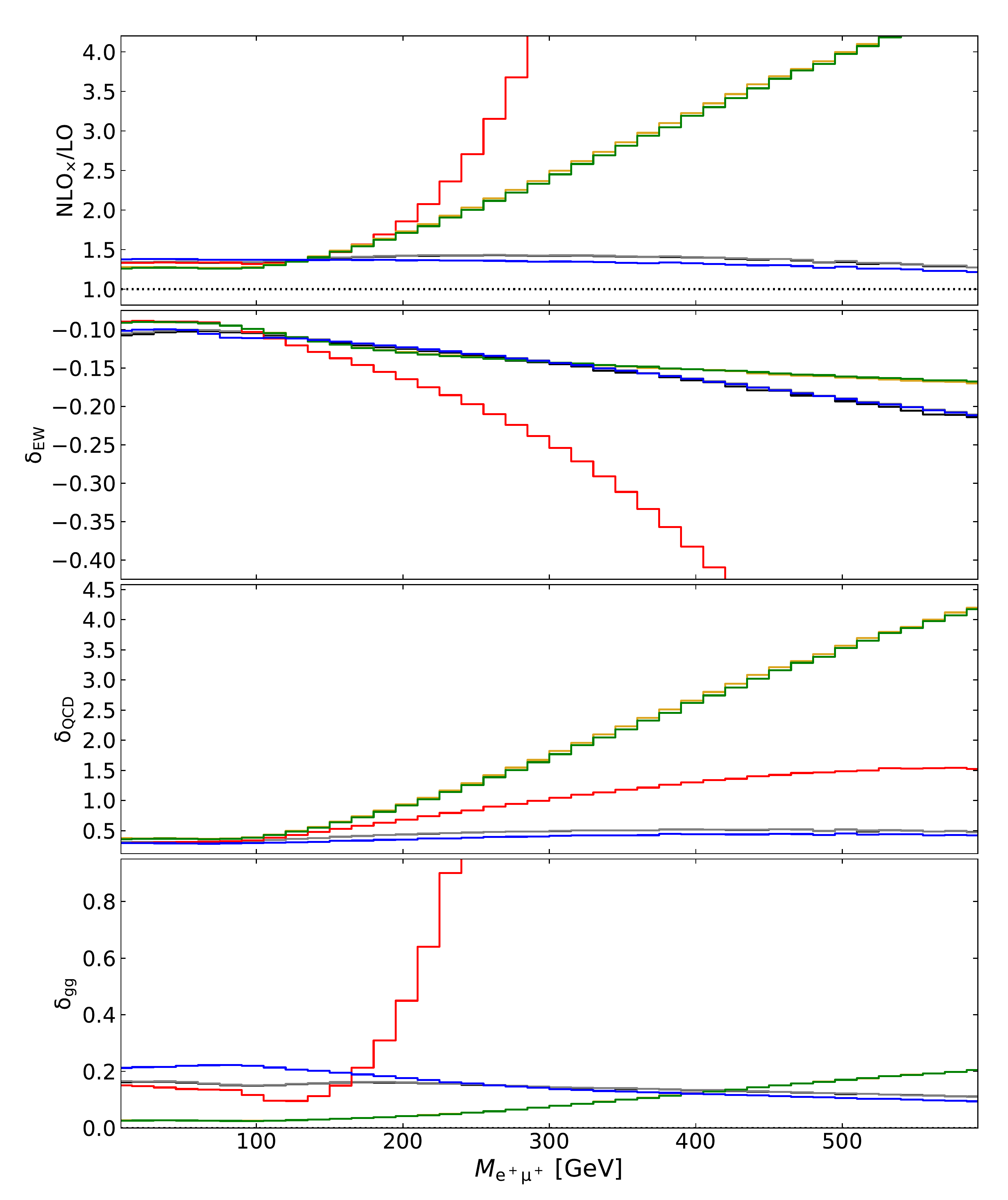}
  \caption{
    Distributions in the invariant mass of the $\Pe^+\mu^-$ system.
 Same structure as in \reffi{fig:fid_first}.   
  }\label{fig:fid_mepmum} 
\end{figure}

\begin{figure}
     \centering
     \includegraphics[scale=0.28]{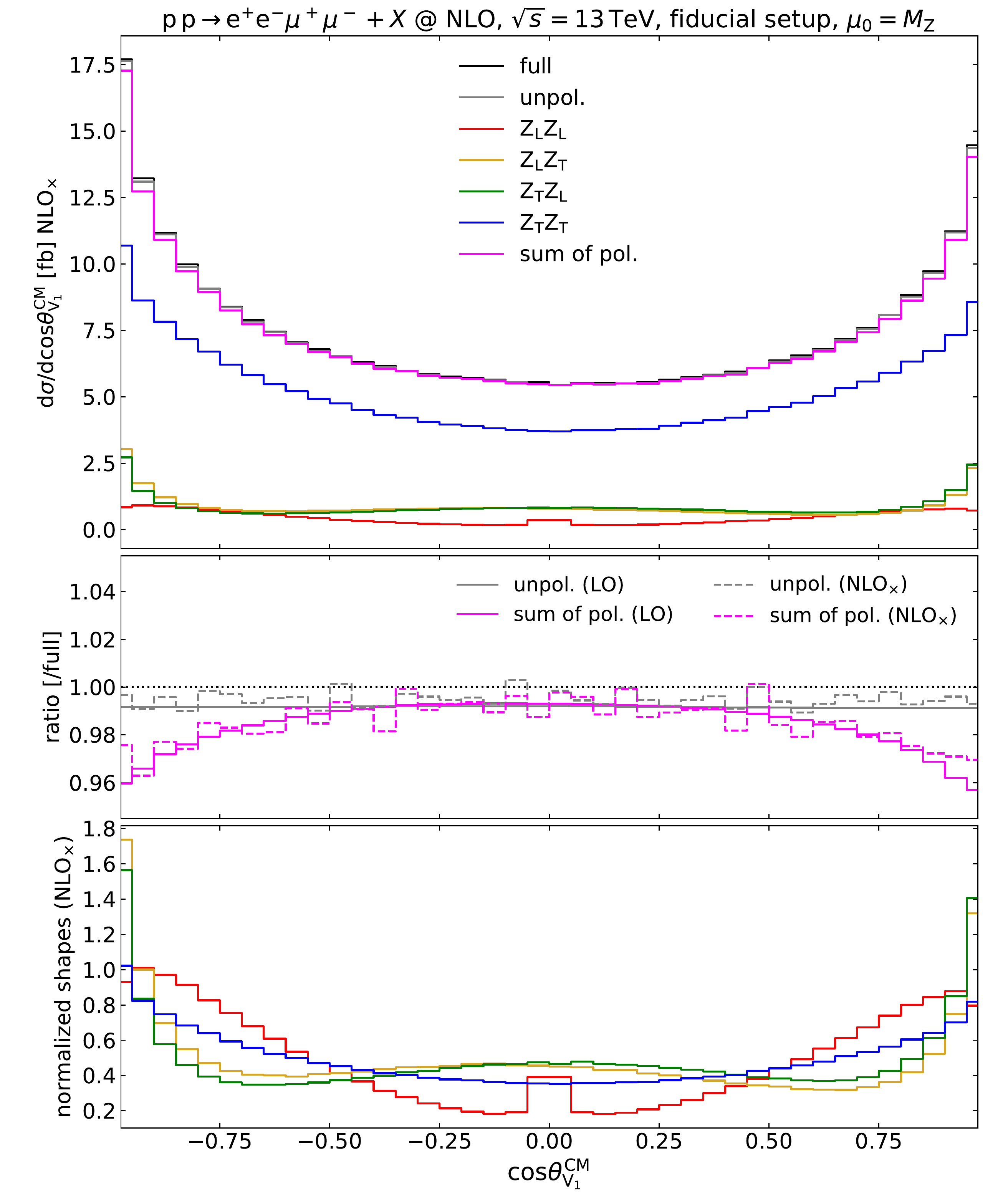}
     \includegraphics[scale=0.28]{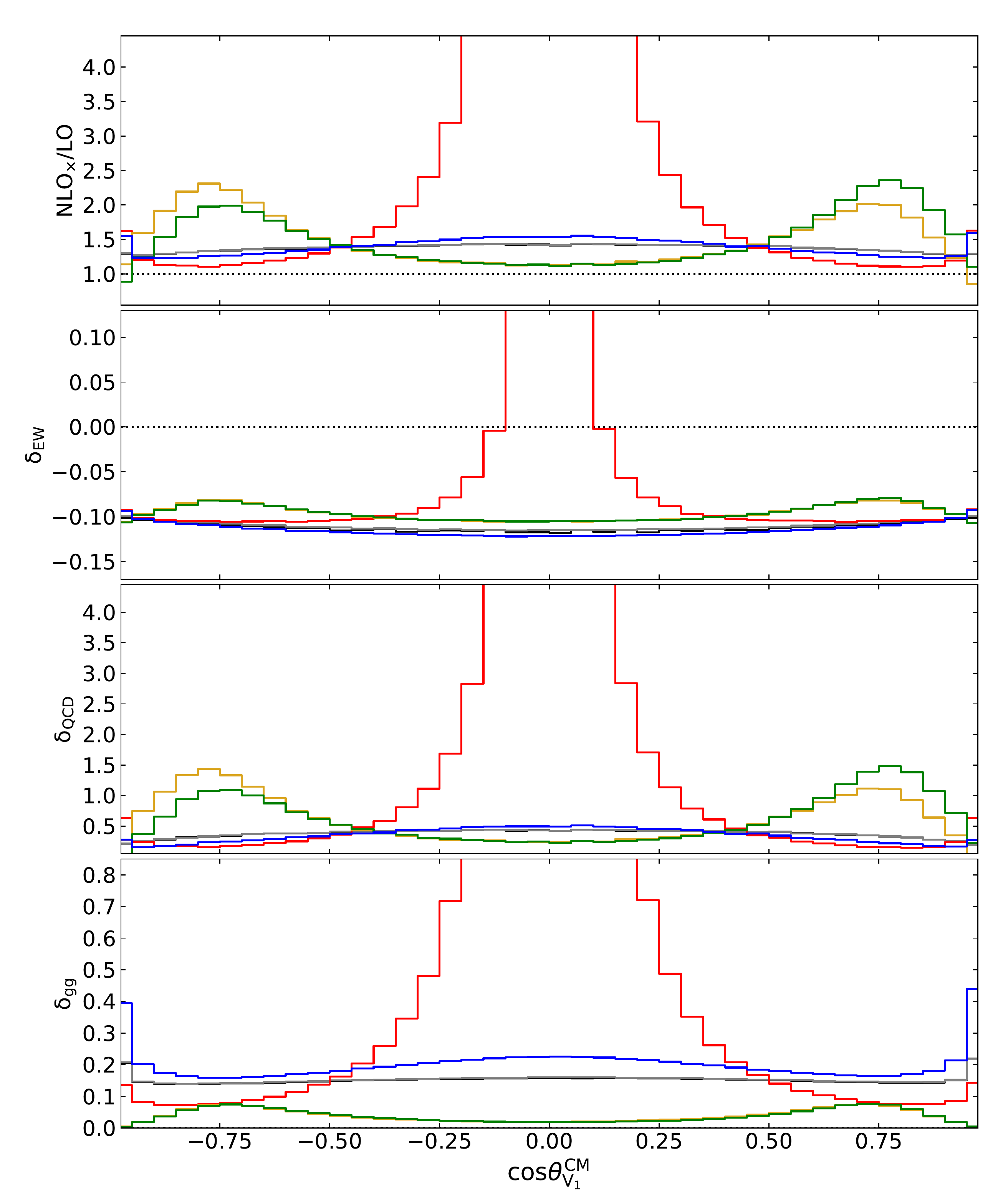}
     \caption{
       Distributions in the scattering angle of the $\PZ$~boson decaying into $\Pe^+\Pe^-$
       in the four-lepton CM frame computed w.r.t.\ the four-lepton--system direction in the
       LAB frame. Same structure as in \reffi{fig:fid_first}.
     }\label{fig:fid_scattTheta} 
   \end{figure}

\begin{figure}
     \centering
     \includegraphics[scale=0.28]{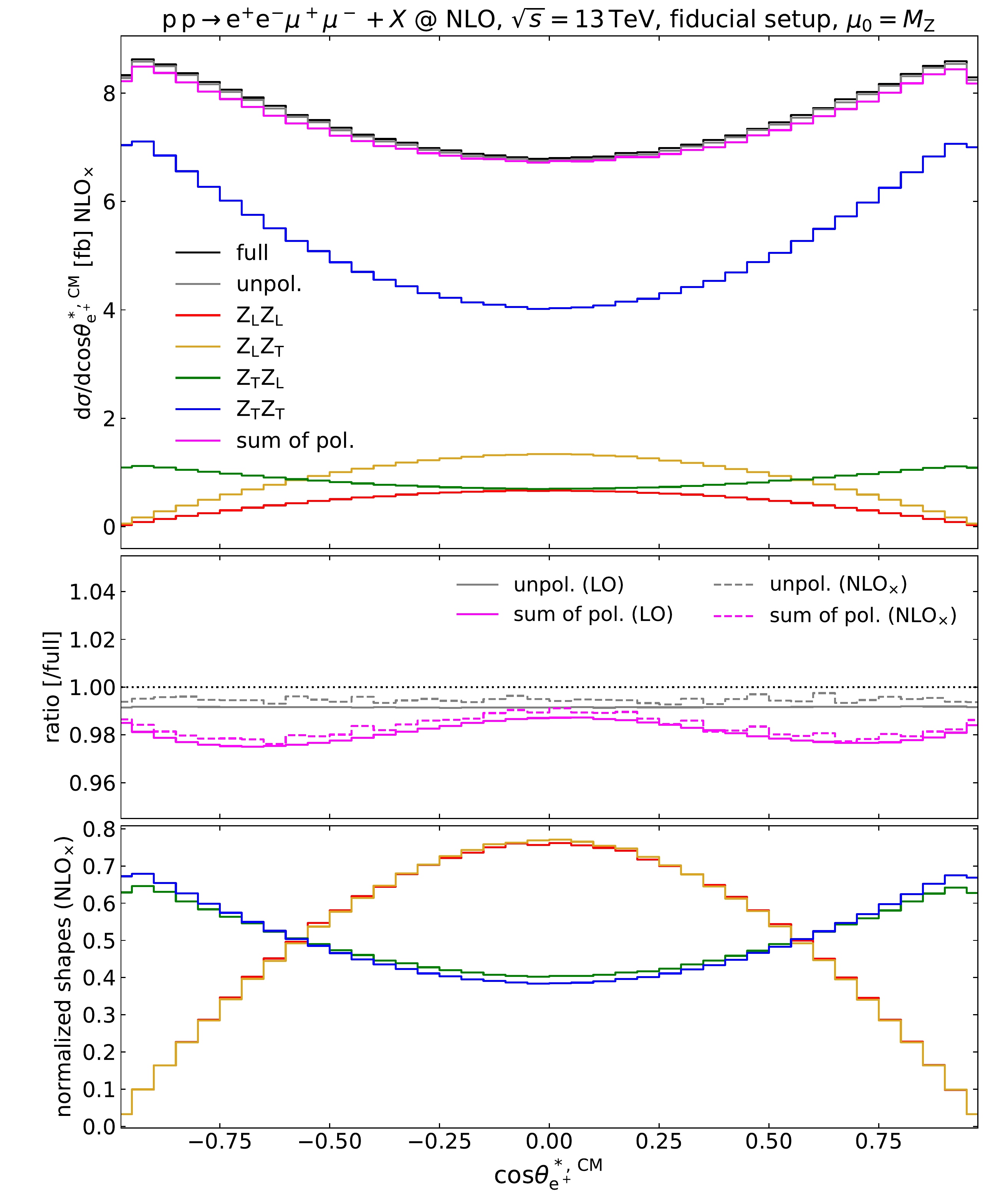}
     \includegraphics[scale=0.28]{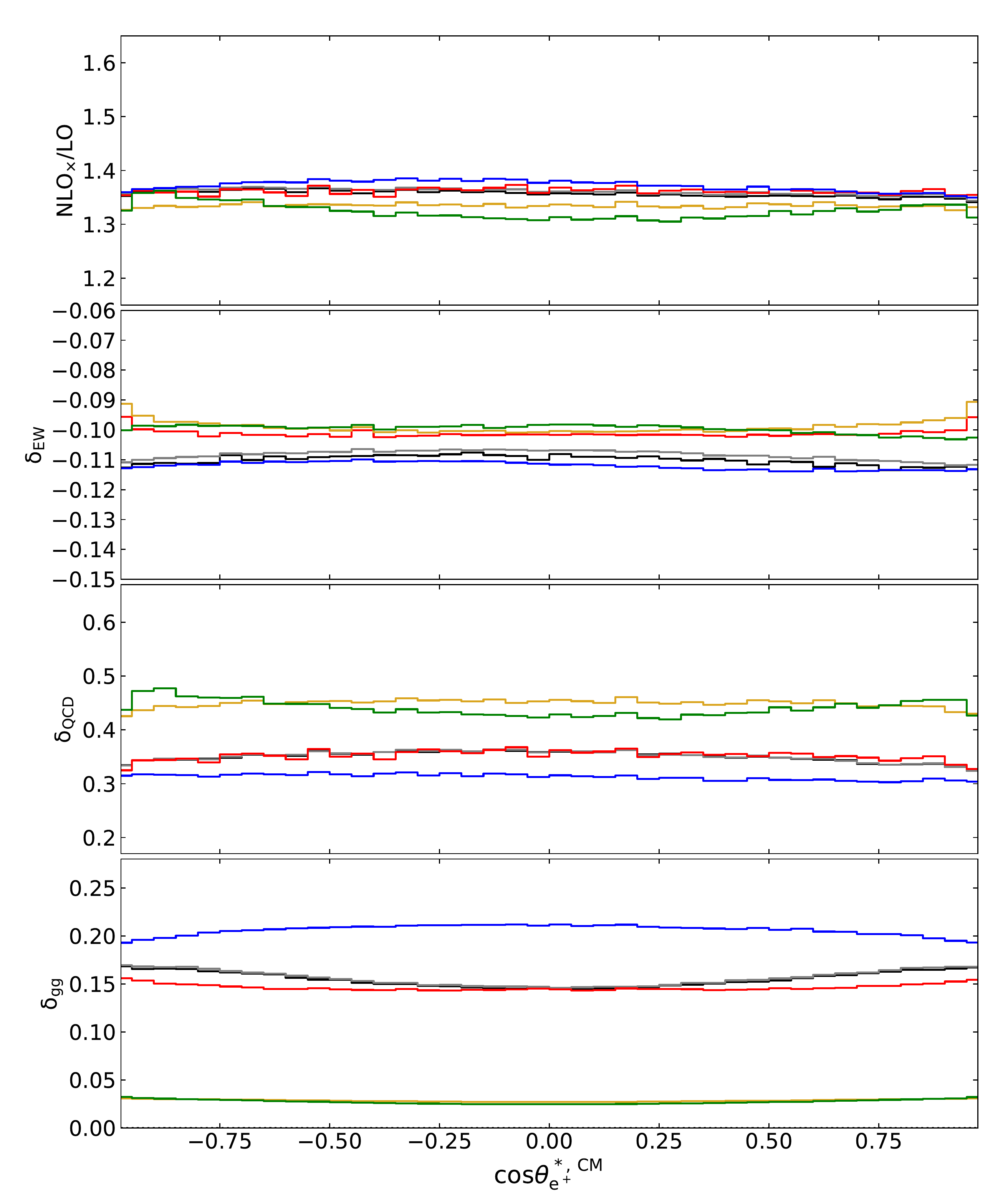}
     \caption{
       Distributions in the cosine of the polar decay angle of the positron
       $\theta^{*,\text{CM}}_{\Pe^+}$ in
      its corresponding $\PZ$-boson rest frame computed w.r.t.\ the $\PZ$-boson spatial direction
      in the CM frame. Same structure as in \reffi{fig:fid_first}.
     }\label{fig:fid_cthstarCMep} 
   \end{figure}

\begin{figure}
  \centering
  \includegraphics[scale=0.28]{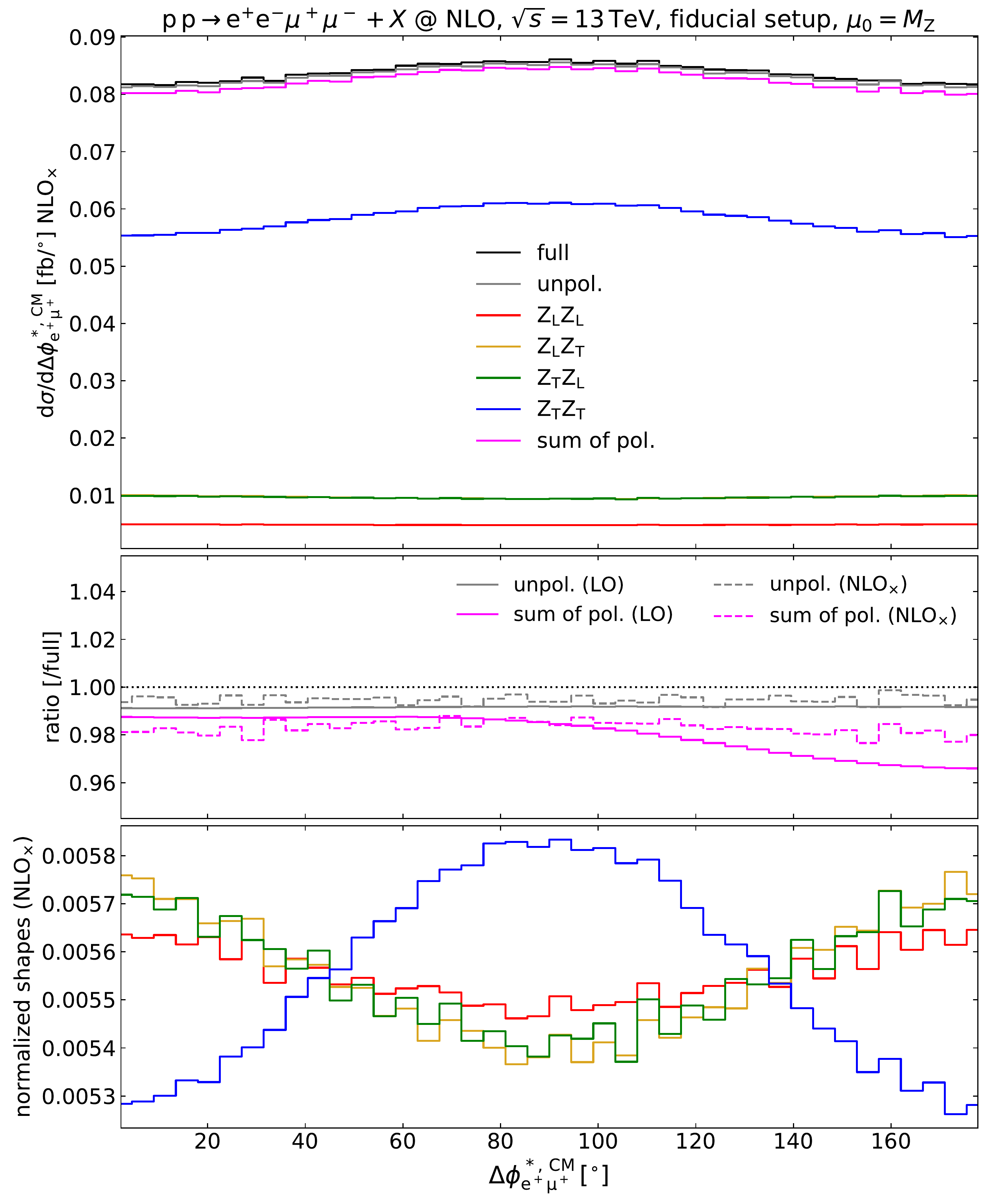}
  \includegraphics[scale=0.28]{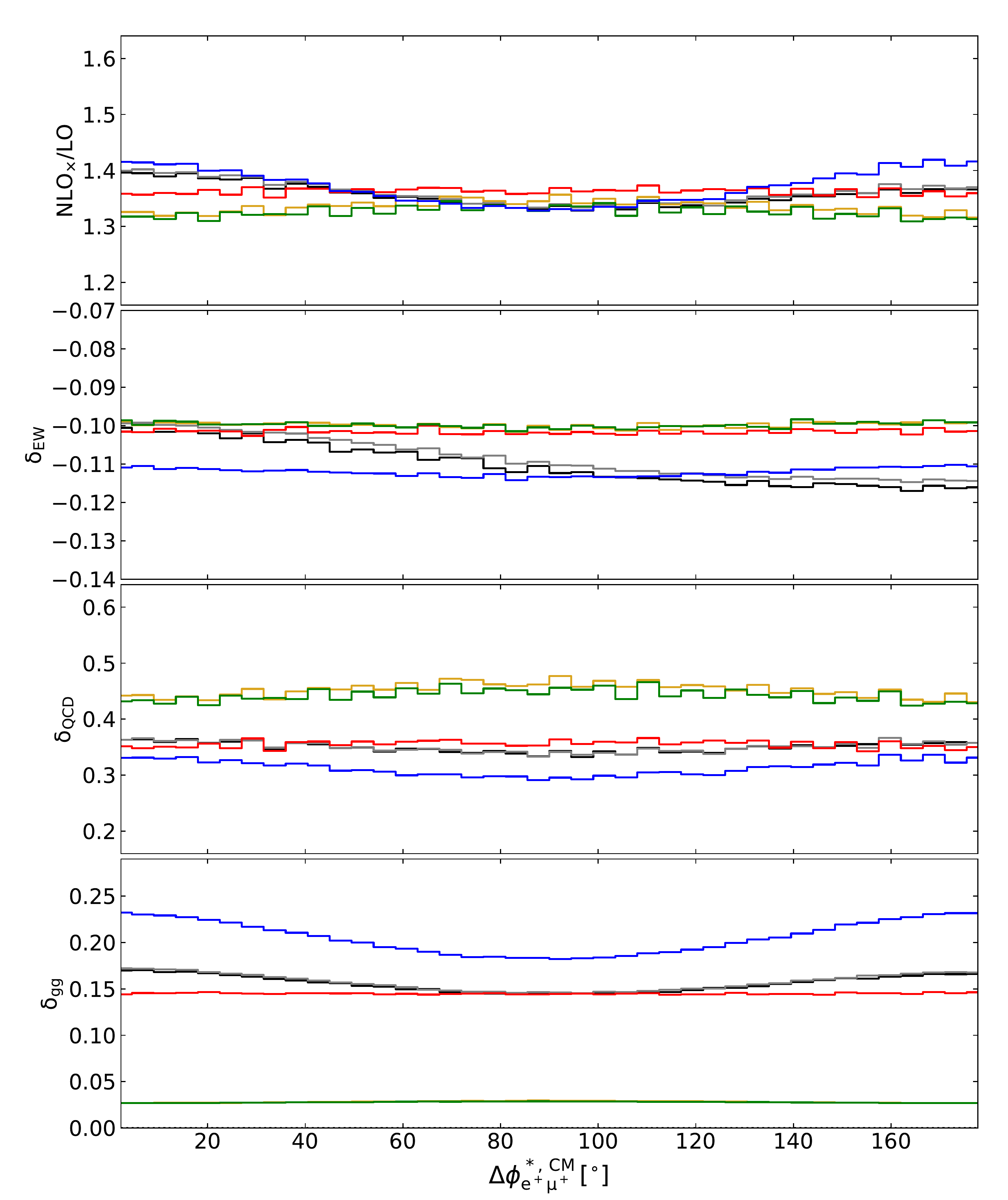}
  \caption{
    Distributions in the difference between azimuthal decay angles $\phi^{*,\text{CM}}_{\Pe^+}$
    and $\phi^{*,\text{CM}}_{\mu^+}$, where $\phi^{*,\text{CM}}_{\ell^+}$ is the decay angle of the 
    positively-charged lepton in its corresponding $\PZ$-boson rest frame computed
    w.r.t.\ the $\PZ$-boson spatial direction in the CM frame. Same
    structure as in \reffi{fig:fid_first}.
  }\label{fig:fid_dphistarpp} 
\end{figure}

\begin{figure}
  \centering
  \includegraphics[scale=0.28]{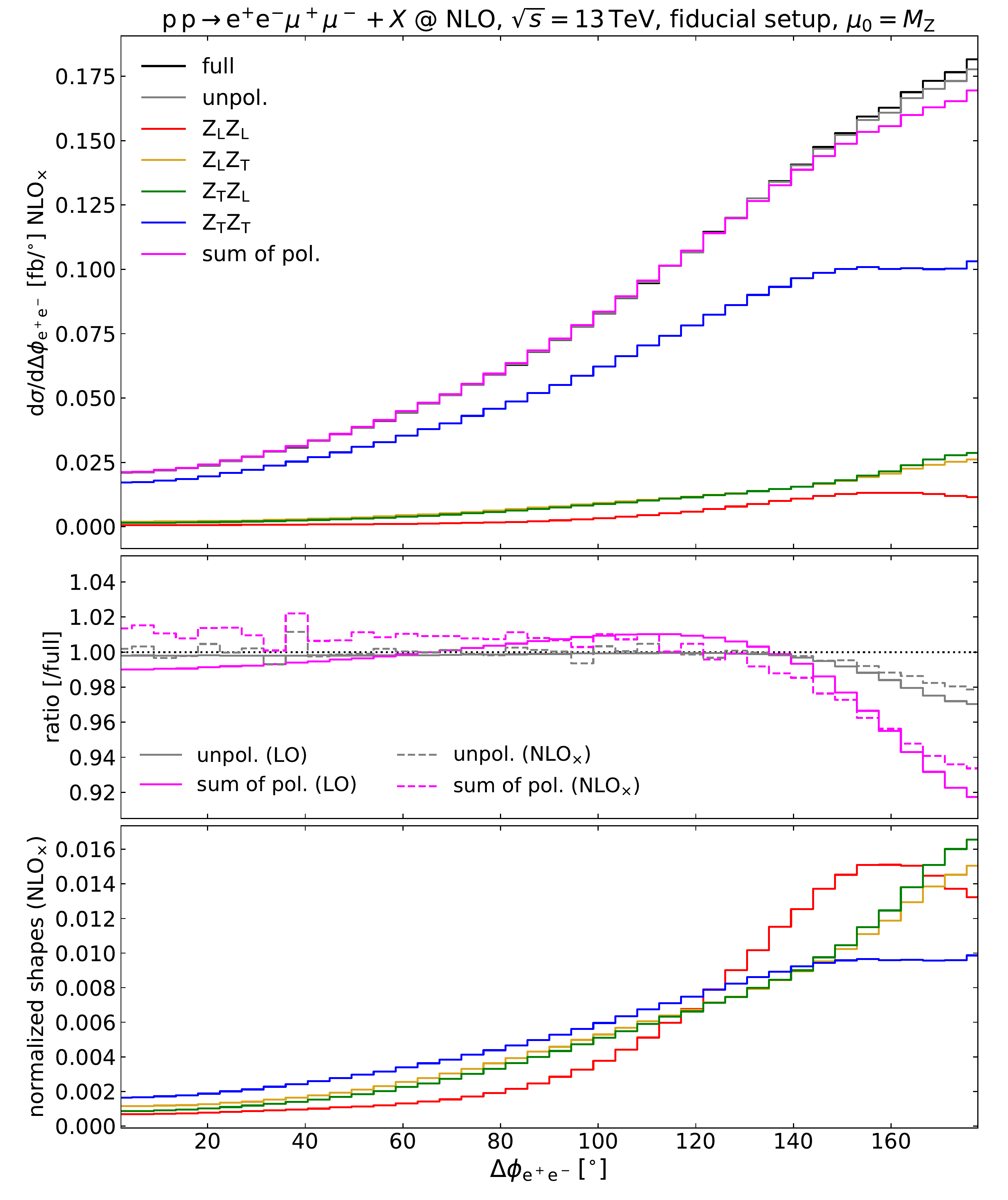}
  \includegraphics[scale=0.28]{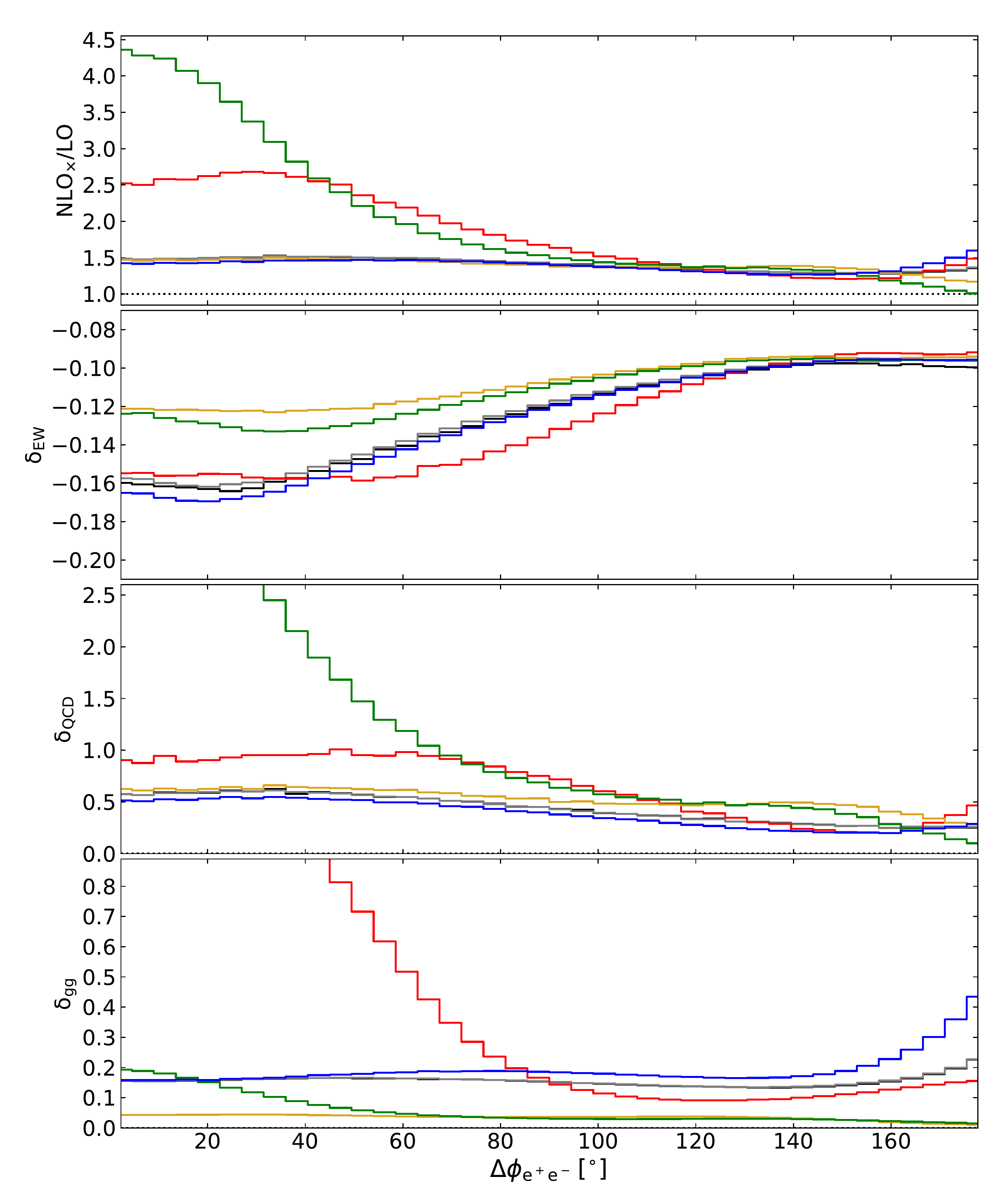}
  \caption{
    Distributions in the azimuthal separation between the positron
    and the electron. Same structure as in \reffi{fig:fid_first}.
  }\label{fig:fid_deltaPhiee} 
\end{figure}

\begin{figure}
  \centering
  \includegraphics[scale=0.28]{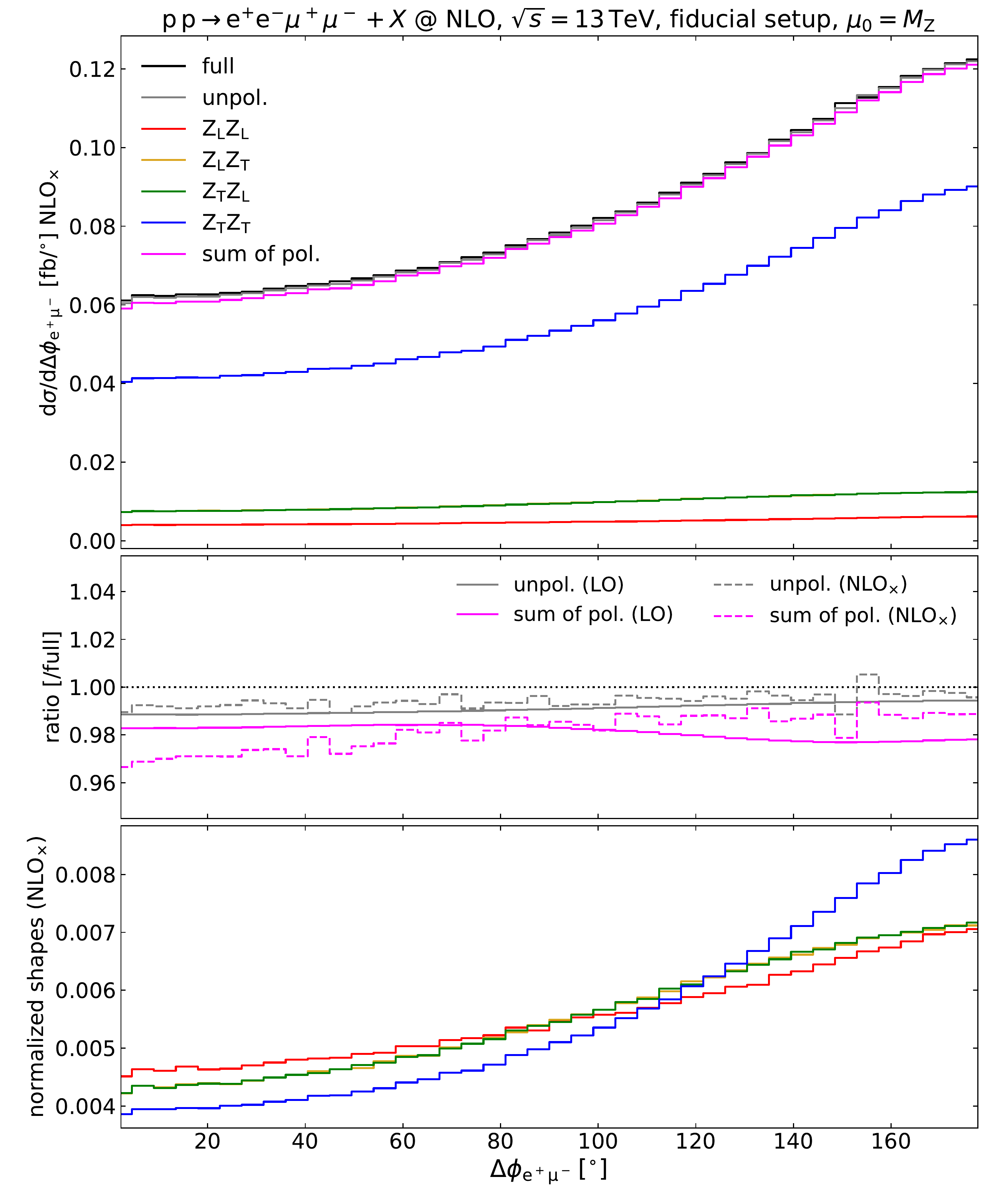}
  \includegraphics[scale=0.28]{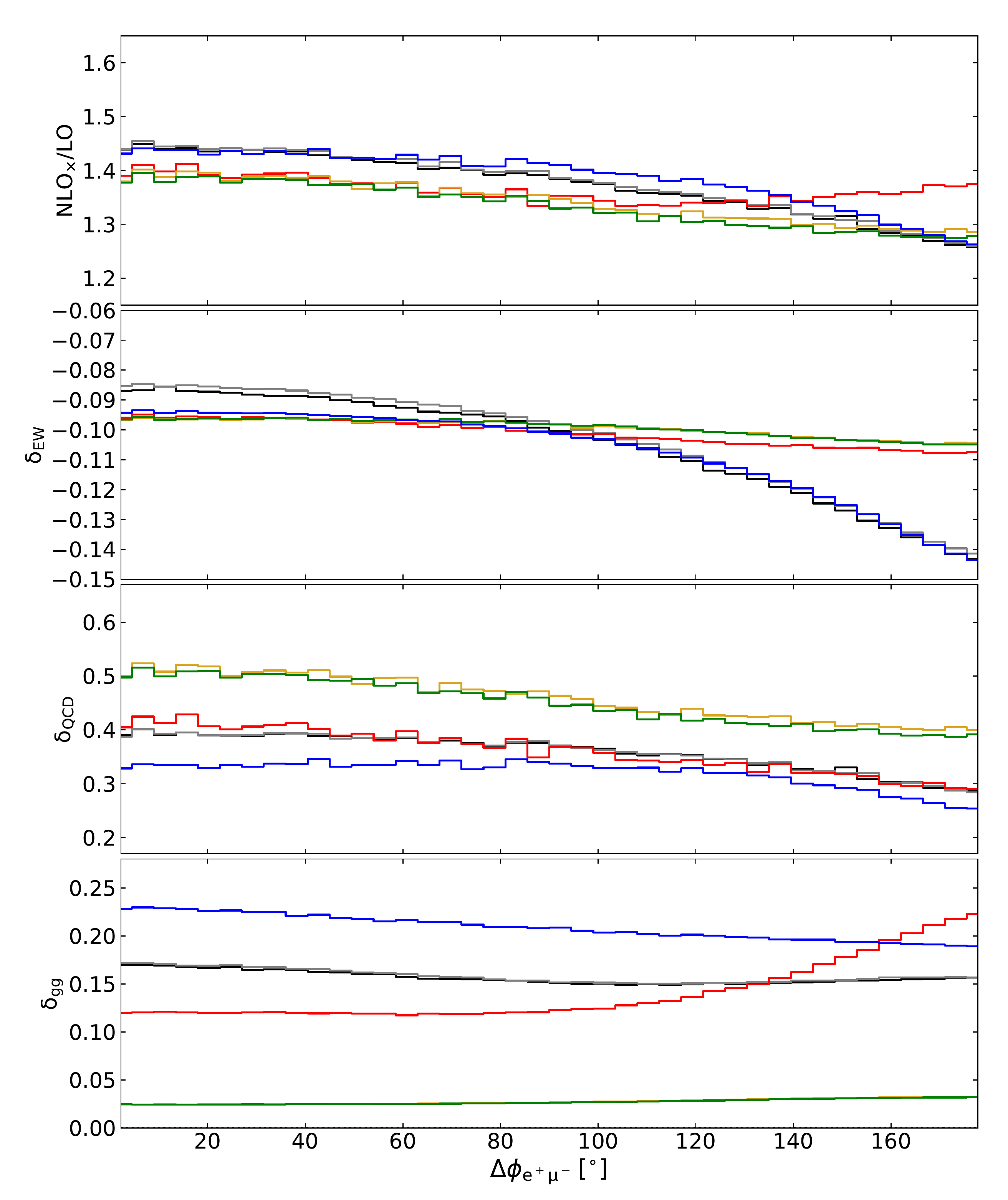}
  \caption{
    Distributions in the azimuthal separation between the positron
    and the muon. Same structure as in \reffi{fig:fid_first}.
  }\label{fig:fid_deltaPhiemu} 
\end{figure}

\begin{figure}
     \centering
     \includegraphics[scale=0.28]{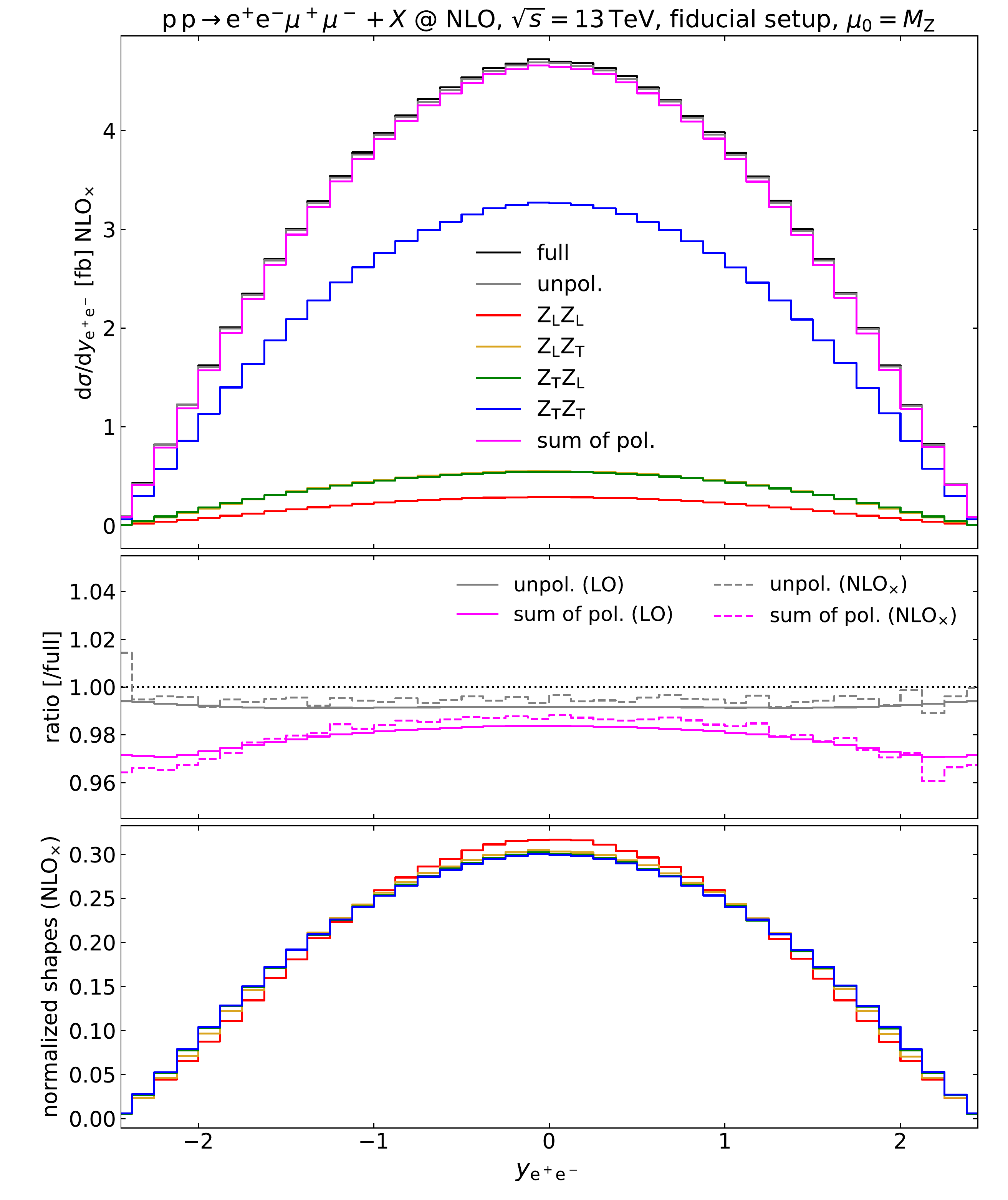}
     \includegraphics[scale=0.28]{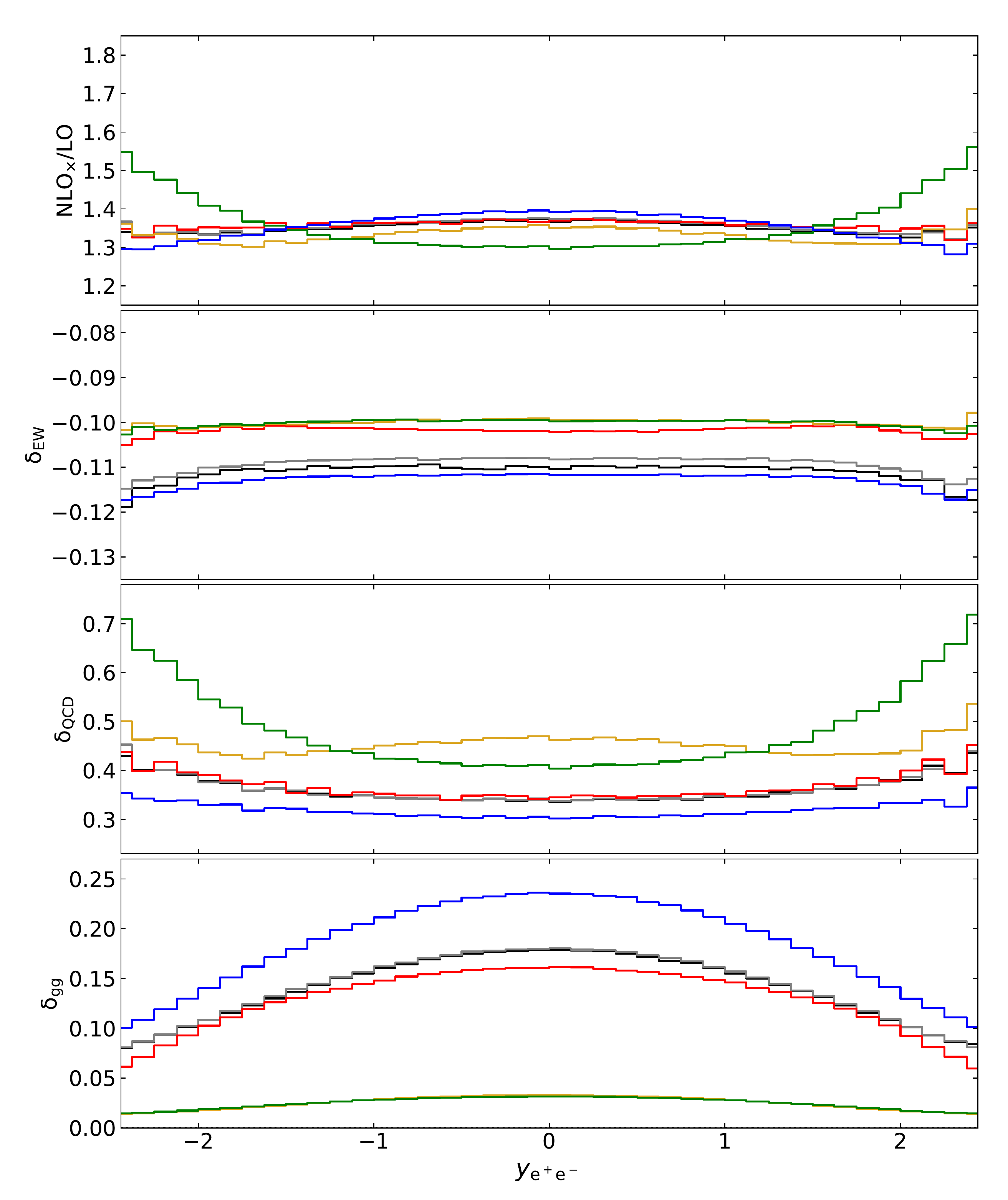}
     \caption{
       Distributions in the rapidity of the $\Pe^+\Pe^-$ pair (single $\PZ$~boson).
       Same structure as in \reffi{fig:fid_first}.
     }\label{fig:fid_yz} 
   \end{figure}

\begin{figure}
  \centering
  \includegraphics[scale=0.28]{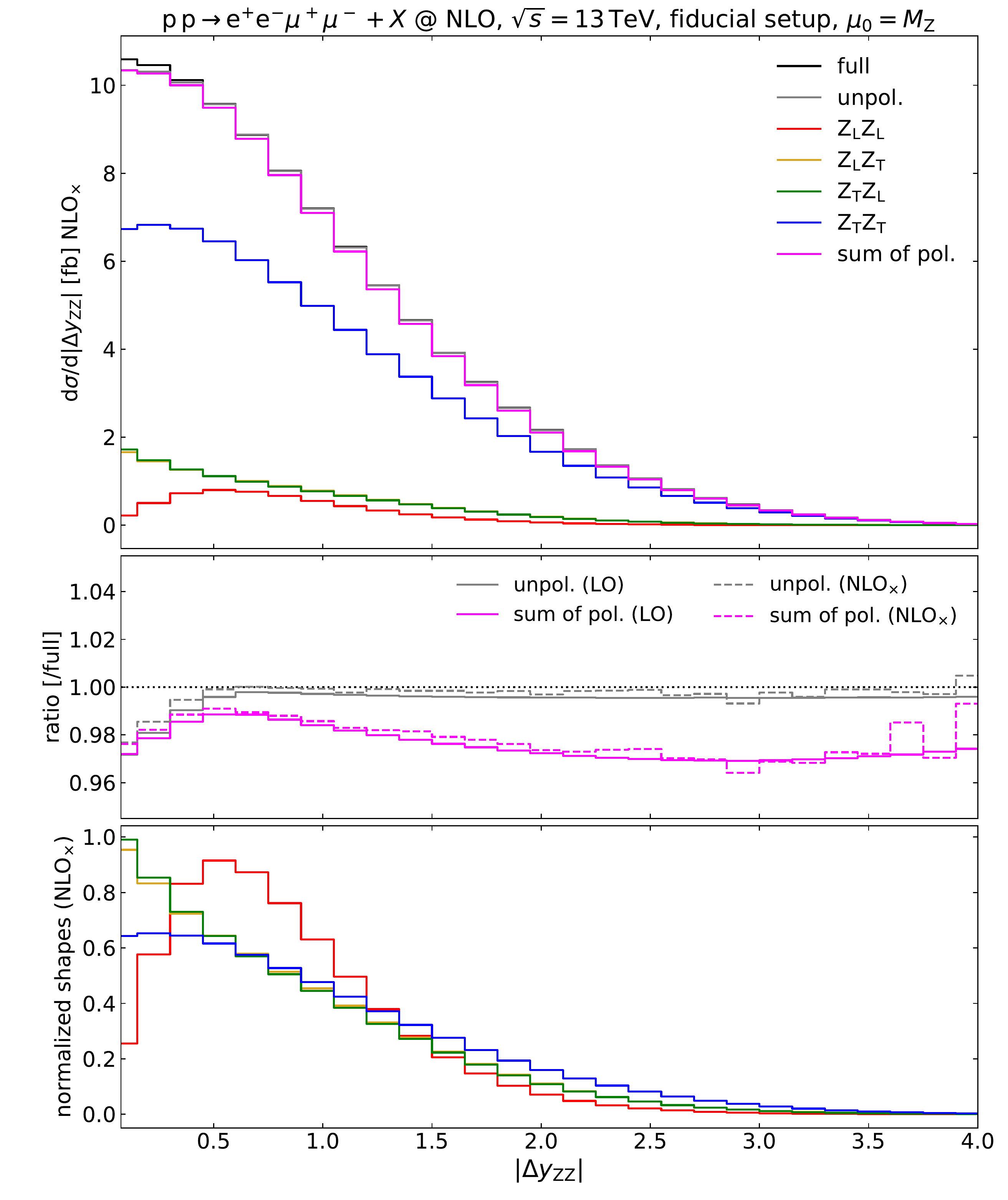}
  \includegraphics[scale=0.28]{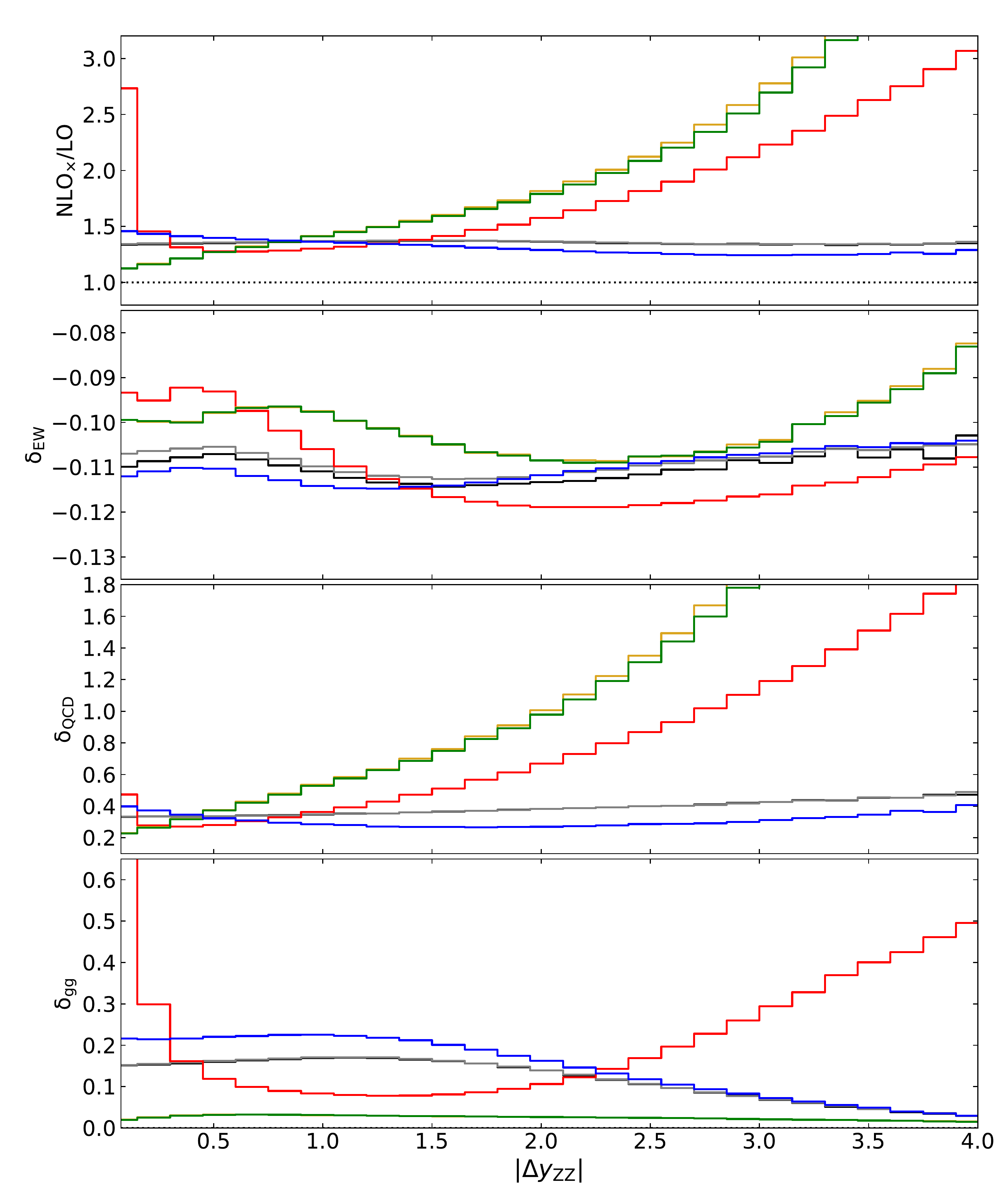}
  \caption{
    Distributions in the absolute value of the rapidity separation between the two $\PZ$~bosons.            
    Same structure as in \reffi{fig:fid_first}.
  }\label{fig:dyzz} 
\end{figure}

We now present differential results in the fiducial region for observables that are
relevant for the discrimination among polarization states of the $\PZ$
bosons using the multiplicative combination of EW and QCD corrections
[Eq.~\eqref{eq:multiplicative}]. 
We have checked numerically that the corresponding differential results obtained
with the additive prescription [Eq.~\eqref{eq:additive}] are very similar.
Non-negligible differences between the two approaches only appear in phase-space regions where
the cross-section for $\PZ\PZ$ production is suppressed.
While the multiplicative combination of NLO corrections is theoretically preferable over
the additive one, it causes artificial effects
in phase-space regions where the LO cross-section is suppressed, particularly for the
$\rL\rL$ polarization state.
In \reffis{fig:fid_first}--\ref{fig:dyzz} we focus 
on the size of interferences, on the relative impact of higher-order corrections,
and on the differences among the shapes of polarized distributions.

As first observable we consider the transverse momentum of the electron--positron system
in \reffi{fig:fid_first}.
In the case of DPA calculations, this observable coincides with the transverse momentum of the $\PZ$~boson
that decays into the two dressed leptons $\Pe^+$ and $\Pe^-$.  In the
soft region of the spectrum, the non-doubly-resonant background
(the deviation of the grey curves from 1 in the left middle panel) 
contributes 6\% (4\%) to the full result at LO (NLO$_\times$). In the same
kinematic region, the interferences (the difference between the grey
and magenta curves in the left middle panel) range between 2\% and 3\%, while
they are below the 1\% level at moderate and large transverse momenta.
The unpolarized distribution is dominated by the $\rT\rT$ polarization mode both at LO and
when including radiative corrections. The $\rL\rL$ and $\rL\rT$ polarized distributions
are strongly suppressed at large transverse momenta, where they are two orders of magnitude smaller
than the unpolarized and purely-transverse ones, while the $\rT\rL$
mode is only suppressed by one order of magnitude. The $\rL\rL$
polarization state, which is strongly suppressed 
at LO w.r.t.\ to all other polarization states, receives a sizeable enhancement from QCD real corrections
and loop-induced contributions, already at moderate transverse momentum.
The two mixed-polarization distributions have similar behaviours
for $\pt{\Pe^+\Pe^-}\lesssim 100\GeV$, while at larger values the
$\rT\rL$ mode receives huge (up to 1000\%) corrections
from real QCD radiation, mostly coming from the $\Pg q,\,\Pg\bar{q}$
partonic processes that open up at NLO QCD.
In the $\rT\rL$ polarization configuration, a large $p_{\rm T}$ of the
transverse boson implies that also the 
longitudinal boson has large $p_{\rm T}$ at LO, giving a suppression
in the cross-section by $1/p^2_{\rm T}$ w.r.t.\ to the TT polarization state. In the presence of 
QCD radiation, the transverse momentum needed to balance a large
$p_{\rm T}$ of the transverse boson is 
shared by the longitudinal boson and the radiated
gluon or quark. Therefore, the longitudinal \PZ~boson carries smaller
$p_{\rm T}$ than at LO giving a much less
severe suppression to the signal.
For the $\rL\rT$ polarization configuration, on the other hand, the
presence of real radiation does not influence the transverse momentum of the
longitudinal boson in the considered distribution. The large
transverse momentum of the longitudinal $\PZ$~boson suppresses the
contributions at LO and NLO QCD alike.
This argument explains a $\rL\rT$ signal that is 10 times smaller
than the $\rT\rL$ one for $\pt{\Pe^+\Pe^-}\gtrsim 500\GeV$ and
the enhancement of the $\rL\rL$ mode by real QCD radiation.
An analogous effect is not present at NLO EW, because the EW virtual corrections are noticeably larger than real ones,
the EW coupling is suppressed w.r.t.\ the QCD case, and the
photon-induced partonic processes ($\gamma q,\,\gamma\bar{q}$) are
less luminous than those with a gluon in the initial state.
The loop-induced gluon--gluon contributions are particularly sizeable
for the $\rT\rT$ signal in the very soft region of the spectrum,
where the quark-induced LO $\rT\rT$ cross-section is suppressed owing
to angular-momentum conservation (the soft-$p_{\rm T}$ region corresponds to a scattering
angle that is close to zero or $\pi$).
Although all polarized distributions are peaked at small transverse
momenta, the transverse momentum of a single $\PZ$~boson
is sensitive to the polarization state of both $\PZ$~bosons thanks to the strong kinematic and spin correlation between
the two bosons. In fact, the normalized shape of the $\rL\rL$ distribution
is much narrower than the $\rT\rT$ one, while the
normalized shapes of mixed-polarization configurations show an intermediate 
behaviour between the $\rL\rL$ and $\rT\rT$ ones. This indicates that the transverse momentum
of a single \PZ~boson provides a handle to separate the $\rL\rL$ polarization state from the others.

In \reffi{fig:fid_mepmum}, we display the distributions in the
invariant mass of the $\Pe^+\mu^+$ pair.
This variable shares strong similarities with the invariant
mass of the four-lepton system (not shown) concerning relative
NLO corrections, but is characterized by an enhanced discrimination power
among polarizations. It has been found to be sensitive to polarizations of
$\PZ$~bosons also in Higgs decays \cite{Maina:2020rgd,Maina:2021xpe}.
The non-doubly-resonant background  is as small as at the integrated level,
and the interferences amount to at most 2\% of
the unpolarized result in the considered range. 
Although the $\rL\rL$ and mixed modes are strongly suppressed w.r.t.\
the $\rT\rT$ mode already
at moderate invariant mass, in the soft region of the spectrum the shapes
of polarized distributions are different. The $\rT\rT$ distribution is characterized by a maximum around $50\GeV$
and a rather large width, while the maximum of the $\rL\rL$ curve is around $80\GeV$ and the
shape is narrower. The results for the mixed states lie in between
those of the $\rT\rT$ and $\rL\rL$ ones.
The suppression of the LO distributions for the polarization states
with one or two longitudinal bosons at high $M_{\Pe^+\mu^+}$ entails a strong impact of NLO and
loop-induced contributions.  As observed in 
invariant-mass and transverse-momentum distributions in the inclusive setup [see
\reffi{fig:pteCM}], the EW corrections grow large and negative for the purely-longitudinal
signal, accounting for $-40\%$ of the LO prediction at $M_{\Pe^+\mu^+}\approx 400\GeV$, while
milder EW corrections are found for other polarization states, ranging between
$-10\%$ and $-20\%$ in the considered range. The QCD corrections are similar
to those at the integrated level in the soft region of the spectrum, while
for $M_{\Pe^+\mu^+}\gtrsim 100\GeV$ the $\rL\rL$ and mixed signals are enhanced
by very large real-radiation contributions both in the $q\bar{q}$
and in the $\Pg q, \Pg\bar{q}$ partonic channels. 
The QCD radiative effects are particularly large for the mixed states,
resulting in NLO QCD cross-sections
that are five times larger than the LO ones at $M_{\Pe^+\mu^+}\approx 600\GeV$.
This effect is due to huge real contributions from the gluon--(anti)quark channels
that enhance the mixed signals  more sizeably than the $\rL\rL$ one.%
\footnote{This enhancement is not directly related to the suppression
    of longitudinal polarizations.}
The purely-longitudinal signal is affected by large gluon-induced contributions that
grow fast towards larger invariant masses, where the LO $q\bar{q}$ contributions are
highly suppressed. 
The multiplicative combination of corrections gives huge and monotonically increasing
$K$-factors for the $\rL\rL$ state, slightly smaller but still very large $K$-factors for the
mixed states, while the $\rT\rT$ and unpolarized results feature decreasing $K$-factors
($+40\%$ in the soft region, $+25\%$ at $M_{\Pe^+\mu^+}\approx
600\GeV$) owing to the
negative EW corrections that are also applied to the flat QCD correction, and to the
decreasing gluon-induced contribution in the high-energy regime.

Very large effects of higher-order corrections to longitudinal signals are not
only found in the tails of invariant-mass and transverse-momentum distributions,
but also in inclusive angular variables like the cosine of the scattering angle
between the two $\PZ$~bosons presented in \reffi{fig:fid_scattTheta}.
The scattering angle $\theta_{V_1}^{\mathrm{CM}}$ is
defined as the angle between the momentum of the $\Pe^+\Pe^-$ pair
computed in the rest frame of the four-lepton system and the
momentum of the four-lepton system computed in the LAB frame.
At LO it coincides with the scattering angle between the two $\PZ$~bosons
that is used (for on-shell bosons) in Eq.~\eqref{eq:analyticZZform}. At NLO it receives
substantial modifications by real-radiation contributions.
The interferences are almost vanishing for $\theta^{\text{CM}}_{V_1}=\pi/2$ , while they reach the 3\%
level in the forward and backward regions, where the unpolarized, transverse--transverse
and mixed-polarization distributions have their maxima. The peaks in these regions are
particularly pronounced for the mixed signals. The $\rL\rL$ signal has maxima at slightly more
central configurations ($\cos\theta^{\text{CM}}_{V_1}\approx \pm 0.9$).
At $\cos\theta^{\text{CM}}_{V_1}=0$ the polarized signals display a rather different behaviour.
The $\rT\rT$ signal has a minimum, the mixed states have a local maximum, while the
purely-longitudinal distribution is characterized by an impressive enhancement, which is an artificial
effect of the multiplicative combination of the NLO corrections. In fact,
the $\rL\rL$ signal vanishes at $\cos\theta^{\text{CM}}_{V_1}=0$, while the NLO corrections
do not, \ie the additional term w.r.t.\ the additive combination
($\Delta\sigma_{\rm EW}\Delta\sigma_{\rm QCD}/\sigma_{\rm LO}$) becomes huge. We have checked that
the additive combination gives a smooth $\rL\rL$ curve around $\cos\theta^{\text{CM}}_{V_1}=0$.
As a consequence, the multiplicative combination of NLO corrections should be applied with a grain
of salt and compared against the results of the additive combination, particularly in those
regions where the LO is very small or vanishing.
The fake effects of the multiplicative combination  at
  $\cos\theta^{\text{CM}}_{V_1}=0$ are also
  present in the $K$-factor shown in the right top panel of
  \reffi{fig:fid_scattTheta}, while 
the huge effects  in the relative corrections $\delta_{\text{EW}}$,
$\delta_{\text{QCD}}$, and $\delta_{\Pg\Pg}$ simply result from the
normalisation to the vanishing LO cross-section.
Also EW corrections to the $\rL\rL$ distribution become large and positive in the central
region, while they are negative and flat towards $\cos\theta^{\text{CM}}_{V_1}=\pm 1$.
The combined ${\rm NLO_\times}$ corrections to the $\rT\rT$ distribution are relatively flat, apart from an
enhancement in the forward and backward regions due to the gluon-induced contribution.
Mixed-polarization distributions receive NLO QCD corrections and gluon-induced contributions
that are maximal around $\cos\theta^{\text{CM}}_{V_1}=\pm 0.75$.
A similar shape in the relative correction is found also for EW corrections.
Up to the effects of the multiplicative combination, the different
shapes for the various polarized signals and the
rather small interference effects make this angular variable very suitable
for the discrimination among polarization states.

The possibility to fully reconstruct the rest frame of each $\PZ$~boson
gives access to the lepton decay angles, which are the most direct
probes of the polarization states of decayed bosons. 
We consider in \reffi{fig:fid_cthstarCMep} the cosine of the polar decay angle
of the positron in the corresponding $\PZ$-boson rest frame. 
The NLO EW distributions in the inclusive setup for the same variable are shown in \reffi{fig:cthcmCM}.
The effect of transverse-momentum and rapidity cuts on this observable is mild and restricted
to $\cos\theta^{*,\text{CM}}_{\Pe^+}=\pm 1$, where the $\rT\rT$ and $\rT\rL$ distributions
do not have a maximum anymore. The size of interferences is similar to the one found at the
integrated level, ranging from 0.3\% of the full result in the central region to 2\% around
$\cos\theta^{*,\text{CM}}_{\Pe^+}\approx\pm 0.7$.
This observable is highly sensitive to the polarization of the $\PZ$
boson that decays to $\Pe^+\Pe^-$, as can be appreciated from the
comparison of normalized shapes for the
various polarized signals. There is no shape difference between the $\rL\rL$ and $\rL\rT$ curves,
despite the different cuts applied on the $\Pe^\pm$ and $\mu^\pm$ kinematics, and very small shape
differences are found between the $\rT\rL$ and $\rT\rT$ curves. Obviously, in order to
discriminate between the longitudinal and transverse polarizations of the $\PZ$~boson decaying into $\mu^+\mu^-$
the $\cos\theta^{*,\text{CM}}_{\Pe^+}$ variable is not helpful.
The distributions are not distorted sizeably by the inclusion of higher-order corrections neither
for polarized signals nor for the unpolarized ones. In fact, the NLO effects and the
loop-induced contribution entail rather flat corrections to this observable, resulting in
combined $K$-factors that are comparable with those of the integrated results in the whole angular range.
Owing to lepton universality, the results for the $\cos\theta^{*,\text{CM}}_{\mu^+}$ variable are
very similar to those for $\cos\theta^{*,\text{CM}}_{\Pe^+}$ up to small effects due to the different
transverse-momentum and rapidity cuts for the two lepton flavours.

It is clear that any experimental analysis that targets the polarization of $\PZ$~bosons should
rely on decay angles as discriminating variables. In the rest frame of a single $\PZ$~boson, the leptonic-decay
kinematics is fully described by the polar decay angle, $\theta^{*,\text{CM}}_{\ell^+}$, and the azimuthal
one, $\phi^{*,\text{CM}}_{\ell^+}$. If two $\PZ$~bosons are produced, it is interesting to investigate
the spin correlation between them by means of the difference between azimuthal decay
angles of the positively-charged leptons, each computed in the corresponding $\PZ$-boson rest frame.
This variable has been proved to be sensitive to the polarization of
weak bosons that come from
a Higgs decay \cite{Maina:2020rgd,Maina:2021xpe}. Since polarizations are defined in the CM frame,
the reference direction for the angle $\phi^{*,\text{CM}}_{\Pe^+}$ coincides with the one for $\phi^{*,\text{CM}}_{\mu^+}$,
up to a minus sign, and both angles are measured relative to the same plane. 
The azimuthal difference is defined as
\beq
\Delta\phi^{*,\text{CM}}_{\Pe^+\mu^+}=
{\rm min}\bigl(\bigl|\phi^{*,\text{CM}}_{\Pe^+}-\phi^{*,\text{CM}}_{\mu^+}\bigr|,
\,2\pi-\bigl|\phi^{*,\text{CM}}_{\Pe^+}-\phi^{*,\text{CM}}_{\mu^+}\bigr|\bigr)\,,
\eeq
and the distributions in this variable are shown in \reffi{fig:fid_dphistarpp}.
The polarized and unpolarized distributions are symmetric about $\Delta\phi^{*,\text{CM}}_{\Pe^+\mu^+}=\pi/2$,
and the non-doubly-resonant background reflects the integrated results with no shape distortion.
The interferences vary between 0.5\% and 3\% at LO, showing a sizeable increase towards $\Delta\phi^{*,\text{CM}}_{\Pe^+\mu^+}=\pi$.
On the contrary, the combined NLO results give rather constant interference effects over the whole
angular range.
The QCD corrections are pretty flat for all polarized and unpolarized signals, while a noticeable
shape difference is found between LO and loop-induced results for the $\rT\rT$ signal.
The EW radiative corrections are constant for polarized signals, while
a 1.5\% variation shows up
in the relative correction for the unpolarized case, which is due to the different interference
patterns at LO and at NLO EW. While an 8\% variation is found in the
combined $K$-factor for the $\rT\rT$ signal, the variations for
signals with one or two longitudinal bosons are much smaller.
The comparison of normalized shapes demonstrates that this variable
is particularly suitable for the separation of the $\rT\rT$ polarization mode. In fact,
the distribution for this mode has a maximum at $\Delta\phi^{*,\text{CM}}_{\Pe^+\mu^+}=\pi/2$
and minima at $\Delta\phi^{*,\text{CM}}_{\Pe^+\mu^+}=0,\,\pi$, while 
the other modes exhibit the opposite behaviour. Furthermore, the $\rL\rL$ and mixed distributions
are somewhat flatter than the $\rT\rT$ one.
Separating the purely-longitudinal state from the mixed ones is less straightforward
with this observable, as the shape differences among these modes are mild.

Also other angular variables are sensitive to the
polarization of $\PZ$~bosons.
In \reffi{fig:fid_deltaPhiee} we present the differential distributions in the
azimuthal separation (computed in the LAB frame) between the positron and the electron.
In the DPA calculations, either polarized or unpolarized, the two leptons are decay products of
a $\PZ$~boson and are typically produced in opposite hemispheres.
The non-doubly-resonant effects are compatible with zero in most of the spectrum of this observable,
while they get slightly larger in the most populated region, reaching 2.5\% and 2\% at LO and
at NLO$_\times$, respectively. Around $\Delta\phi_{\Pe^+\Pe^-}=\pi$ also large interferences show up,
contributing up to 5\% to the full results, both in LO and in NLO$_\times$ predictions.
The NLO EW corrections are at the $-10\%$ level in the most populated region, while they become
more sizeable at $\Delta\phi_{\Pe^+\Pe^-}=0$ ($-16\%$ for diagonal polarization states,
$-12\%$ for mixed ones). Huge QCD corrections (more than 300\%) are
found for the $\rT\rL$ signal in the regime
of small azimuthal separation,  which result from real corrections, in
particular from those with initial-state gluons.
The QCD corrections are smaller, but still large, for the $\rL\rL$ signal
(around 100\%), while for other polarized and for the unpolarized
signals QCD corrections never exceed 60\% in the whole range.
Large gluon-induced contributions to the purely-longitudinal signal are found
for $\Delta\phi_{\Pe^+\Pe^-}<\pi/2$.
The combination of all corrections gives large $K$-factors for the $\rT\rL$ signal (up to 4.5) and to the
$\rL\rL$ one (up to 2.7).
The comparison of polarized signals reveals different distribution shapes, with monotonically-increasing
curves for the mixed states, a plateau in the region $\Delta\phi_{\Pe^+\Pe^-}>5\pi/6$ for the $\rT\rT$
state, and a maximum near $\Delta\phi_{\Pe^+\Pe^-}=8\pi/9$ for the $\rL\rL$
state. This renders this observable suitable for the discrimination among polarization modes in LHC data,
provided that also the interference contribution and the irreducible
background are properly accounted for in a SM-template fit of the data.

Another variable that is suitable for the polarization discrimination
is the azimuthal separation between the positron and the muon considered in \reffi{fig:fid_deltaPhiemu}.
Since the two particles originate from different $\PZ$-boson decays, the distributions are less peaked
at $\pi$ than the $\Delta\phi_{\Pe^+\Pe^-}$ ones.
Both LO and combined NLO results give positive interferences ranging between 1\% and 2\% of the full
distribution in the whole angular range.
The NLO EW corrections cause sizeable distortions only to the $\rT\rT$ polarization mode, as already
observed in the inclusive setup [see \reffi{fig:incAziCM}]. The impact of QCD corrections diminishes
monotonically towards the most populated region, with a 10\% variation
for the $\rL\rL$ and mixed signals and a
slightly smaller for the $\rT\rT$ one. The gluon-induced process enhances the LO distributions in
a rather uniform way for most polarization modes, apart from a 10\% increase for the $\rL\rL$ mode.
The combined NLO distributions suggest that this angular variable is suitable for the separation
of the $\rT\rT$ signal from the others. In fact, at the level of normalized shapes, the
$\rT\rT$ curve increases by 100\% going from $\Delta\phi_{\Pe^+\mu^-}=0$ to $\Delta\phi_{\Pe^+\mu^-}=\pi$,
while other polarized curves increase by roughly 50\% only.

In \reffi{fig:fid_yz} we present the distributions in the rapidity of the electron--positron system.
Owing to the selection cut on the rapidities of both $\Pe^+$ and $\Pe^-$, $|y_{\Pe^\pm}|<2.47$, also
the combination of the two momenta, \ie{the reconstructed $\PZ$~boson}, is central.
The non-doubly-resonant effects are flat and equal to those in the total cross-section.
A positive $3.5\%$ interference contribution appears in the most
forward and backward configurations allowed
by the selections, while in the central region, which is the most populated one, the interferences
account for less than 1\%.
The EW corrections are pretty flat for all polarization modes, while
some distortions of the LO curves are introduced by QCD corrections and by loop-induced
contributions. The NLO QCD corrections enhance in particular the $\rT\rL$ distribution by
up to $70\%$ in the most forward and backward regions, while milder effects are found for other
polarization states.
The loop-induced contributions sizeably enhance the $\rL\rL$ and the
$\rT\rT$ distributions, with maximal effects in the central region (23\% for $\rT\rT$, 15\% for $\rL\rL$,
at $y_{\Pe^+\Pe^-}=0$).
The multiplicative combination of corrections results in $K$-factors varying between 1.3 and 1.4 (with a similar
pattern) for all polarized signals but the $\rT\rL$ one, which varies from 1.3 in the central region to 1.55
in the most forward and backward regions.
All combined distributions have a maximum at zero rapidity, and are characterized by a negative convexity
in the whole available range, with very small shape differences among different polarization states.
This shows that the rapidity of a single $\PZ$~boson is hardly sensitive to the polarization of the decayed boson,
as it has been already found in $\PW\PZ$ production with polarizations
defined in the di-boson CM frame \cite{Denner:2020eck}.

Finally, we consider distributions in the absolute value of the
rapidity separation between the two $\PZ$~bosons in \reffi{fig:dyzz}.
The momenta of the $\PZ$~bosons are determined from the momenta of
their decay products after possible photon recombination.
Owing to the rapidity cuts on all charged leptons, the rapidity distance
between the two opposite-flavour lepton pairs is implicitly cut such that
$|\Delta y_{\PZ\PZ}|\lesssim 5$, but already at $|\Delta y_{\PZ\PZ}|\approx 4$
all cross-sections are more than 3 orders of magnitude lower
than those at the maximum of the distributions.
The region of small rapidity separation is the most populated one for all polarized
and unpolarized signals. In this configuration, the events are characterized by slightly
larger non-doubly-resonant effects (2\%) than those found at the integrated level, and by
almost vanishing interferences, both at LO and at NLO.
For larger rapidity separations, the non-doubly-resonant background becomes smaller
than 0.5\% of the full result, while interferences increase up to 3\%. 
The NLO EW corrections, despite the different patterns for various polarization modes,
induce only mild shape distortions of the distributions with at most 3\% variations in the
considered range for the $\rL\rL$ and mixed polarization states. Even flatter EW corrections
are found for the purely-transverse and for the unpolarized distributions.
At variance with the EW effects, the NLO QCD corrections sizeably increase from small
to large rapidity separations for the $\rL\rL$ and mixed polarization
states reaching $120\%$ and $180\%$, respectively, at $|\Delta y_{\PZ\PZ}|=3$. A
$50\%$ enhancement results for the $\rL\rL$
state at zero rapidity separation, due to the fact that this polarization state is
strongly suppressed in such configuration at LO. The QCD corrections for the purely-transverse
state turn out to be relatively flat.  The loop-induced process enhances
the suppressed LO cross-section with both longitudinally-polarized
bosons by 150\% at  $|\Delta y_{\PZ\PZ}|\approx 0$ and causes  effects
of up to $50\%$ 
for large rapidity separations for the same polarization mode. 
About 20\% loop-induced contributions
characterize the $\rT\rT$ signal at small rapidity separation.
The normalized $\rT\rT$ distribution has a flat maximum at $|\Delta y_{\PZ\PZ}|\approx 0.2$,
while mixed contributions have pronounced maxima at zero rapidity separation. The $\rL\rL$ distribution
is maximal at $|\Delta y_{\PZ\PZ}|\approx 0.5$ with a shape that is noticeably different
from the other three polarization states. Therefore, using this observable in a fit to LHC data
is expected to improve the capability of isolating the purely-longitudinal signal from the other ones.

\section{Conclusion}\label{sub:con}

Extending the methods of \citeres{Denner:2020bcz,Denner:2020eck} to
NLO EW corrections and in particular to the emission of photons off
the decay products of $\PZ$~bosons, we have presented a general
procedure to compute cross-sections for the production of $\PZ$~bosons
in definite polarization states valid up to NLO accuracy both in the
EW and in the QCD coupling. The definition of polarized signals relies
on the application of a double-pole approximation (DPA) to
contributions with two resonant $\PZ$~bosons and on the separation of
polarization states at the amplitude level.
While the method is applicable to any LHC process with neutral resonances, 
even including identical particles in the final state,
we have focused on inclusive $\PZ$-pair production in the 
different-flavour
four-charged-lepton
channel at complete NLO accuracy, including
also gluon-initiated, loop-induced contributions.

We have first applied the newly-introduced DPA techniques to the
unpolarized process at NLO EW for very inclusive selection cuts,
finding a small contribution ($\approx 1\%$) of the
non-doubly-resonant background to the full computation. Beyond its
phenomenological importance, this result provides a validation of
the DPA method for complete NLO corrections, and in particular of the
well-behaved definition of subtraction counterterms
that are needed for the cancellation of IR singularities in the
DPA calculation.

In the same inclusive setup, the separation of polarizations for both $\PZ$~bosons has been investigated,
comparing the results of two different definitions of polarization vectors, one in
the di-boson centre-of-mass (CM) frame and one in the laboratory (LAB) frame.
Despite the absence of cuts on single leptons, non-negligible
interference patterns are found in some differential distributions both at LO
and at NLO EW. Sizeable differences show up between the CM- and LAB-frame
definitions of polarizations, but smaller interference effects are
present when defining the polarizations in the 
CM frame, which is also more natural for $\PZ\PZ$ production.

In the presence of realistic fiducial cuts, inspired by the
latest ATLAS analysis, we have combined NLO EW and QCD corrections and gluon-induced contributions (formally of NNLO QCD accuracy)
both in a multiplicative and in an additive way. While the
multiplicative combination is better motivated from the theory point
of view in general, it introduces artificial effects in regions of
phase space where the purely-longitudinal signal is suppressed at LO.

Scale uncertainties are at the 5\% level in the combined NLO predictions both for unpolarized
and polarized signals. A substantial fraction of the NLO scale uncertainties is due
to the gluon-induced process, in particular for the polarization modes
with equally polarized Z~bosons (transverse--transverse and
longitudinal--longitudinal).

In spite of sizeable and negative EW corrections of about $-10\%$, the largest radiative effects
originate from NLO QCD corrections which cause an increase of the LO cross-section of about 35\%.
Also the gluon-induced process contributes noticeably to $\PZ\PZ$ production ($\approx 15$--$20\%$
for unpolarized  $\PZ\PZ$ pairs and those with equal polarizations).
Polarization fractions are roughly conserved going from LO to NLO accuracy, but sizeable
distortions are found at the differential level with rather different $K$-factors for
the various polarization states.

At variance with di-boson processes with neutrinos as decay products, the possibility to fully
reconstruct the final state allows for access to a large set of observables.
We have presented a number of differential results for fiducial event selections,
focusing on those variables that are particularly suited for the separation of polarized signals.
The non-doubly-resonant background (0.5\% at the integrated level) never exceeds 3\% of the full result
for all presented differential distributions.
Interferences, which are at the 1\% level in the integrated cross-section, can reach
up to $5\%$  in more exclusive phase-space regions.
Several angular variables, including decay angles as well as azimuthal and
rapidity separations are well suited for polarization discrimination. However, also some
invariant-mass distributions show a fair discrimination power among different polarization states.
A combination of the variables considered in this paper, either in a multivariate analysis or
as input to machine-learning techniques, is expected to allow for the separation of polarized states
in the LHC data despite the small rate of the unpolarized process and the strongly-suppressed
cross-section for longitudinal bosons.

\section*{Acknowledgements}
We thank Jean-Nicolas Lang for supporting \recola and Mathieu Pellen
for help with \mocanlo. GP thanks Alessandro Ballestrero and Ezio Maina
for useful discussion.
The authors acknowledge financial support by the German
Federal Ministry for Education and Research (BMBF) under contract
no.~05H18WWCA1.

\bibliographystyle{JHEPmod}
\bibliography{polvv}

\end{document}